\documentclass[preprint]{article}
\usepackage{authblk}
\usepackage{graphicx}
\usepackage{dcolumn}
\usepackage{bm}
\usepackage{amsmath}
\usepackage{amssymb}
\usepackage{multirow}
\usepackage{caption}
\usepackage{subcaption}
\usepackage{epstopdf}
\usepackage{hyperref}

\title{\bf Constraining model parameters in $f(Q,C)$ gravity: Observational analysis and geometric diagnostics}
\author[1]{Amit Samaddar\thanks{samaddaramit4@gmail.com}}
\author[2]{S. Surendra Singh\thanks{ssuren.mu@gmail.com}}
\affil[1,2]{Department of Mathematics, National Institute of Technology Manipur, Imphal-795004,India.}
\begin{document}
\maketitle

\textbf{Abstract}: We investigate the cosmological implications of $f(Q,C)$ gravity with $f(Q,C)=\alpha Q+\beta C$, where $Q$ is the non-metricity scalar and $C$ encapsulates cosmological expansion terms. Three parameterizations of the EoS for dark energy, $\omega=\omega_{0}+\omega_{1}z$, $\omega=\omega_{0}+\frac{\omega_{1}z(1+z)}{1+z^{2}}$ and $\omega=\omega_{0}+\frac{\omega_{1}z^{2}}{1+z^{2}}$ are tested using the Hubble, Hubble+BAO, and Hubble+BAO+Pantheon datasets to constrain model parameters. The resulting Hubble and deceleration parameters reveal a transition from deceleration to acceleration, supporting current cosmic acceleration observations. Analysis of the energy density and pressure confirms positive energy density and a negative pressure for dark energy, potentially driving the late-time acceleration. We examine energy conditions, showing compliance with NEC, WEC and DEC, while SEC remains negative, supporting an accelerated expansion. Statefinder diagnostics suggest that two of the EoS parameterizations lead to Quintessence-like behavior with a time-varying dark energy component, while the third closely approaches $\Lambda$CDM showing slight deviations consistent with recent observations. Sound speed analysis demonstrates the physical stability of all parameterizations. Overall, our findings support $f(Q,C)$ gravity as a viable framework for describing diverse dark energy dynamics, providing insights into the Universe’s accelerated expansion. \\


\section{Introduction}\label{sec1}
\hspace{0.6cm} The accelerating expansion of the Universe stands as one of the most profound discoveries in modern cosmology, reshaping our understanding of cosmic dynamics. Observational evidence from Type Ia Supernovae (SNeIa), Cosmic Microwave Background (CMB) radiation and Baryon Acoustic Oscillations (BAO) \cite{Misner73,Clifton12,Riess19} has consistently pointed to this acceleration, suggesting that a form of energy with strong negative pressure – commonly referred to as dark energy (DE) – dominates the Universe’s energy budget. Recent CMB data suggest that dark energy constitutes nearly $76\%$ of the total energy density of the Universe, exerting a repulsive force that drives accelerated expansion despite the gravitational pull of ordinary matter. Within the context of General Relativity (GR), dark energy is often represented by the cosmological constant ($\Lambda$), a simple solution with an equation of state parameter $\omega=-1$ that provides a constant energy density throughout space and time \cite{Komatsu11}. However, while $\Lambda$-based models are effective at describing current observations, they leave critical theoretical questions unanswered, including the ``cosmological constant problem," where the predicted quantum field theoretical value of $\Lambda$ is vastly larger than observed, necessitating extreme fine-tuning. This has led to the emergence of the ``coincidence problem," which questions why the effects of dark energy and matter are comparable at the present cosmic epoch \cite{Copeland06}.

These challenges have driven considerable interest in exploring modified gravity theories, which offer an alternative path to understanding cosmic acceleration by altering the underlying gravitational dynamics rather than introducing exotic energy components. In modified gravity, the acceleration of the Universe may arise from extensions or modifications to the Einstein-Hilbert action, suggesting that gravity itself could behave differently on cosmological scales. Models such as $f(R)$ gravity \cite{Nojiri11} generalize the Einstein-Hilbert action by allowing the gravitational Lagrangian to be a function of the Ricci scalar, $R$. Other models, like $f(G)$ gravity \cite{Odintsov07}, incorporate the Gauss-Bonnet invariant $G$ which introduces higher-order curvature terms, while theories such as Lovelock gravity \cite{T13}, scalar-tensor theories \cite{Luca99} and braneworld scenarios \cite{Gia00} extend GR in various other dimensions. Each of these approaches seeks to provide a framework that not only accounts for the accelerated expansion but also addresses some of the theoretical issues arising from $\Lambda$-based models. The latest developments in modified gravity theories have led to a significant expansion of our understanding of gravity, introducing new geometric concepts such as torsion and non-metricity. These innovative ideas enable us to explore gravitational effects beyond the traditional curvature of spacetime, paving the way for alternative geometric frameworks. In symmetric teleparallel gravity, the non-metricity scalar $Q$ takes center stage, capturing the essence of distance preservation failures during parallel transport. This breakthrough leads to the $f(Q)$ gravity framework, which revolutionizes our understanding of gravitational interactions by prioritizing non-metricity over curvature and torsion \cite{J99, Jimenez18}. This novel approach offers a fresh perspective on cosmic evolution, inviting us to reexamine our understanding of the Universe's dynamics and the enigmatic dark energy phenomenon.

The $f(Q,C)$ gravity model is an extension of the $f(Q)$ gravity framework, which builds upon the non-metricity approach by incorporating a second scalar term $C$ that depends on both the Hubble parameter $H$ and its time derivative $\dot{H}$. The additional term $C$, defined as $C=6(3H^{2}+\dot{H})$ is related to the dynamical aspects of cosmic expansion, specifically the rate of change of the Hubble parameter over time \cite{S23, De52, V0}. This makes $f(Q,C)$ gravity distinct because it merges the geometrical property of non-metricity (from $Q$) with the evolution of cosmic expansion (through $C$). The unique combination of $Q$ and $C$  within the gravitational action means that $f(Q,C)$ gravity can effectively model both decelerating and accelerating phases of the Universe. In comparison to simpler models, it can inherently incorporate dynamic transitions in cosmic expansion, potentially explaining observations without the extreme fine-tuning required by the cosmological constant $\Lambda$. Furthermore, the explicit time-dependence in $C$ allows the model to align with observational data on cosmic expansion more flexibly, providing new insights into the interplay between geometry and expansion dynamics.

Parameterization is a crucial approach in cosmological models, especially for probing the underlying properties of dark energy and evaluating modified gravity theories like $f(Q,C)$ gravity. By parameterizing cosmological quantities, researchers gain a structured framework to study dark energy, the enigmatic driver behind the Universe’s accelerated expansion. This approach enables a detailed analysis of how dark energy influences cosmic dynamics across different epochs, thereby reconstructing the Universe’s expansion history and potentially revealing signatures unique to modified gravity models. In the $f(Q,C)$ gravity framework,parameterizing the equation of state (EoS) of dark energy is particularly advantageous for exploring the theory’s implications for cosmic acceleration. Unlike the static cosmological constant model, which relies on precise fine-tuning, $f(Q,C)$ gravity incorporates both non-metricity and dynamical terms that vary with cosmic time. This combination allows the model to naturally describe transitions between deceleration and acceleration phases in the Universe’s expansion. Explicitly parameterizing the EoS of dark energy within $f(Q,C)$ gravity provides a systematic method for comparing theoretical predictions with empirical data from multiple cosmological probes, such as Type Ia supernovae,CMB and BAO. This data-driven approach allows researchers to rigorously test the validity of $f(Q,C)$ gravity by determining whether it can reproduce observed cosmic acceleration trends without requiring unnatural fine-tuning. Various researchers have previously performed Equation of State (EoS) parameterization in distinct modified gravity theories, exploring the implications of different EoS forms within these alternative gravitational frameworks \cite{M24,Kou24,Arora21,J24,Garg24}. Motivated by these works, investigating EoS parameterization within $f(Q,C)$ gravity offers a promising avenue to understand whether this modified gravity theory can provide a more natural and fundamental explanation for the Universe’s accelerated expansion. Through this study, scientists can assess the viability of $f(Q,C)$ gravity as an alternative to the cosmological constant model, potentially uncovering new insights into the true nature of dark energy and the underlying forces shaping cosmic evolution.

The article is structured as follows: We commence by deriving the field equations in the context of $f(Q,C)$ gravity in Section \ref{sec2}. Section \ref{sec3} delves into the $f(Q,C)$ gravity model while Section \ref{sec4} provides a detailed description of the datasets utilized. Section \ref{sec5} focuses on the EoS parameterization and the constrained values of the parameters, followed by a discussion on the cosmological implications of these parameters, including the deceleration parameter, energy density, pressure, statefinder parameters and sound speed in Section \ref{sec6}. Ultimately, Section \ref{sec7} presents a summary of our findings and the conclusions drawn from this study.
\section{An overview of $f(Q,C)$ gravity}\label{sec2}
\hspace{0.6cm} The Levi-Civita connection $(\overset{\circ}{\Gamma^{\eta}}_{\mu\nu})$, is an affine connection uniquely characterized by being both torsion-free and compatible with the metric, meaning it preserves the inner product structure of the spacetime metric under parallel transport. In contrast, symmetric teleparallel gravity broadens this framework by introducing a different connection, $\Gamma^{\eta}_{\mu\nu}$, which does not require metric compatibility. Here, the connection is still symmetric with respect to its lower indices, hence ``symmetric,” and remains torsion-free but is not constrained to maintain the metric structure, leading to the appearance of the non-metricity tensor $Q_{\alpha\mu\nu}$. This tensor, defined by
\begin{equation}\label{1}
Q_{\alpha\mu\nu}=\nabla_{\alpha}g_{\mu\nu}=\partial_{\alpha}g_{\mu\nu}-\Gamma^{\beta}_{\alpha\mu}g_{\beta\nu}
 \Gamma^{\beta}_{\alpha\nu}g_{\beta\mu}\neq 0.
\end{equation}
measures the failure of the connection to maintain metric compatibility. Here, $\Gamma^{\alpha}_{\mu\nu}$ can be decomposed as $\Gamma^{\alpha}_{\mu\nu}=\overset{\circ}{\Gamma^{\alpha}}_{\mu\nu}+L^{\alpha}_{\mu\nu}$, where $\Gamma^{\alpha}_{\mu\nu}$ encapsulates the deviation from the Levi-Civita connection. This implies that the tensor $\Gamma^{\alpha}_{\mu\nu}$ can be expressed as
\begin{equation}\label{2}
 L^{\alpha}_{\mu\nu}=\frac{1}{2}(Q^{\alpha}_{\mu\nu}-Q_{\mu\;\;\nu}^{\;\;\alpha}-Q_{\nu\;\;\mu}^{\;\;\alpha}).
 \end{equation}
Two different representations of non-metricity vectors can be derived:
\begin{equation}\label{3}
 Q_{\mu}=g^{\nu\alpha}Q_{\mu\nu\alpha}=Q_{\mu\;\;\nu}^{\;\;\nu},\hspace{0.5cm} \tilde{Q}_{\mu}=g^{\nu\alpha}Q_{\nu\mu\alpha}=Q_{\nu\mu}^{\;\;\;\;\nu}.
 \end{equation}
 The superpotential tensor $P^{\alpha}_{\mu\nu}$ is a mathematical construction that captures the essence of non-metricity in a connection, defined as:
 \begin{equation}\label{4}
 P^{\alpha}_{\mu\nu}=\frac{1}{4}\bigg[-2L^{\alpha}_{\mu\nu}+Q^{\alpha}g_{\mu\nu}-\tilde{Q}^{\alpha}g_{\mu\nu}
 -\delta^{\alpha}_{\mu} Q_{\nu}\bigg].
 \end{equation}
Assuming that both torsion and curvature vanish, specific field equations emerge. These equations incorporate the Ricci tensor $\overset{\circ}R_{\mu\nu}$, the divergence of the distortion tensor, and terms involving products of the distortion tensor itself. Additionally, a second equation links the Ricci scalar, the divergence of the non-metricity vectors and the non-metricity scalar $Q$, establishing a relationship that defines the underlying geometry of the theory. Consequently, these equations are derived from an alternative interpretation of the connection and geometry in the absence of torsion and curvature as follows:
\begin{equation}\label{5}
 \overset{\circ}R_{\mu\nu}+\overset{\circ}\nabla_{\eta}L^{\eta}_{\mu\nu}-\overset{\circ}\nabla_{\nu}\tilde{L}_{\mu}
 +\tilde{L}_{\eta}LL^{\eta}_{\mu\nu}-L_{\eta\beta\nu}L^{\beta\eta}_{\mu}=0,
 \end{equation}
 and
 \begin{equation}\label{6}
 \overset{\circ}R+\overset{\circ}\nabla_{\eta}(L^{\eta}-\tilde{L}^{\eta})-Q=0.
 \end{equation} 
Using the connections that were already established, the boundary term $C$ can be interpreted as follows: 
\begin{equation}\label{7}
 C=\overset{\circ}R-Q=-\overset{\circ}\nabla_{\eta}(Q^{\eta}-\tilde{Q}^{\eta}).
 \end{equation}
Here, the expression $Q^{\eta}-\tilde{Q}^{\eta}==L^{\eta}-\tilde{L}^{\eta}$ represents the difference between the non-metricity vectors. This formulation indicates that the boundary term $C$ not only captures the discrepancy between the Ricci scalar $\overset{\circ}R$ and the non-metricity scalar $Q$, but it also emphasizes how these quantities relate through the divergence of the difference between the non-metricity vectors. Essentially, this reveals a connection between the geometric properties of the underlying space and the behavior of the non-metricity vectors, reflecting how deviations in non-metricity contribute to the overall curvature of the theory.

In the framework of $f(Q,C)$ gravity, the gravitational action is expressed as follows:
\begin{equation}\label{8}
 S=\; \int \bigg(\frac{1}{2k}f(Q,C)+\mathcal{L}_{m}\bigg)\sqrt{-g}d^{4}x,
 \end{equation}
where the function $f(Q,C)$ encapsulates the interaction between gravitational dynamics and matter fields. It delineates how the non-metricity scalar $Q$ and the boundary term $C$ influence the behavior of the matter fields described by the matter Lagrangian $\mathcal{L}_{m}$. By performing a variation of the action with respect to the metric, the resulting field equations can be derived:
\begin{align}\label{9}
&\kappa T_{\mu\nu}=-\frac{f}{2}g_{\mu\nu}+\frac{2}{\sqrt{-g}}\partial_{\alpha}\bigg(\sqrt{-g}f_{Q}P^{\alpha}_{\mu\nu}\bigg)
+\bigg(P_{\mu\eta\beta}Q_{\nu}^{\eta\beta}-2P_{\eta\beta\nu}Q^{\eta\beta}_{\mu}\bigg)f_{Q}\\\nonumber
&+\bigg(\frac{C}{2}g_{\mu\nu}-\overset{\circ}\nabla_{\mu}\overset{\circ}{\nabla_{\nu}}+g_{\mu\nu}\overset{\circ}{\nabla^{\eta}}\overset{\circ}\nabla_{\eta}-2P^{\alpha}_{\mu\nu}\partial_{\alpha}\bigg)f_{C},
 \end{align}
This formulation reveals that the left-hand side, representing the energy-momentum tensor $T_{\mu\nu}$, is related to the modified gravitational dynamics on the right-hand side, which incorporates contributions from both the function $f(Q,C)$ and its derivatives $f_{Q}$ and $f_{C}$. The term $-\frac{f}{2}g_{\mu\nu}$ indicates how the modified gravitational effects are scaled by the metric, while the terms involving the non-metricity tensor and the boundary term $C$ highlight the impact of the geometric structure on matter fields. This highlights a deeper relationship between the geometric properties of spacetime, encoded in $f(Q,C)$ and the physical behavior of matter, establishing a comprehensive framework for understanding gravity's influence in this theoretical context. The covariant form of the gravitational field equation is given by:
\begin{align}\label{10}
&\kappa T_{\mu\nu}=-\frac{f}{2}g_{\mu\nu}+2P^{\alpha}_{\mu\nu}\nabla_{\alpha}(f_{Q}-f_{C})+\bigg(\overset\circ G_{\mu\nu}+\frac{Q}{2}g_{\mu\nu}\bigg)f_{Q}\\\nonumber
&+\bigg(\frac{C}{2}g_{\mu\nu}-\overset{\circ}\nabla_{\mu}\overset{\circ}\nabla_{\nu}+g_{\mu\nu}\overset{\circ}{\nabla^{\eta}}\overset{\circ}\nabla_{\eta}\bigg)f_{C}.
\end{align}
This expression explains the energy-momentum tensor $T_{\mu\nu}$ interacts with the geometric structure defined by the function $f(Q,C)$. The right-hand side includes various contributions that account for modifications to the standard gravitational dynamics introduced by the non-metricity scalar $Q$ and the boundary term $C$. This yields the effective energy-momentum tensor:
\begin{align}\label{11}
&T_{\mu\nu}^{eff}=T_{\mu\nu}+\frac{1}{k}\bigg(\frac{f}{2}g_{\mu\nu}-2P^{\alpha}_{\mu\nu}\nabla_{\alpha}(f_{Q}-f_{C})
-\frac{Qf_{Q}}{2}g_{\mu\nu}-\bigg[\frac{C}{2}g_{\mu\nu}-\overset{\circ}\nabla_{\mu}\overset{\circ}\nabla_{\nu}+g_{\mu\nu}\overset{\circ}{\nabla^{\eta}}\overset{\circ}\nabla_{\eta}\bigg]f_{C}\bigg).
\end{align}
This equation encapsulates the modifications to the energy-momentum tensor that arise due to the presence of the $f(Q,C)$ framework, incorporating terms that involve the geometric characteristics of the spacetime, such as the distortion tensor $P^{\alpha}_{\mu\nu}$ and the divergences associated with $f_{Q}$ and $f_{C}$. 

We can derive an equation that mirrors the structure of General Relativity (GR) using the previously established field equation:
\begin{equation}\label{12}
\overset{\circ}G_{\mu\nu}=\frac{k}{f_{Q}}T_{\mu\nu}^{eff}.
\end{equation} 
For a perfect fluid, the energy-momentum tensor is formulated as:
\begin{equation}\label{13}
T_{\mu\nu}=pg_{\mu\nu}+(\rho+p)u_{\mu}u_{\nu}.
\end{equation}
In this expression, the term $(\rho+p)u_{\mu}u_{\nu}$ represents the energy and momentum carried by the fluid's motion, where $u_{\mu}$ is the four-velocity of the fluid. The term $pg_{\mu\nu}$ describes the isotropic pressure exerted by the fluid, with $p$ being the pressure and $g_{\mu\nu}$ representing the metric tensor. This formulation highlights how the energy density $\rho$ and pressure $p$ of the fluid contribute to the overall dynamics governed by $f(Q,C)$ gravity, reflecting the interplay between matter and the geometry of spacetime.

In this framework, the Universe is modeled as a spatially flat, homogeneous, and isotropic entity characterized by the Friedmann–Robertson–Walker (FRW) metric:
\begin{equation}\label{14}
  ds^{2}=-dt^{2}+a^{2}(t)(dx^{2}+dy^{2}+dz^{2}),
\end{equation}
where $a(t)$ is the scale factor dictating how the Universe expands over time, while $dt$ represents the infinitesimal time interval and $dx, dy$ and $dz$ represent the spatial intervals in the respective directions. To simplify calculations, we assume a ``flat" connection, where all affine connection coefficients are zero ($\Gamma^{\eta}_{\mu\nu}=0$). This leads to important results regarding the geometric quantities of the Universe, given by:
\begin{equation}\label{15}
\overset{\circ}R=6(2H^{2}+\dot{H}), \hspace{0.7cm} Q=-6H^{2}, \hspace{0.7cm} C=6(3H^{2}+\dot{H}).
\end{equation}
where $H(t)$ is the Hubble parameter and $\dot{H}$ indicates the change of the Hubble parameter over time. From these quantities, the field equations for $f(Q,C)$ gravity can be derived, yielding expressions for the energy density $\rho$ and pressure $p$ of the Universe:
\begin{equation}\label{16}
\kappa\rho=\frac{f}{2}+6H^{2}f_{Q}-(9H^{2}+3\dot{H})f_{C}+3H\dot{f}_{C},
\end{equation}
\begin{equation}\label{17}
\kappa p=-\frac{f}{2}-(6H^{2}+2\dot{H})f_{Q}-2H\dot{f}_{Q}+(9H^{2}+3\dot{H})f_{C}-\ddot{f}_{C}.
\end{equation}
Overall, these field equations in $f(Q,C)$ gravity demonstrate how the interplay between non-metricity, curvature and the expansion history can significantly alter the dynamics governing the Universe's evolution, thereby offering a framework for exploring cosmological scenarios beyond standard GR.
\section{$f(Q,C)$ model}\label{sec3}
\hspace{0.6cm} We consider a linear form of the function $f(Q,C)$ as follows:
\begin{equation}\label{18}
f(Q,C)=\alpha Q+\beta C,
\end{equation}
where the constants $\alpha$ and $\beta$ can be interpreted as characterizing the sensitivity of the gravitational field to changes in the non-metricity scalar and the boundary term. For instance, a larger $\alpha$ might suggest that the dynamics of the Universe are more significantly influenced by non-metricity. This assumption simplifies the analysis and allows us to explore the effects of $Q$ and $C$ on the energy density and pressure of the Universe. By varying $\alpha$ and $\beta$, one can explore different models that may yield insights into the current understanding of cosmic acceleration and the nature of dark energy. This linear form lends itself well to numerical simulations, enabling easier implementation in computational cosmology studies aimed at exploring the evolution of structure in the Universe. The choice of linearity can lead to specific cosmological scenarios, such as tracking the behavior of dark energy or modifications to cosmic expansion. For $\beta=0$, the equation (\ref{18}) reduces to $f(Q,C)=\alpha Q$, which is similar to $f(Q)$ gravity. By substituting this linear form in the equations (\ref{16}) and (\ref{17}), we obtain the expression of energy density and pressure as,
\begin{equation}\label{19}
\kappa\rho=3\alpha H^{2},
\end{equation}
\begin{equation}\label{20}
\kappa p=-3\alpha H^{2}-2\alpha\dot{H}.
\end{equation}
The EoS parameter's expression is derived by using the formula $(\omega=\frac{p}{\rho})$ in the following manner:
 \begin{equation}\label{21}
 \omega=-1-\frac{2}{3}\frac{\dot{H}}{H^{2}}.
 \end{equation}
 \section{Observational data}\label{sec4}
 \hspace{0.6cm} This section summarizes the most recent observational evidence and methodologies used to analyze it. The Cosmic Microwave Background (CMB), Baryon Acoustic Oscillations (BAO) and Type Ia Supernovae datasets are widely regarded as crucial tools for testing and refining cosmological theories, providing valuable insights into the Universe's origins, structure and destiny. We explore the Universe's evolution using a model-independent approach, leveraging $H(z)$ datasets to trace the expansion history. The ages of the oldest and most quiescent galaxies serve as cosmic clocks, enabling direct measurements of $H(z)$ across different redshifts. This method provides a robust standard probe for cosmology, allowing for an accurate reconstruction of the Universe's expansion history. This study employs a multi-faceted approach to constrain the parameters of our cosmological model, leveraging data from two primary methods for determining $H(z)$: radial BAO size and galaxy differential age (Cosmic Chronometer) \cite{BK03}. Additionally, we incorporate Supernovae observations and utilize cutting-edge statistical tools, including MCMC, Bayesian analysis and the emcee Python library, to analyze the data and gain a deeper understanding of the Universe's expansion history. We employ a comprehensive dataset, comprising Hubble Space Telescope observations, Type Ia Supernovae distance modulus data and additional datasets, to constrain our model's parameters and describe various cosmic epochs. By combining $46$ Hubble data points, $15$ BAO data points and $1408$ data points from the Pantheon, we create a robust and accurate picture of the Universe's evolution.
 
MCMC is a statistical approach that has become a cornerstone in cosmology, enabling researchers to navigate complex model parameter spaces and derive probability distributions for cosmological parameters. Especially useful when confronted with large parameter spaces and non-Gaussian or non-linear likelihood functions, MCMC generates a Markov chain that methodically explores the parameter space of a model, guided by a probability distribution. This chain consists of a series of parameter values, each generated from the previous one using transition rules influenced by a proposal distribution, which proposes new parameter values that are accepted or rejected based on their posterior probability, considering the data and prior probability distribution.
\subsection{Hubble data}\label{sec4.1}
\hspace{0.6cm} Cosmic chronometers, specifically elliptical galaxies, provide essential age data at various redshifts, helping unravel the Universe's expansion history. Despite their simplicity and early formation, they are challenging to observe and analyze, making their use an active research area \cite{Verde10}. To overcome this, astronomers employ advanced telescopes and spectrographs to precisely measure the spectra and colors of these galaxies, collecting critical data for cosmic chronometry. By applying cutting-edge statistical techniques, the ages of the galaxies and the Universe's expansion history are then determined, offering profound insights into the Universe's evolution \cite{Scol22}. The data from cosmic chronometers is vital for understanding the universe and the underlying physical laws. By utilizing the differential age (DA) method, researchers have analyzed $46$ Hubble data points within the redshift range $0\leq z \leq 2.36$, shedding light on the Universe's expansion history. This method enables the calculation of the expansion rate at a given redshift, allowing for the determination of $H(z)$ using the equation $H(z)$ $=$ -$\frac{1}{(1+z)}\frac{dz}{dt}$. Additionally, by optimizing the chi-square ($\chi^{2}$) for the Hubble datasets, the mean values of the model parameters can be determined, providing valuable insights into the Universe's expansion history and the fundamental laws of physics. The chi-square is defined as:
\begin{equation}\label{22}
\chi^{2}_{H}=\sum_{i=1}^{46}\frac{[H_{th}(p,z_{i})-H_{obs}(z_{i})]^{2}}{[\sigma_{H}(z_{i})]^{2}}.
\end{equation}
Here, $H_{th}(p,z_{i})$ represents the theoretical Hubble parameter predicted by the model at redshift $z_{i}$, $p$ mentions the set of model parameters being constrained, $H_{obs}$ represents the Hubble parameter's observed value, $\sigma_{H}^{2}$ represents the standard errors in the $H(z)$'s observed values. These data points, along with their references, are summarized in Table \ref{Tab:T1}.
\begin{table}
\centering
\caption{$46$ datasets of $H(z)$}
\begin{tabular}{||p{1.3cm}|p{1.3cm}|p{1.3cm}|p{0.6cm}||p{1.3cm}|p{1.6cm}|p{0.6cm}|p{1cm}||}
\hline
 $\hspace{0.4cm}z_{i}$ & $\hspace{0.2cm} H_{obs}$ & $\hspace{0.3cm}\sigma_{H}$ & Ref. & $\hspace{0.4cm}z_{i}$ & $\hspace{0.2cm}H_{obs}$ & $\hspace{0.2cm}\sigma_{H}$ & Ref. \\
\hline
$\hspace{0.4cm}0$ & $\hspace{0.2cm}67.77$ & $\hspace{0.2cm}1.30$ & \cite{Nichol19} & $\hspace{0.2cm}0.4783$ & $\hspace{0.2cm}80.9$ & $\hspace{0.2cm}9$ & \cite{Moresco16}\\
\hline
$\hspace{0.2cm}0.07$ & $\hspace{0.4cm}69$ & $\hspace{0.2cm} 19.6$ & \cite{Zhang14} & $\hspace{0.2cm}0.48$ & $\hspace{0.2cm}97$ & $\hspace{0.2cm} 60$ & \cite{Stern10} \\
\hline
$\hspace{0.2cm}0.09$ & $\hspace{0.4cm}69$ & $\hspace{0.3cm} 12$ & \cite{Simon05} & $\hspace{0.2cm}0.51$ & $\hspace{0.2cm}90.4$ & $\hspace{0.2cm} 1.9$ & \cite{Alam16} \\
\hline
$\hspace{0.2cm}0.01$ & $\hspace{0.4cm}69$ & $\hspace{0.3cm} 12$ & \cite{Stern10} & $\hspace{0.2cm}0.57$ & $\hspace{0.2cm}97$ & $\hspace{0.2cm} 3.4$ & \cite{Cimatti12} \\
\hline
$\hspace{0.2cm}0.12$ & $\hspace{0.4cm}68.6$ & $\hspace{0.2cm} 26.2$ & \cite{Zhang14} & $\hspace{0.2cm}0.59$ & $\hspace{0.2cm}104$ & $\hspace{0.2cm} 13$ & \cite{Alam16}\\
\hline
$\hspace{0.2cm}0.17$ & $\hspace{0.4cm}83$ & $\hspace{0.4cm} 8$ & \cite{Stern10} & $\hspace{0.2cm}0.60$ & $\hspace{0.2cm}87.6$ & $\hspace{0.2cm} 6.1$ & \cite{Blake12}\\
\hline
$\hspace{0.2cm}0.179$ & $\hspace{0.4cm}75$ & $\hspace{0.4cm} 4$ & \cite{Cimatti12} & $\hspace{0.2cm}0.61$ & $\hspace{0.2cm}97.3$ & $\hspace{0.2cm} 2.1$ & \cite{Alam16} \\
\hline
$\hspace{0.2cm}0.1993$ & $\hspace{0.4cm}75$ & $\hspace{0.4cm} 5$ & \cite{Cimatti12} & $\hspace{0.2cm}0.68$ & $\hspace{0.2cm}92$ & $\hspace{0.2cm} 8$ & \cite{Cimatti12} \\
\hline
$\hspace{0.2cm}0.20$ & $\hspace{0.4cm}72.9$ & $\hspace{0.2cm} 29.6$ & \cite{Zhang14} & $\hspace{0.2cm}0.73$ & $\hspace{0.2cm}97.3$ & $\hspace{0.2cm} 7$ & \cite{Blake12} \\
\hline
$\hspace{0.2cm}0.24$ & $\hspace{0.4cm}79.7$ & $\hspace{0.2cm} 2.7$ & \cite{Cabre09} & $\hspace{0.2cm}0.781$ & $\hspace{0.2cm}105$ & $\hspace{0.2cm} 12$ & \cite{Cimatti12} \\
\hline
$\hspace{0.2cm}0.27$ & $\hspace{0.4cm}77$ & $\hspace{0.2cm} 14$ & \cite{Stern10} & $\hspace{0.2cm}0.875$ & $\hspace{0.2cm}125$ & $\hspace{0.2cm} 17$ & \cite{Cimatti12} \\
\hline
$\hspace{0.2cm}0.28$ & $\hspace{0.4cm}88.8$ & $\hspace{0.2cm} 36.6$ & \cite{Zhang14} & $\hspace{0.2cm}0.881$ & $\hspace{0.2cm}90$ & $\hspace{0.2cm} 40$ & \cite{Stern10} \\
\hline
$\hspace{0.2cm}0.35$ & $\hspace{0.4cm}82.7$ & $\hspace{0.2cm} 8.4$ & \cite{Wang13} & $\hspace{0.2cm}0.9$ & $\hspace{0.2cm}117$ & $\hspace{0.2cm} 23$ & \cite{Stern10} \\
\hline
$\hspace{0.2cm}0.352$ & $\hspace{0.4cm}83$ & $\hspace{0.2cm} 14$ & \cite{Cimatti12} & $\hspace{0.2cm}1.037$ & $\hspace{0.2cm}154$ & $\hspace{0.2cm} 20$ & \cite{Cimatti12} \\
\hline
$\hspace{0.2cm}0.38$ & $\hspace{0.4cm}81.5$ & $\hspace{0.2cm} 1.9$ & \cite{Alam16} & $\hspace{0.2cm}1.3$ & $\hspace{0.2cm}168$ & $\hspace{0.2cm} 17$ & \cite{Stern10} \\
\hline
$\hspace{0.2cm}0.3802$ & $\hspace{0.4cm}88.8$ & $\hspace{0.2cm} 36.6$ & \cite{Moresco16} & $\hspace{0.2cm}1.363$ & $\hspace{0.2cm}160$ & $\hspace{0.2cm} 33.6$ & \cite{Michele16} \\
\hline
$\hspace{0.2cm}0.4$ & $\hspace{0.4cm}95$ & $\hspace{0.2cm} 17$ & \cite{Simon05} & $\hspace{0.2cm}1.43$ & $\hspace{0.2cm}177$ & $\hspace{0.2cm} 18$ & \cite{Stern10} \\
\hline
$\hspace{0.2cm}0.4004$ & $\hspace{0.4cm}77$ & $\hspace{0.2cm} 10.2$ & \cite{Moresco16} & $\hspace{0.2cm}1.53$ & $\hspace{0.2cm}140$ & $\hspace{0.2cm} 14$ & \cite{Stern10} \\
\hline
$\hspace{0.2cm}0.4247$ & $\hspace{0.4cm}87.1$ & $\hspace{0.2cm} 11.2$ & \cite{Moresco16} & $\hspace{0.2cm}1.75$ & $\hspace{0.2cm}202$ & $\hspace{0.2cm} 40$ & \cite{Stern10} \\
\hline
$\hspace{0.2cm}0.43$ & $\hspace{0.4cm}86.5$ & $\hspace{0.2cm} 3.7$ & \cite{Cabre09} & $\hspace{0.2cm}1.965$ & $\hspace{0.2cm}186.5$ & $\hspace{0.2cm} 50.4$ & \cite{Michele16} \\
\hline
$\hspace{0.2cm}0.44$ & $\hspace{0.4cm}82.6$ & $\hspace{0.2cm} 7.8$ & \cite{Blake12} & $\hspace{0.2cm}2.3$ & $\hspace{0.2cm}224$ & $\hspace{0.2cm} 8$ & \cite{Rich13} \\
\hline
$\hspace{0.2cm}0.44497$ & $\hspace{0.4cm}92.8$ & $\hspace{0.2cm} 12.9$ & \cite{Moresco16} & $\hspace{0.2cm}2.34$ & $\hspace{0.2cm}222$ & $\hspace{0.2cm} 7$ & \cite{Delubac15} \\
\hline
$\hspace{0.2cm}0.47$ & $\hspace{0.4cm}89$ & $\hspace{0.2cm} 49.6$ & \cite{Cress17} & $\hspace{0.2cm}2.36$ & $\hspace{0.2cm}226$ & $\hspace{0.2cm} 8$ & \cite{Andreu14} \\
\hline
\end{tabular}
\label{Tab:T1}
\end{table}
\subsection{BAO datasets}\label{sec4.2}
\hspace{0.6cm} To probe oscillations in the early Universe caused by cosmic perturbations, we study a fluid of photons, baryons and dark matter that closely interact through Thomson scattering. We leverage data from three main sources - 6dFGS, BOSS and high-resolution SDSS - to measure Baryon Acoustic Oscillations (BAO) \cite{W10}. In order to ensure accuracy and avoid potential errors stemming from data correlations, we selectively use $15$ data points from a total of $333$ BAO data points. This cautious approach enables us to limit the impact of data correlations on our analysis. From transverse BAO studies and the comoving angular diameter distance, we calculate the values for $\frac{D_{H}(z)}{r_{d}}=\frac{c}{H(z)r_{d}}$ as follows:
\begin{equation}\label{23}
D_{M}=c\int_{0}^{z}\frac{dz'}{H(z')},
\end{equation}
where $D_{A}=\frac{D_{M}}{1+z}$ represents the angular diameter distance and $\frac{D_{V}(z)}{r_{d}}$ gives the ratio with $D_{A}$ and $D_{V}$ marking the BAO peak positions and $r_{d}$ is the sound horizon. The sound horizon $r_{d}$ is calculated via:
\begin{equation}\label{24}
D_{V}(z)\equiv \bigg(zD_{H}(z)D_{M}^{2}(z)\bigg )^{\frac{1}{3}}.
\end{equation} 
For this BAO analysis, the $\chi^{2}$ function is expressed as:
\begin{equation}\label{25}
\chi^{2}_{BAO}=\sum_{i=1}^{15}\bigg[\frac{D_{obs}-D_{th}(z_{i})}{\Delta D_{i}}\bigg]^{2}.
\end{equation}
with the redshift range $0.24<z<2.36$. Here, $D_{th}$ represents the theoretical distance modulus from our model, compared to the observed distance modulus $D_{obs}$ at redshift $z_{i}$ for $i$ $=$ $1$ to $n$. Table \ref{Tab:T2} lists these $15$ selected BAO data points and their respective sources.
\begin{table}
\centering
\caption{$15$ BAO datasets along with data from other measurement techniques.}
\begin{tabular}{||p{1.3cm}|p{1.3cm}|p{1.3cm}|p{0.6cm}||}
\hline
 $\hspace{0.5cm}z_{i}$ & $\hspace{0.3cm} H_{obs}$ & $\hspace{0.3cm}\sigma_{H}$ & Ref.\\
\hline\hline
$\hspace{0.4cm}0.30$ & $\hspace{0.2cm}81.77$ & $\hspace{0.2cm}6.22$ & \cite{Oka14}\\
\hline
$\hspace{0.4cm}0.31$ & $\hspace{0.2cm}78.18$ & $\hspace{0.2cm} 4.74$ & \cite{E09}\\
\hline
$\hspace{0.4cm}0.34$ & $\hspace{0.2cm}83.8$ & $\hspace{0.2cm} 3.66$ & \cite{E09}\\
\hline
$\hspace{0.4cm}0.36$ & $\hspace{0.2cm}79.94$ & $\hspace{0.2cm} 3.38$ & \cite{Y17} \\
\hline
$\hspace{0.4cm}0.40$ & $\hspace{0.2cm}82.04$ & $\hspace{0.2cm} 2.03$ & \cite{Y17}\\
\hline
$\hspace{0.4cm}0.43$ & $\hspace{0.2cm}86.45$ & $\hspace{0.2cm} 3.97$ & \cite{E09}\\
\hline
$\hspace{0.4cm}0.44$ & $\hspace{0.2cm}84.81$ & $\hspace{0.2cm} 1.83$ & \cite{Blake12}\\
\hline
$\hspace{0.4cm}0.48$ & $\hspace{0.2cm}87.79$ & $\hspace{0.2cm} 2.03$ & \cite{Blake12} \\
\hline
$\hspace{0.4cm}0.52$ & $\hspace{0.2cm}94.35$ & $\hspace{0.2cm} 2.64$ & \cite{Blake12} \\
\hline
$\hspace{0.4cm}0.56$ & $\hspace{0.2cm}93.34$ & $\hspace{0.2cm} 2.3$ & \cite{Blake12} \\
\hline
$\hspace{0.4cm}0.57$ & $\hspace{0.2cm}87.6$ & $\hspace{0.2cm} 7.8$ & \cite{C13} \\
\hline
$\hspace{0.4cm}0.59$ & $\hspace{0.2cm}98.48$ & $\hspace{0.2cm} 3.18$ & \cite{L14} \\
\hline
$\hspace{0.4cm}0.61$ & $\hspace{0.2cm}97.3$ & $\hspace{0.2cm} 2.1$ & \cite{The17} \\
\hline
$\hspace{0.4cm}0.64$ & $\hspace{0.2cm}98.82$ & $\hspace{0.2cm} 2.98$ & \cite{Busca13} \\
\hline
$\hspace{0.4cm}2.33$ & $\hspace{0.2cm}224$ & $\hspace{0.4cm} 8$ & \cite{The17} \\
\hline\hline
\end{tabular}
\label{Tab:T2}
\end{table}
\subsection{Combining Hubble and BAO Data for MCMC analysis}\label{sec4.3}
\hspace{0.6cm} To improve the accuracy and dependability of parameter estimation in the context of  $f(Q,C)$ gravity, we conduct a joint analysis of Hubble and BAO datasets through the MCMC approach. By merging these complementary datasets, we utilize the distinct advantages each provides—Hubble data offers direct measurements of the Universe's expansion rate across various epochs, while BAO data imposes essential constraints on the scale of cosmic structures. This combined strategy allows for more precise and reliable model parameter refinement. We perform the joint likelihood analysis by minimizing a total chi-square function, given by
\begin{equation}\label{26}
 \chi^{2}=\chi^{2}_{H}+\chi^{2}_{BAO}.
 \end{equation}
where $\chi^{2}_{H}$ and $\chi^{2}_{BAO}$ represent the chi-square contributions from the Hubble and BAO datasets, respectively. This unified approach allows us to capitalize on the distinct advantages of each observational dataset, integrating their complementary constraints to derive the most probable values for our model parameters and enhance the robustness of our findings.
\subsection{Pantheon dataset}\label{sec4.4}
\hspace{0.6cm} The Pantheon dataset provides $1048$ supernova observations at redshifts from $0.01$ to $2.26$, allowing researchers to explore the Universe's expansion history \cite{Ak20,Dm18}. Through meticulous analysis of luminosity and comoving distance measurements, the dataset provides a unique opportunity to test cosmological models and unravel the enigmatic forces propelling cosmic acceleration, ultimately illuminating the Universe's intricate evolution. By incorporating the Pantheon dataset into our analysis, we significantly enhance the predictive power of our $f(Q,C)$ gravity model, exploiting its broad coverage and precision to constrain model parameters and deepen our understanding of cosmological phenomena. Comoving distance, represented by $D_{C}(z)$, is a central concept in cosmology, quantifying the distance between two points in the Universe at a particular moment, while accounting for the expansion of space. The mathematical expression for comoving distance is:
\begin{equation}\label{27}
D_{C}(z)=c\int_{0}^{z}\frac{dz'}{H(z')},
\end{equation}  
where $c$ indicates the speed of light. This metric plays a significant role in understanding the expansion rate of the Universe, providing valuable insights into the interplay of different components such as dark matter and dark energy. In addition to comoving distance, another essential measure in cosmology is luminosity distance, represented by $D_{L}(z)$ which governs how bright distant objects appear to us. The mathematical relationship between comoving distance and luminosity distance is expressed as:
\begin{equation}\label{28}
D_{L}(z)=(1+z)D_{C}(z),
\end{equation}
This metric is essential for determining the intrinsic luminosity of supernovae. The observed distance modulus $\mu$ (which quantifies the difference between an object's apparent and intrinsic brightness) is related to the luminosity distance through the following equation: 
\begin{equation}\label{29}
\mu(z)=5\log_{10}\bigg(\frac{D_{L}}{H_{0}Mpc}\bigg)+25.
\end{equation}
To evaluate the consistency of our model with the Pantheon dataset, we formulate the chi-square statistic as follows:
\begin{equation}\label{30}
\chi^{2}_{Pantheon}=\sum_{i=1}^{1048}\bigg[\frac{\mu_{obs}(z_{i})-\mu_{th}(z_{i})}{\sigma_{\mu} (z_{i})}\bigg]^{2}.
\end{equation}
In this analysis,$\mu_{obs}(z_{i}$ denotes the observed distance modulus at redshift $z_{i}$, $\mu_{th}(z_{i})$ is the theoretical prediction from our model, and $\sigma_{\mu} (z_{i})$ represents the uncertainty in the observations. A comprehensive overview of the Phantom datasets is presented in Table \ref{Tab:T3} \cite{K22}. This study not only advances our understanding of cosmic expansion but also strengthens the validation of our $f(Q,C)$ gravity model against robust observational evidence.
\begin{table}[h!]
\centering
\caption{Pantheon dataset summary }
\begin{tabular}{||p{1.2cm}|p{5.0cm}|p{3.0cm}||}
\hline\hline
Samples & The quantity of SNeIa observations for every sample & Range of Redshifts \\
\hline\hline
CfA 1-4 & \hspace{2.4cm}$147$ & \hspace{0.5cm}$0.01-0.07$\\[1pt]
\hline
CSP & \hspace{2.5cm}$25$ & \hspace{0.4cm} $0.01-0.06$\\[1pt]
\hline
SDSS & \hspace{2.3cm} $335$ & \hspace{0.4cm} $0.03-0.40$\\[1pt]
\hline
SNLS & \hspace{2.3cm} $236$ & \hspace{0.4cm} $0.12-1.06$\\[1pt]
\hline
PS1 & \hspace{2.3cm} $279$ & \hspace{0.4cm} $0.02-0.63$\\[1pt]
\hline
high-z & \hspace{2.3cm} $26$ & \hspace{0.4cm} $0.73-2.26$\\[1pt]
\hline
Total & \hspace{2.2cm} $1048$ & \hspace{0.4cm} $0.01-2.26$\\[1pt]
\hline\hline
\end{tabular}
\label{Tab:T3}
\end{table}
\subsection{H(z)+BAO+Pantheon datasets for MCMC analysis}\label{sec4.4}
\hspace{0.6cm} To improve the accuracy of our parameter estimation and deepen our understanding of the $f(Q,C)$ gravity, we merge the Hubble, BAO and Pantheon datasets.  This integrated approach harnesses the unique strengths of each dataset, yielding complementary constraints that facilitate a more robust and dependable determination of model parameters. The chi-square function for this analysis is as follows:
\begin{equation}\label{31}
 \chi^{2}=\chi^{2}_{H}+\chi^{2}_{BAO}+\chi^{2}_{Pantheon}.
\end{equation}
Here, $\chi^{2}_{H}$ represents the impact made by the Hubble dataset, $\chi^{2}_{BAO}$ signified the involvement of the BAO dataset and $\chi^{2}_{Pantheon}$. denotes that from the Pantheon dataset.
\section{EoS parameter models and their parameterizations}\label{sec5}
\hspace{0.6cm} In cosmological modeling, the equation of state (EoS) parameter $\omega$ is crucial for characterizing the properties of dark energy, particularly in understanding how it influences the expansion dynamics of the Universe. The EoS parameter $\omega$ is defined as the ratio of dark energy pressure $p$ to its energy density $\rho$, given by $\omega=\frac{p}{\rho}$. This parameter dictates whether the dark energy density remains constant over time, as in the cosmological constant $(\omega=-1)$, or varies, as in more dynamic dark energy models. In our analysis of $f(Q,C)$ gravity model, we introduce three parameterized forms of the EoS to account for potential evolution in the dark energy component. By examining variations in $\omega$ across different epochs, we can assess whether a simple constant $\omega$ suffices or if an evolving dark energy model better describes the observed data. This parameterizations allows for exploration of how dark energy’s contribution to the cosmic acceleration might vary with redshift, offering a deeper understanding of the model’s viability and its implications for cosmic evolution.
\subsection{Model 1}\label{sec5.1}
\hspace{0.5cm} To incorporate an evolving dark energy component, we assume a linear parameterization of the EoS parameter, defined by \cite{Hut01,J02}:
\begin{equation}\label{32}
\omega=\omega_{0}+\omega_{1}z,
\end{equation}
where $\omega_{0}$ and $\omega_{1}$ are constants representing the present-day value of $\omega$ and its rate of change with redshift $z$, respectively. This form is a simple yet effective way to capture potential deviations from a cosmological constant $(\omega=-1)$ by allowing $\omega$ to evolve over time, reflecting possible dynamics in dark energy. The linear term $\omega_{1}z$ allows for a range of dark energy behaviors across cosmic epochs. If $\omega_{1}>0$, dark energy could evolve to behave less like a cosmological constant in the past, with a potential transition to phantom behavior $(\omega<-1)$ at low redshifts. Conversely, $\omega_{1}<0$, suggests a decreasing impact of dark energy over time, with its effects becoming less pronounced as the universe expands. The linear form streamlines the integration of the Friedmann equations, facilitating a more efficient and straightforward computational analysis of dark energy dynamics, while avoiding unnecessary complexity.

By inserting the above linear form into equation (\ref{21}) and using the formula $\frac{dH}{dt}=-H(z)(1+z)\frac{dH}{dz}$, we obtain the explicit form of $H(z)$ as
\begin{equation}\label{33}
H(z)=H_{0}(1+z)^{\frac{3}{2}(1+\omega_{0}-\omega_{1})}exp\bigg[\frac{3\omega_{1}z}{2}\bigg],
\end{equation}
Here, $H_{0}$ represents the present-day Hubble parameter. The power term $(1+z)^{\frac{3}{2}(1+\omega_{0}-\omega_{1})}$ reflects a scaling behavior of the Hubble parameter with redshift. This term becomes particularly significant at high redshifts (early times), where $(1+z)$ is large, amplifying or diminishing $H(z)$ based on the combination of $\omega_{0}$ and $\omega_{1}$. The exponential term $exp\bigg[\frac{3\omega_{1}z}{2}\bigg]$ introduces an additional modification that depends solely on $\omega_{1}$, specifically influencing how the Hubble parameter behaves at higher redshifts (larger $z$ values). For positive $\omega_{1}$, this term accelerates the growth of $H(z)$, while for negative $\omega_{1}$, it causes a suppression.

\begin{figure}[htbp]
    \centering
    \begin{subfigure}{0.45\textwidth}
        \includegraphics[width=\linewidth]{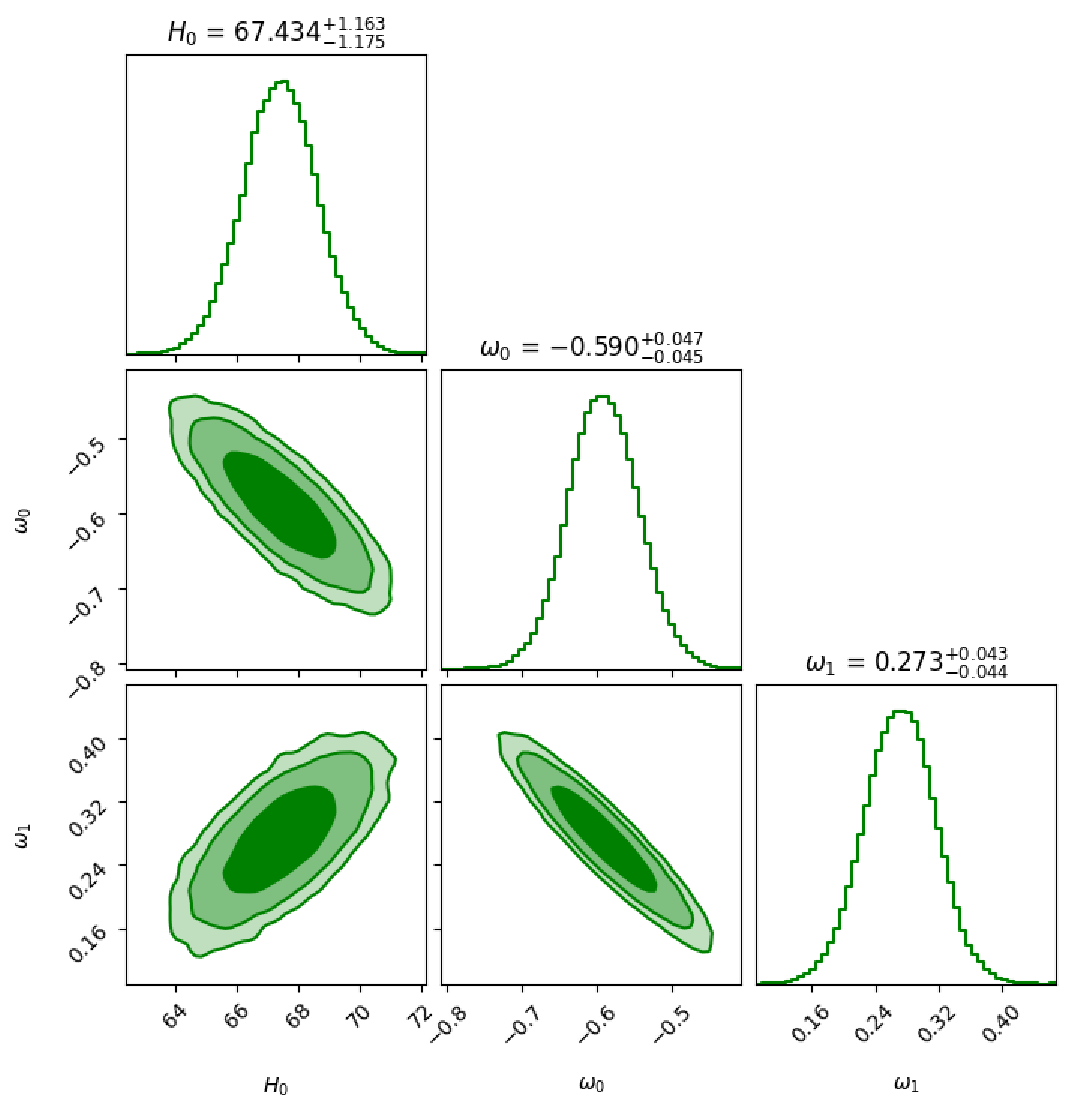} 
        \caption{}
        \label{fig:subfig1}
    \end{subfigure}
    \hfill
    \begin{subfigure}{0.45\textwidth}
        \includegraphics[width=\linewidth]{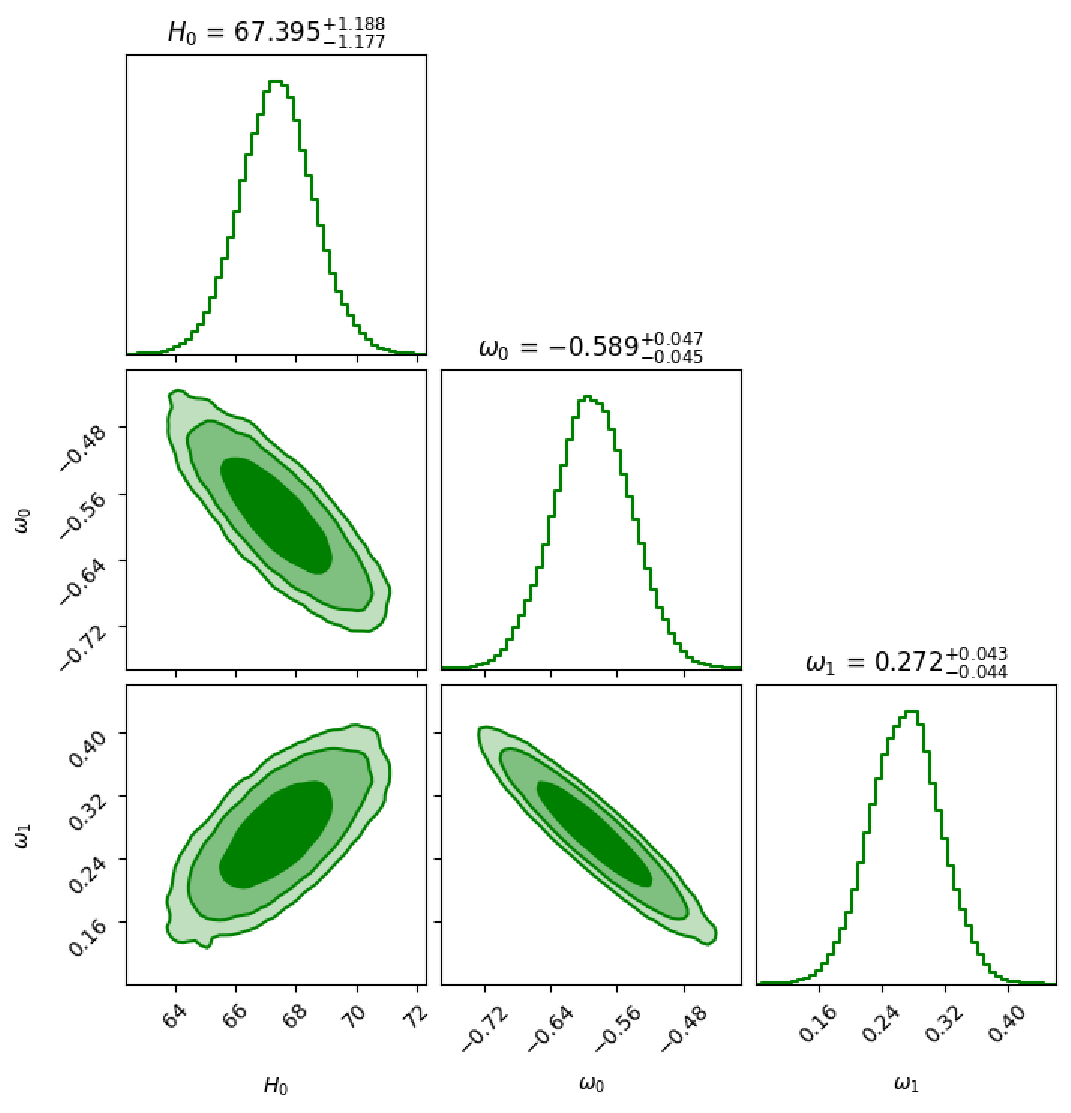} 
        \caption{}
        \label{fig:subfig2}
    \end{subfigure}

    \begin{subfigure}{0.45\textwidth}
        \includegraphics[width=\linewidth]{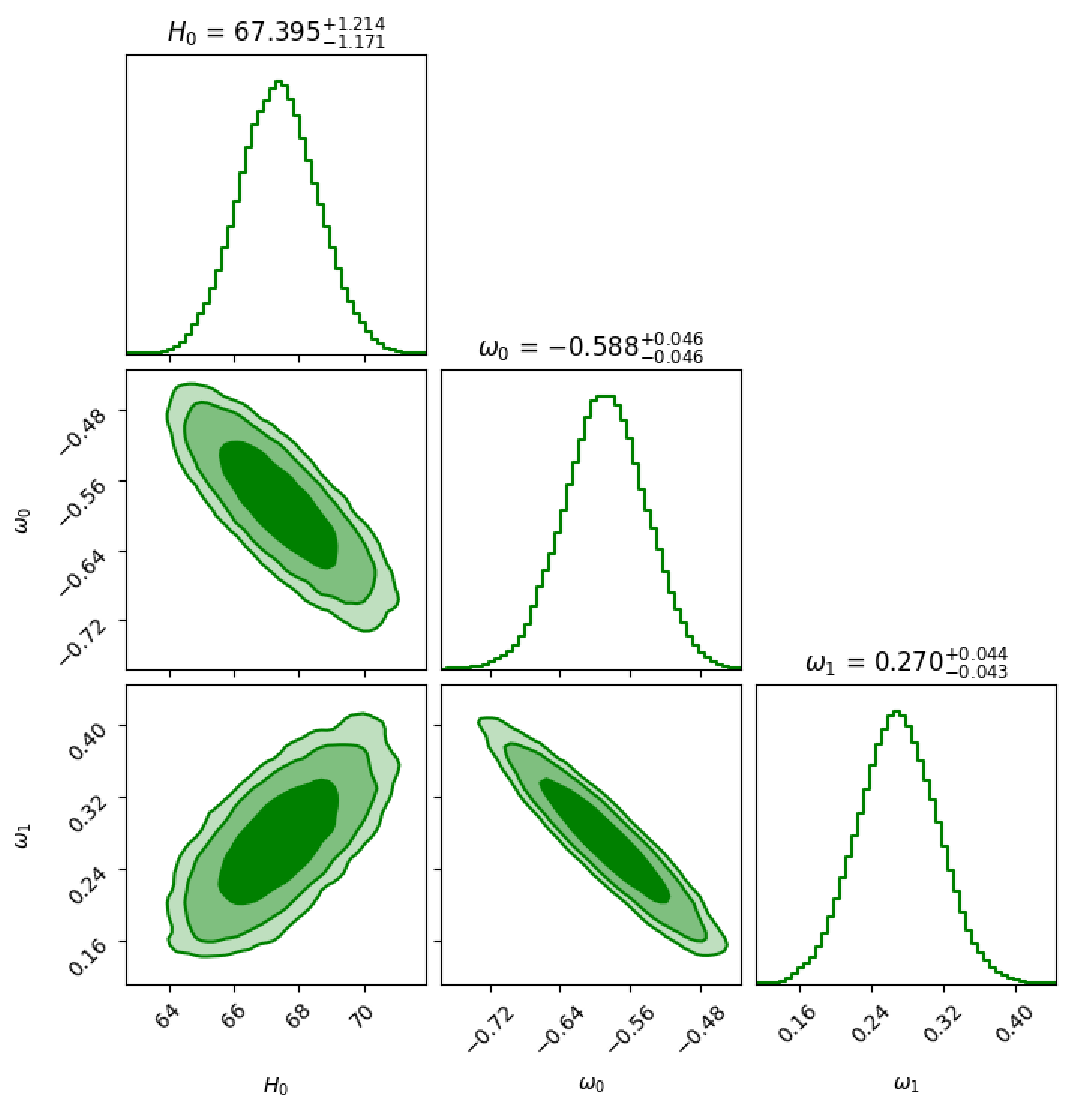} 
        \caption{}
        \label{fig:subfig3}
    \end{subfigure}

    \caption{ $1-\sigma$ and $2-\sigma$ likelihood contours from the analysis of (a) Hubble, (b) Hubble+BAO and (c) Hubble+BAO+Pantheon datasets.}
    \label{fig:f1}
\end{figure}

\begin{figure}[hbt!]
    \centering
    \begin{subfigure}[b]{0.45\textwidth}
        \includegraphics[scale=0.45]{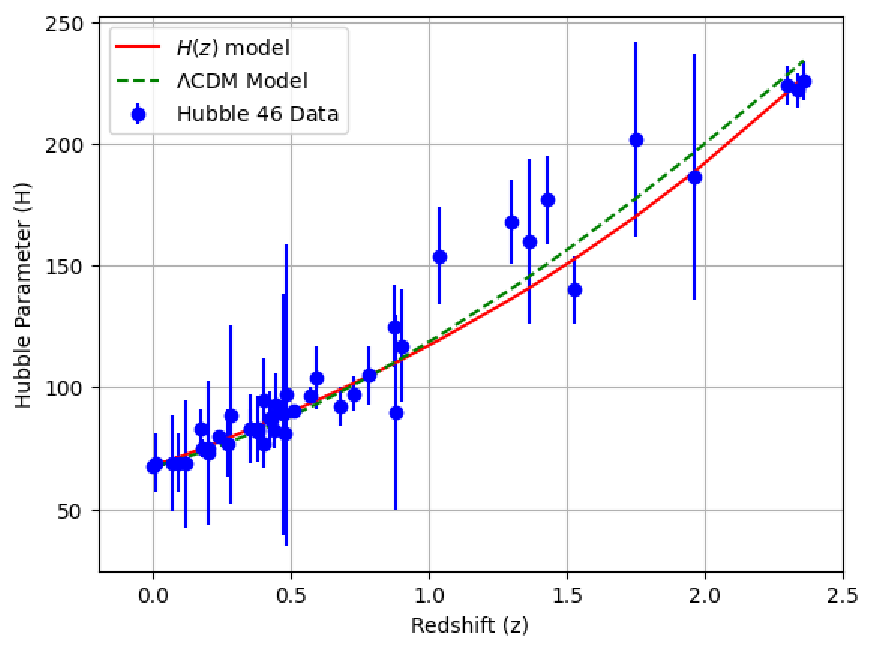}
        \caption{}
        \label{fig:subfig1}
    \end{subfigure}
    \hfill
    \begin{subfigure}[b]{0.45\textwidth}
        \includegraphics[scale=0.45]{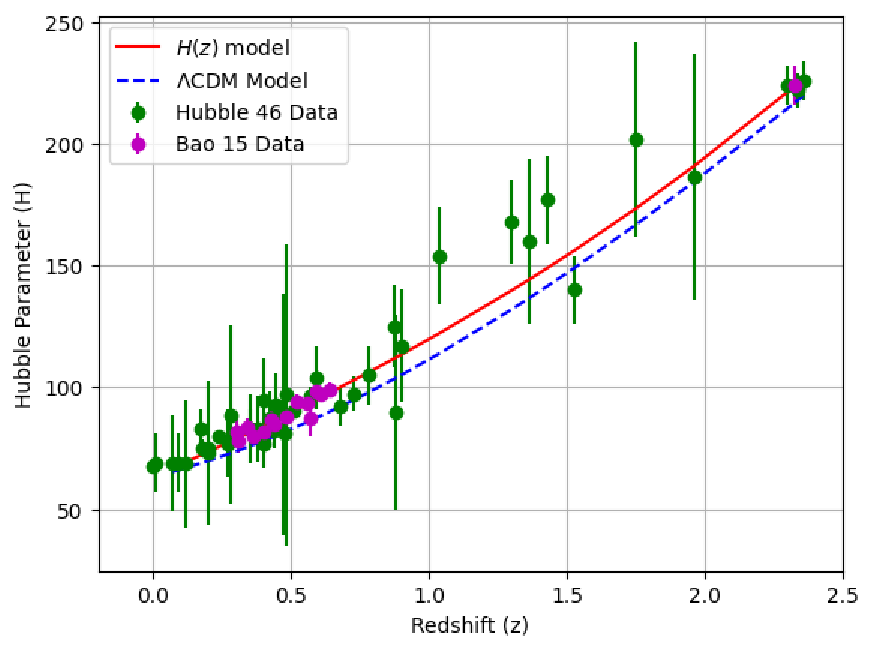}
        \caption{}
        \label{fig:subfig2}
    \end{subfigure}

    \caption{The error bar plot for our $H(z)$ model by using (a) Hubble and (b) Hubble+BAO datasets.}
    \label{fig:f2}
\end{figure}

At $z=0$, the expression simplifies to $H(z)=H_{0}$, ensuring the model is consistent with the present-day expansion rate. For $z>0$ (past), the behavior of $H(z)$ depends upon $\omega_{0}$ and $\omega_{1}$. If $\omega_{0}-\omega_{1}>-1$, the model will predict a slower expansion rate in the past compared to the present. Conversely, if $\omega_{0}-\omega_{1}<-1$, it suggests a faster expansion rate in the past, which could lead to scenarios with rapid early Universe expansion. By substituting the derived form of the Hubble parameter in the equations (\ref{22}), (\ref{26}) and (\ref{31}), we draw the contour plots. The contour plots illustrate the confidence intervals for the parameters $H_{0}$, $\omega_{0}$ and $\omega_{1}$ using three different combinations of datasets: Hubble, Hubble+BAO and Hubble+BAO+Pantheon are given in figure \ref{fig:f1}. These plots enable a comparative analysis of how each dataset combination refines the parameter constraints and enhances our understanding of the model. The error bar plots in \ref{fig:f2} serve to visually assess the compatibility of the model’s predictions with the observational data across redshift. The error bars indicate the range of expected values for $H(z)$ based on the best-fit parameters from each dataset, providing insight into the model’s accuracy in matching observed data points.
\begin{table}[h!]
\centering
\caption{Observationally derived the estimated values of model parameters $H_{0}$, $\omega_{0}$ and $\omega_{1}$ }
\begin{tabular}{||p{4.5cm}|p{2.0cm}|p{2.0cm}|p{2.0cm}||}
\hline\hline
 Parameters & \hspace{0.6cm}$H_{0}$ & \hspace{0.9cm}$\omega_{0}$ & \hspace{0.9cm}$\omega_{1}$ \\
\hline\hline
$\hspace{0.3cm}Hubble$ & $67.434^{+1.163}_{-1.175}$ & $-0.590^{+0.047}_{-0.045}$ & $0.273^{+0.043}_{-0.044}$\\[1pt]
\hline
$Hubble+BAO$ & $67.395^{+1.188}_{-1.177}$ & $-0.589^{+0.047}_{-0.045}$ & $0.272^{+0.043}_{-0.044}$\\ [1pt]
\hline
$Hubble+BAO+Pantheon$ & $67.395^{+1.214}_{-1.171}$ & $-0.588^{+0.046}_{-0.046}$ & $0.270^{+0.044}_{-0.043}$\\ [1pt]
\hline\hline
\end{tabular}
\label{Tab:T4}
\end{table}
\\

The table \ref{Tab:T4} presents the best-fit values for the model parameters $H_{0}$, $\omega_{0}$ and $\omega_{1}$ derived from three different combinations of observational datasets: Hubble, Hubble+BAO and Hubble+BAO+Pantheon. The values of $H_{0}$ range from approximately $67.395$ to $67.434$ across the datasets, with uncertainties of about $\pm 1.18$ $km/s/Mpc$. These findings are in harmony with the Planck mission's results, which indicate $H_{0}\approx67.4$ $km/s/Mpc$ \cite{Planck20} and are in line with current research in the field \cite{Sam24,Amit24}. The dark energy EoS parameter, $\omega_{0}$, is consistently negative across all dataset combinations, with a value of approximately $-0.59$, indicating that dark energy behaves similarly to a cosmological constant $(\omega\approx-1)$. However, the slight deviation from $-1$ suggests that dark energy may exhibit dynamical behavior, rather than being a pure cosmological constant. The parameter $\omega_{1}$, which captures the redshift-dependent evolution of the EoS, takes values around $0.27$, suggesting a small positive evolution in the EoS with redshift. This hints that dark energy could become less negative (or weaker) in the early Universe.
\subsection{Model 2}\label{sec5.2}
\hspace{0.6cm} We consider a specific dynamical model where the EoS parameter $\omega(z)$ is not constant but evolves with redshift, offering a more flexible representation of dark energy than the standard $\Lambda$CDM model. The EoS parameter in this model is defined as \cite{Barboza12}:
\begin{equation}\label{34}
\omega=\omega_{0}+\frac{\omega_{1}z(1+z)}{1+z^{2}},
\end{equation}
where $\omega_{0}$ represents the present-day value of the EoS and $\omega_{1}$ characterizes the rate of change in dark energy’s influence over time. This model is designed to allow a gradual change in the EoS from low redshifts $(z\approx0)$ to high redshifts. When $z$ is close to zero, the EoS parameter reduces to $\omega(z)\approx\omega_{0}$, resembling the behavior of a cosmological constant if $\omega_{0}=-1$. The model's ability to mimic the behavior of a cosmological constant at low redshifts, where dark energy is observed to be uniform and unchanging, reinforces its compatibility with current observational data. As we look further back in time (higher redshifts), the term $\frac{\omega_{1}z(1+z)}{1+z^{2}}$ approaches to $\omega_{1}$ allowing the model to reach a different asymptotic state. This adaptability enables dark energy's role in the early Universe to be more versatile, either being more dominant or less significant, depending on the parameters.

The specific form of $H(z)$ can be obtained by utilising the formula $\frac{dH}{dt}=-H(z)(1+z)\frac{dH}{dz}$ and inserting the previously discussed form into equation (\ref{21}) as follows:
\begin{equation}\label{35}
H(z)=H_{0}(1+z)^{\frac{3(1+\omega_{0})}{2}}(1+z^{2})^{\frac{3\omega_{1}}{4}},
\end{equation}

\begin{figure}[htbp]
    \centering
    \begin{subfigure}{0.45\textwidth}
        \includegraphics[width=\linewidth]{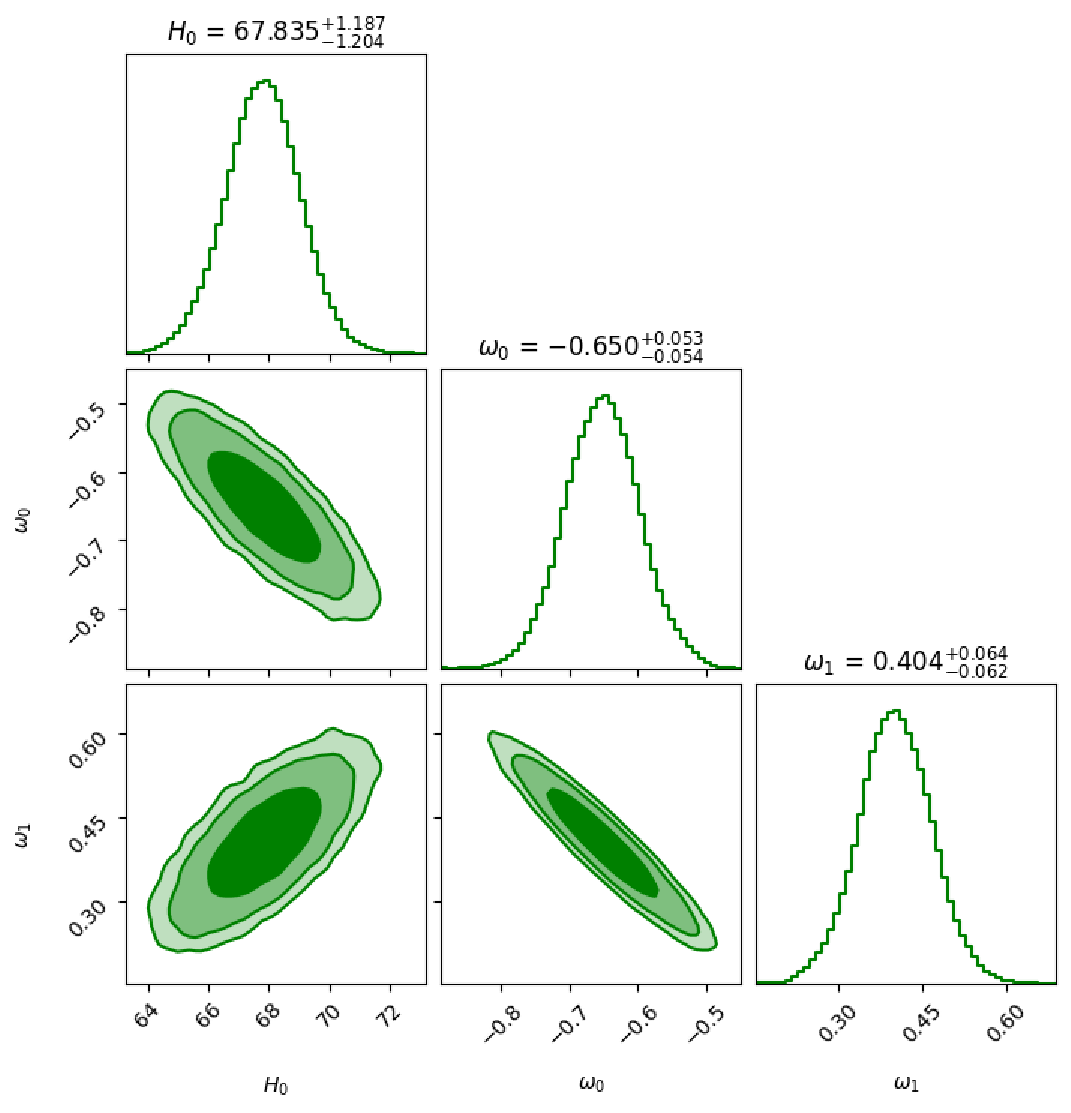} 
        \caption{}
        \label{fig:subfig6}
    \end{subfigure}
    \hfill
    \begin{subfigure}{0.45\textwidth}
        \includegraphics[width=\linewidth]{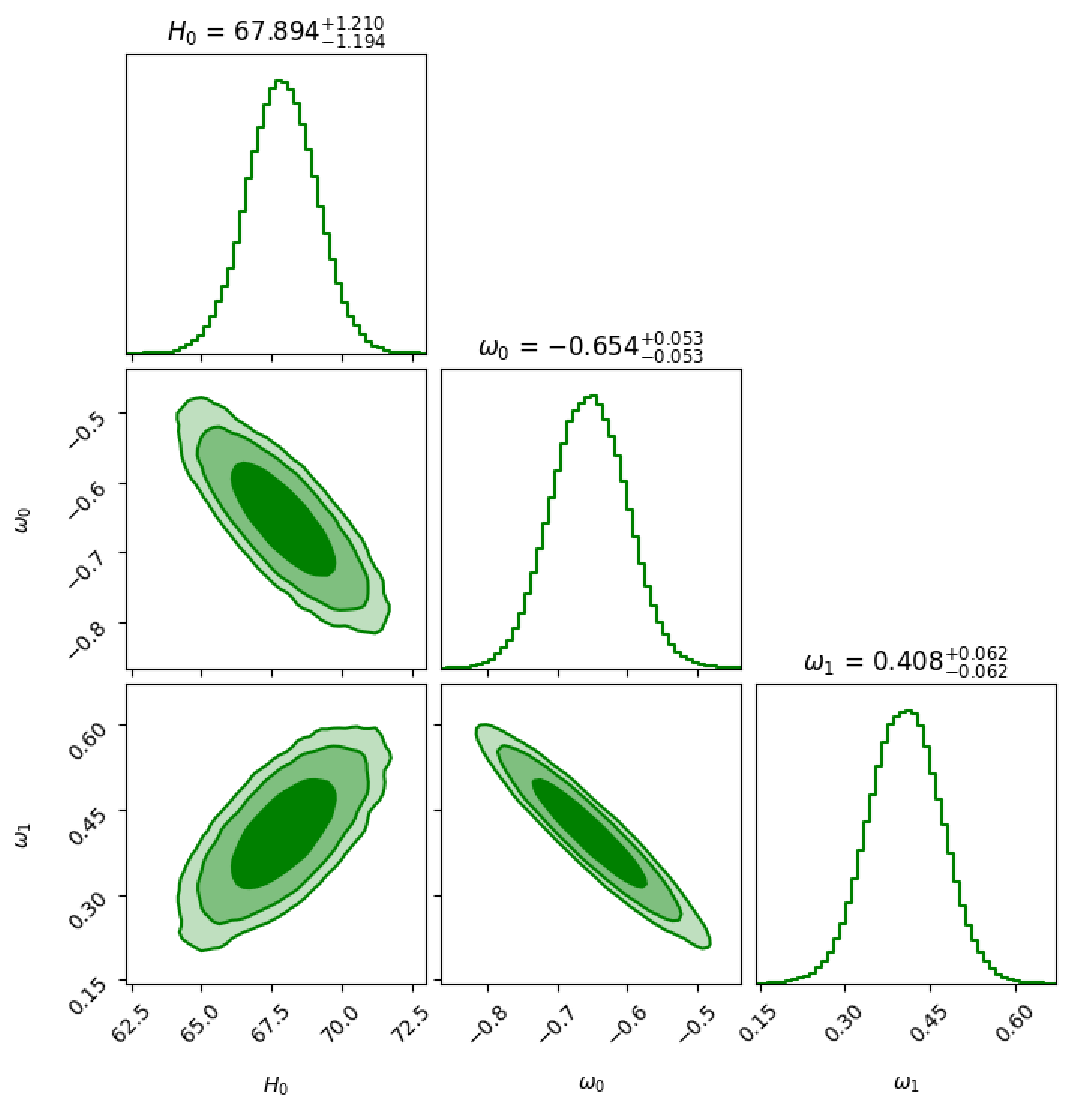} 
        \caption{}
        \label{fig:subfig7}
    \end{subfigure}

    \begin{subfigure}{0.45\textwidth}
        \includegraphics[width=\linewidth]{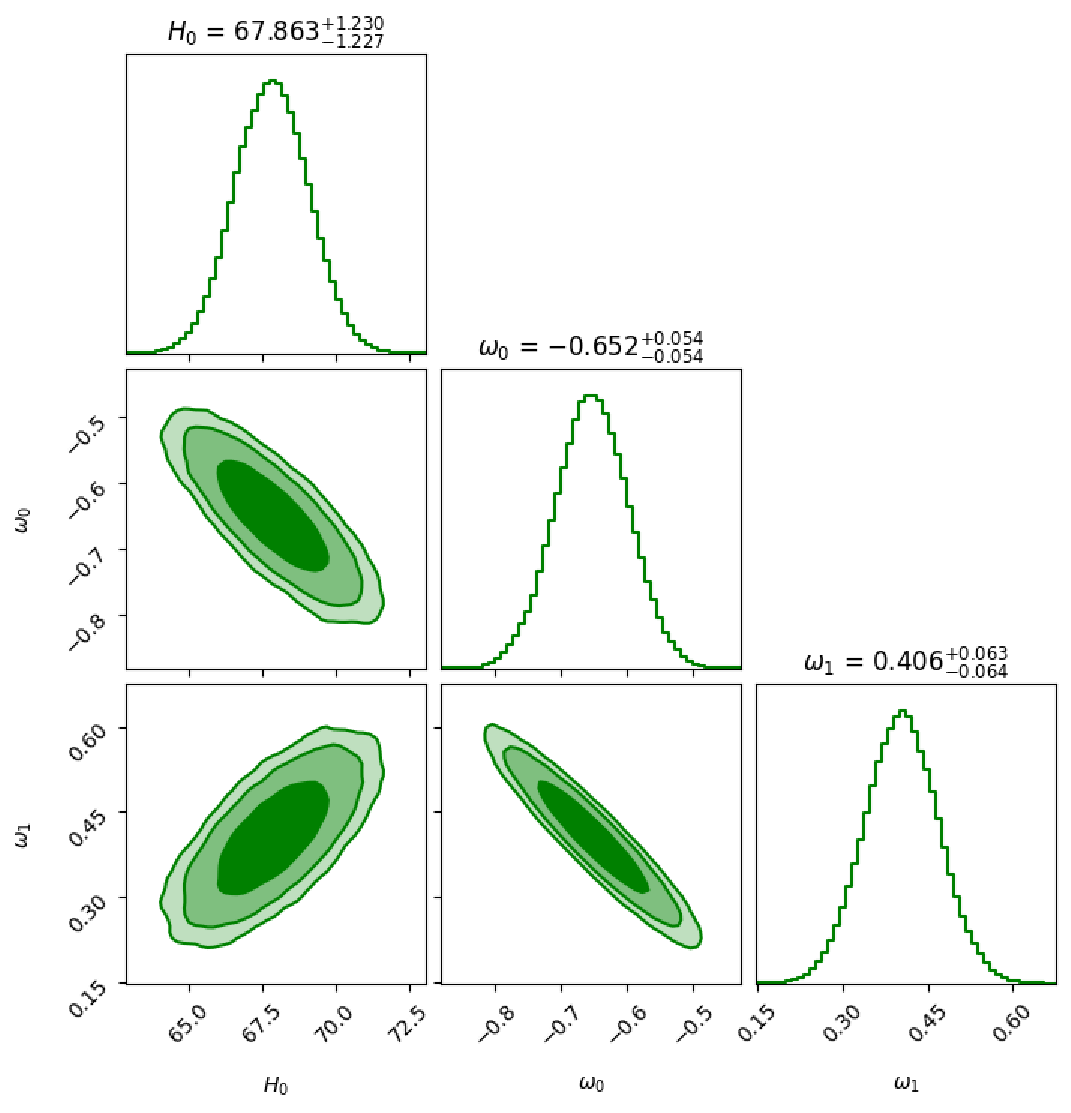} 
        \caption{}
        \label{fig:subfig8}
    \end{subfigure}

    \caption{ $1-\sigma$ and $2-\sigma$ likelihood contours from the analysis of (a) Hubble, (b) Hubble+BAO and (c) Hubble+BAO+Pantheon datasets.}
    \label{fig:f3}
\end{figure}

\begin{figure}[hbt!]
    \centering
    \begin{subfigure}[b]{0.45\textwidth}
        \includegraphics[scale=0.45]{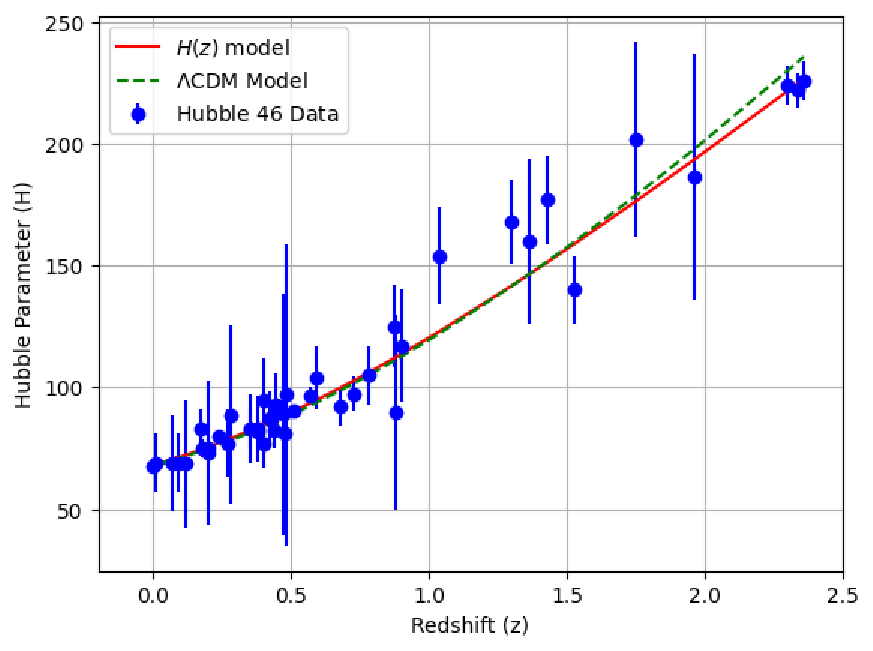}
        \caption{}
        \label{fig:subfig9}
    \end{subfigure}
    \hfill
    \begin{subfigure}[b]{0.45\textwidth}
        \includegraphics[scale=0.45]{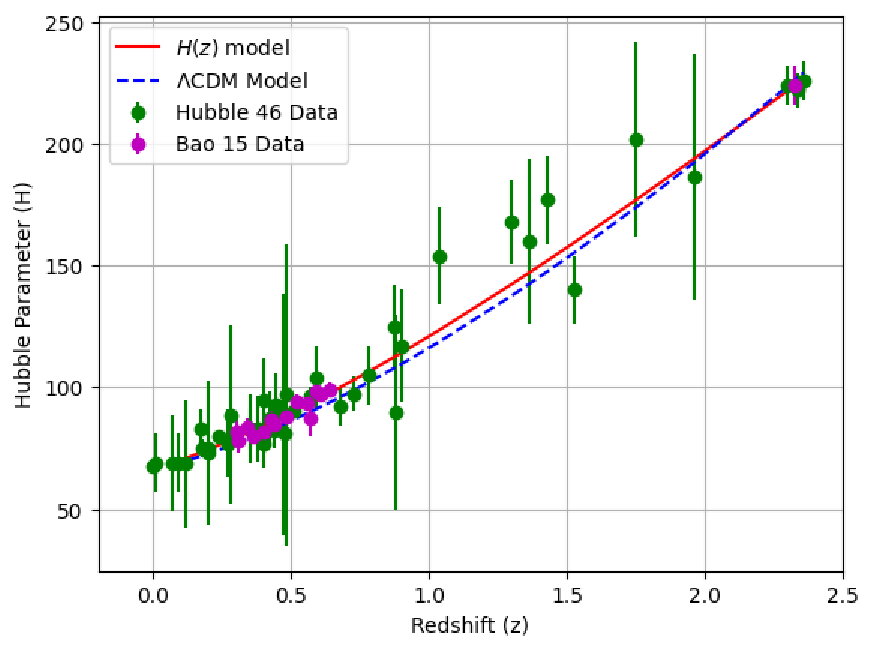}
        \caption{}
        \label{fig:subfig10}
    \end{subfigure}

    \caption{The error bar plot for our $H(z)$ model by using (a) Hubble and (b) Hubble+BAO datasets.}
    \label{fig:f4}
\end{figure}

The model introduces two terms $(1+z)^{\frac{3(1+\omega_{0})}{2}}$ and $(1+z^{2})^{\frac{3\omega_{1}}{4}}$, which influence the expansion rate at different stages of cosmic history. When $z$ is close to zero, the term $(1+z)^{\frac{3(1+\omega_{0})}{2}}$ term dominates, with $(1+z^{2})^{\frac{3\omega_{1}}{4}}\approx1$. This implies that the model behaves similarly to a $\Lambda$CDM model with a small modification dependent on $\omega_{0}$, consistent with current observations of dark energy in the late Universe. As $z$ increases, both terms begin to contribute more significantly to the behavior of $H(z)$, introducing a smooth transition from the low-redshift behavior (dominated by $\omega_{0}$) to a high-redshift behavior influenced by $\omega_{1}$. At high redshifts, the term $(1+z^{2})^{\frac{3\omega_{1}}{4}}$ term becomes prominent. The dependence on $z^{2}$ allows the model to evolve more gently than models with linear $z$ terms, which can prevent extreme behavior at early times while still capturing potential deviations from the standard model. By integrating the derived Hubble parameter form into the equations, we generate contour plots that visualize the confidence intervals for parameters $H_{0}$, $\omega_{0}$ and $\omega_{1}$ in Figure \ref{fig:f3}. These plots are based on three dataset combinations, enabling a comparative analysis of how each combination constrains parameter values and enhances our understanding of the model. Additionally, error bar plots in Figure \ref{fig:f4} provide a visual assessment of the model's consistency with observational data across redshift, showcasing the expected range of $H(z)$ values based on best-fit parameters from each dataset.
\begin{table}[h!]
\centering
\caption{Observationally derived the estimated values of model parameters $H_{0}$, $\omega_{0}$ and $\omega_{1}$ }
\begin{tabular}{||p{4.5cm}|p{2.0cm}|p{2.0cm}|p{2.0cm}||}
\hline\hline
 Parameters & \hspace{0.6cm}$H_{0}$ & \hspace{0.9cm}$\omega_{0}$ & \hspace{0.9cm}$\omega_{1}$ \\
\hline\hline
$\hspace{0.3cm}Hubble$ & $67.835^{+1.187}_{-1.204}$ & $-0.650^{+0.053}_{-0.054}$ & $0.404^{+0.064}_{-0.062}$\\[1pt]
\hline
$Hubble+BAO$ & $67.894^{+1.210}_{-1.194}$ & $-0.654^{+0.053}_{-0.053}$ & $0.408^{+0.062}_{-0.062}$\\ [1pt]
\hline
$Hubble+BAO+Pantheon$ & $67.863^{+1.230}_{-1.227}$ & $-0.652^{+0.054}_{-0.054}$ & $0.406^{+0.063}_{-0.064}$\\ [1pt]
\hline\hline
\end{tabular}
\label{Tab:T5}
\end{table}
\\

The table \ref{Tab:T5} summarizes the optimal parameter values for $H_{0}$, $\omega_{0}$ and $\omega_{1}$, obtained from a fitting process that utilized three different datasets: Hubble data alone, Hubble data paired with BAO and Hubble data paired with BAO and Pantheon. The estimated values for the Hubble constant $H_{0}\approx67.8$ $km/s/Mpc$, with slight discrepancies between them. The agreement between the derived Hubble constant values and the Planck satellite's CMB measurements provides strong evidence for the model's accuracy in describing the early Universe's dynamics \cite{Planck20}. This consistency demonstrates that the model parameters are well-determined and align with the Universe's evolution during the CMB era, boosting confidence in the model's reliability \cite{A24}. 

The present day EoS parameter $\omega_{0}$ is estimate $-0.65$ across all datasets. The values of $\omega_{0}$ here are slightly greater than $-1$ suggesting a deviation from a pure cosmological constant and it also indicates a form of dark energy where the EoS varies with time. This behavior is consistent with models that allow for a dark energy density that evolves gradually, possibly in response to changes in cosmic expansion. In dynamical dark energy models, such as quintessence, $\omega_{0}$ can be greater than $-1$ (for a slowly evolving field), while phantom models have $\omega_{0}<-1$. Here, $\omega_{0}$ values suggest scenarios like quintessence over a cosmological constant. Within a dark energy model featuring a redshift-dependent equation of state, $\omega_{1}$ determines the pace at which the EoS parameter $\omega(z)$ changes over cosmic time, thus impacting the Universe's expansion history. Here, a positive $\omega_{1}$ suggests that dark energy may have been more ``phantom-like" $(\omega<-1)$ in the past but is trending closer to the cosmological constant value $(\omega=-1)$ or beyond $z$ decreases toward the present time. The values clustering around 0.4 reveal a moderate evolution in the dark energy's EoS, pointing to a gradual transformation of the dark energy density that is neither too subtle nor too dramatic, but rather a balanced and measured progression. This outcome enhances the model's credibility as a plausible substitute for the cosmological constant scenario, offering fresh perspectives on the Universe's progressive transformation and the future expansion pathway, potentially illuminating the cosmic horizon.\\

\subsection{Model 3}\label{sec5.3}
\hspace{0.6cm} To describe dark energy’s dynamic behavior, we adopt a specific parameterization of the equation of state \cite{Feng12}:
\begin{equation}\label{36}
\omega=\omega_{0}+\frac{\omega_{1}z^{2}}{1+z^{2}},
\end{equation}
At low redshift $(z\approx0)$, the model simplifies to $\omega(z)\approx\omega_{0}$, which gives the current value of the dark energy equation of state. This suggests that near the present day, the model behaves similarly to a constant EoS. At high redshift $(z\gg1)$, the term $\frac{z^{2}}{1+z^{2}}$ approaches $1$, so the EoS asymptotically approaches $\omega(z)\approx\omega_{0}+\omega_{1}$. This indicates that in the early Universe, dark energy could behave differently, with the EoS deviating from its current value by an additional amount $\omega_{1}$. The model implies that dark energy behaves similarly to a cosmological constant at present if $\omega_{0}\approx1$ and evolves smoothly as $z$ increases. If $\omega_{1}>0$, the dark energy EoS moves closer to zero at high redshifts, indicating a shift towards more ``matter-like" behavior, potentially reducing dark energy’s dominance in the early Universe. Conversely, if $\omega_{1}<0$, dark energy could appear more ``phantom-like" $(\omega<-1)$ at high redshifts, which can suggest even stronger acceleration in the early Universe. 

The precise form of $H(z)$ can be deduced by utilizing the differential equation $\frac{dH}{dt}=-H(z)(1+z)\frac{dH}{dz}$ and substituting the previously derived form into equation (\ref{21}), resulting in:
\begin{equation}\label{37}
H(z)=H_{0}(1+z)^{\frac{3(2+2\omega_{0}+\omega_{1})}{2}}(1+z^{2})^{\frac{3\omega_{1}}{8}}exp\bigg[-\frac{3\omega_{1}}{4}tan^{-1}z\bigg].
\end{equation}

The Hubble parameter function assumes a sophisticated form, incorporating both power-law and exponential elements, indicating a rich and non-trivial evolution of dark energy, which unfolds in a more intricate manner than a straightforward, monotonous progression. The factor $(1+z)^{\frac{3(2+2\omega_{0}+\omega_{1})}{2}}$ encapsulates the redshift dependence of matter and dark energy in a framework where dark energy evolves dynamically. The exponent is modulated by $\omega_{0}$ and $\omega_{1}$, which govern the effects of the dark energy equation of state at present and its variation over cosmic time. The term $(1+z^{2})^{\frac{3\omega_{1}}{8}}$ captures modifications due to $\omega_{1}$ at high redshift. As $z$ increases, this term approaches $z^{\frac{3\omega_{1}}{4}}$, reflecting how the dark energy contribution changes for large $z$. This high-redshift limit impacts the evolution of the Universe at earlier times, making the model sensitive to observational data from epochs such as those probed by high-redshift supernovae and BAO measurements. The factor $exp\bigg[-\frac{3\omega_{1}}{4}tan^{-1}z\bigg]$ represents a smooth, asymptotic evolution of the dark energy EoS over time. The $tan^{-1}z$ term asymptotes to a constant value for large $z$, meaning that dark energy stabilizes at higher redshifts rather than diverging or changing abruptly.

\begin{figure}[htbp]
    \centering
    \begin{subfigure}{0.45\textwidth}
        \includegraphics[width=\linewidth]{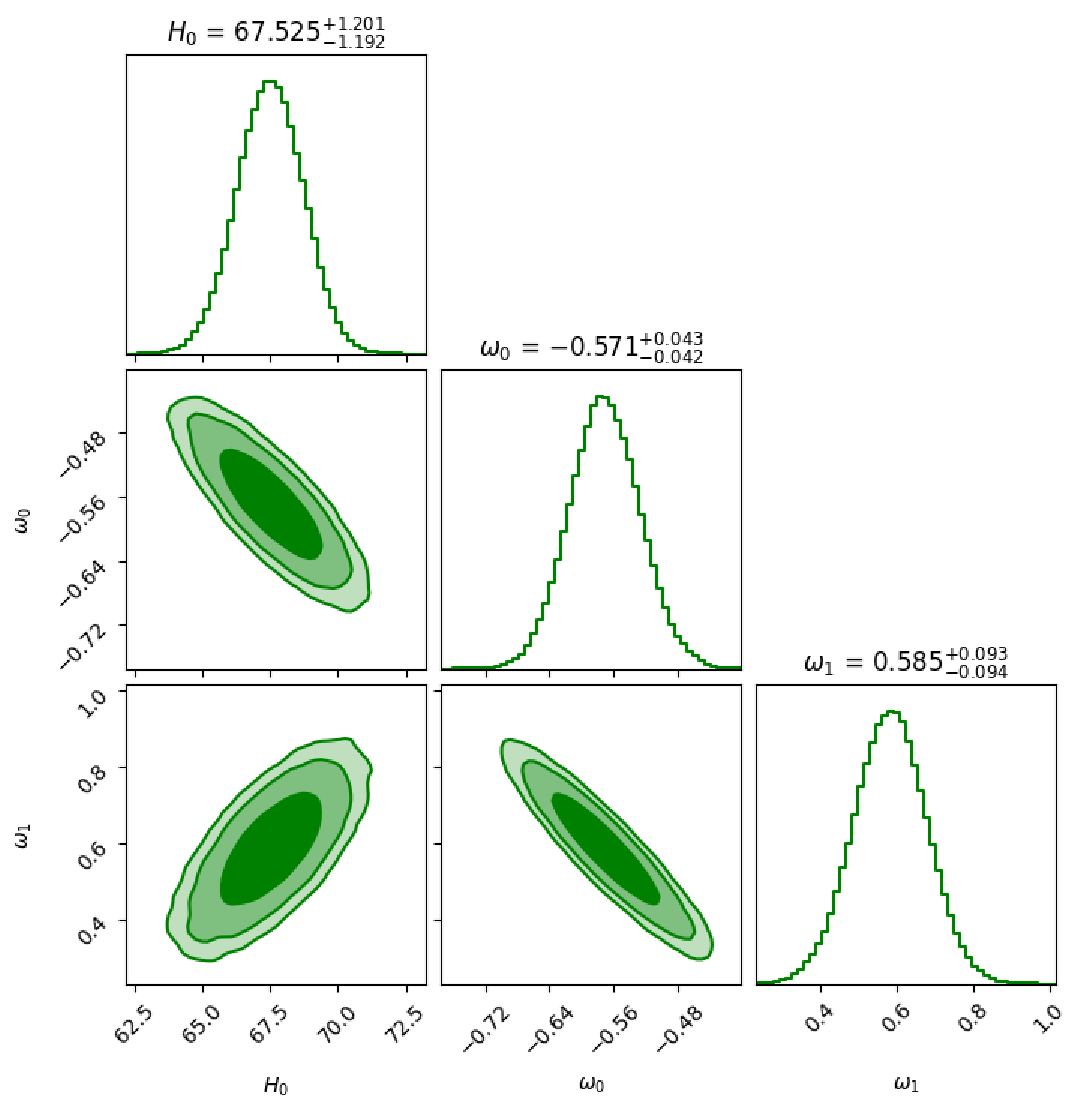} 
        \caption{}
        \label{fig:subfig11}
    \end{subfigure}
    \hfill
    \begin{subfigure}{0.45\textwidth}
        \includegraphics[width=\linewidth]{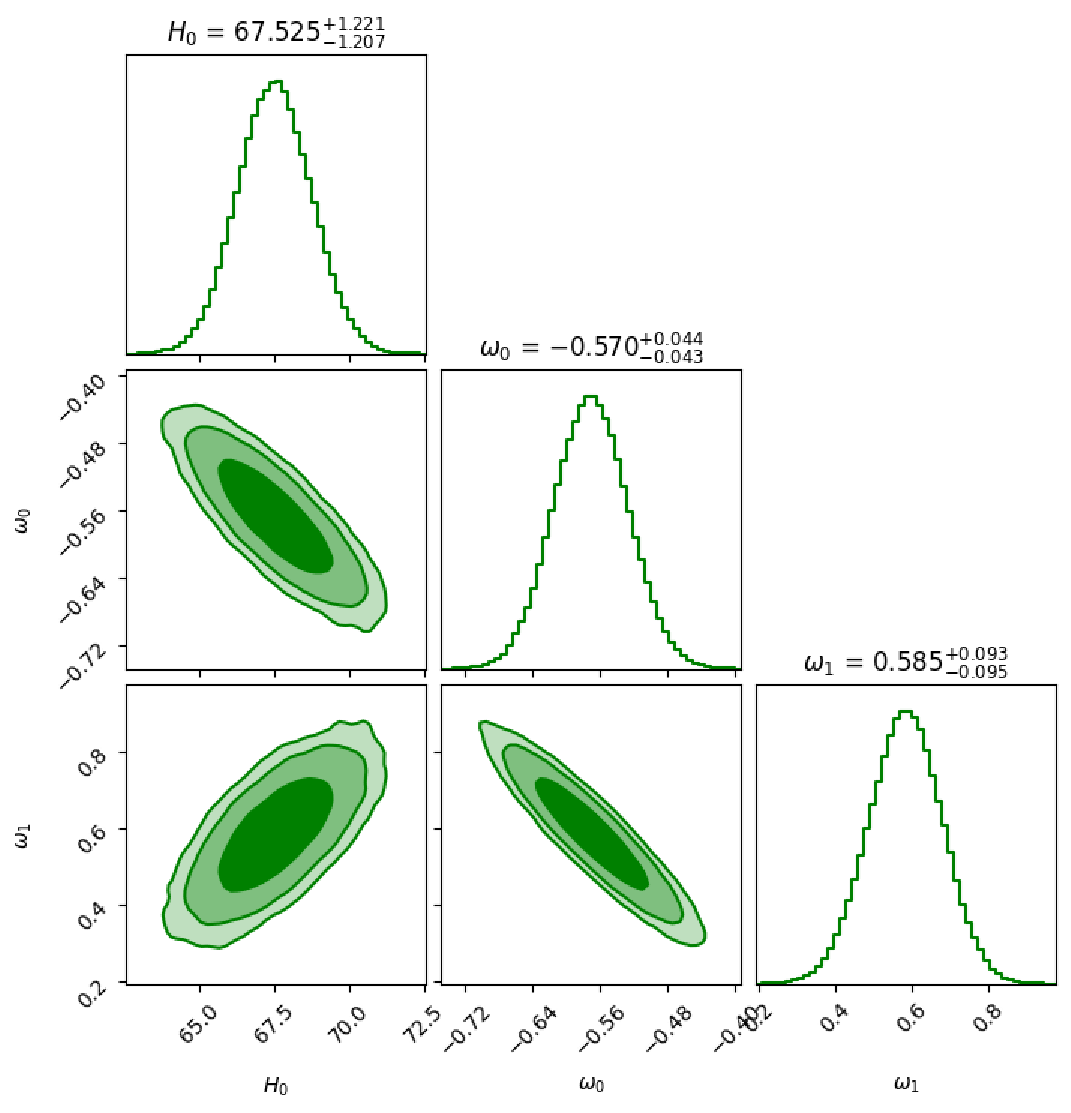} 
        \caption{}
        \label{fig:subfig12}
    \end{subfigure}

    \begin{subfigure}{0.45\textwidth}
        \includegraphics[width=\linewidth]{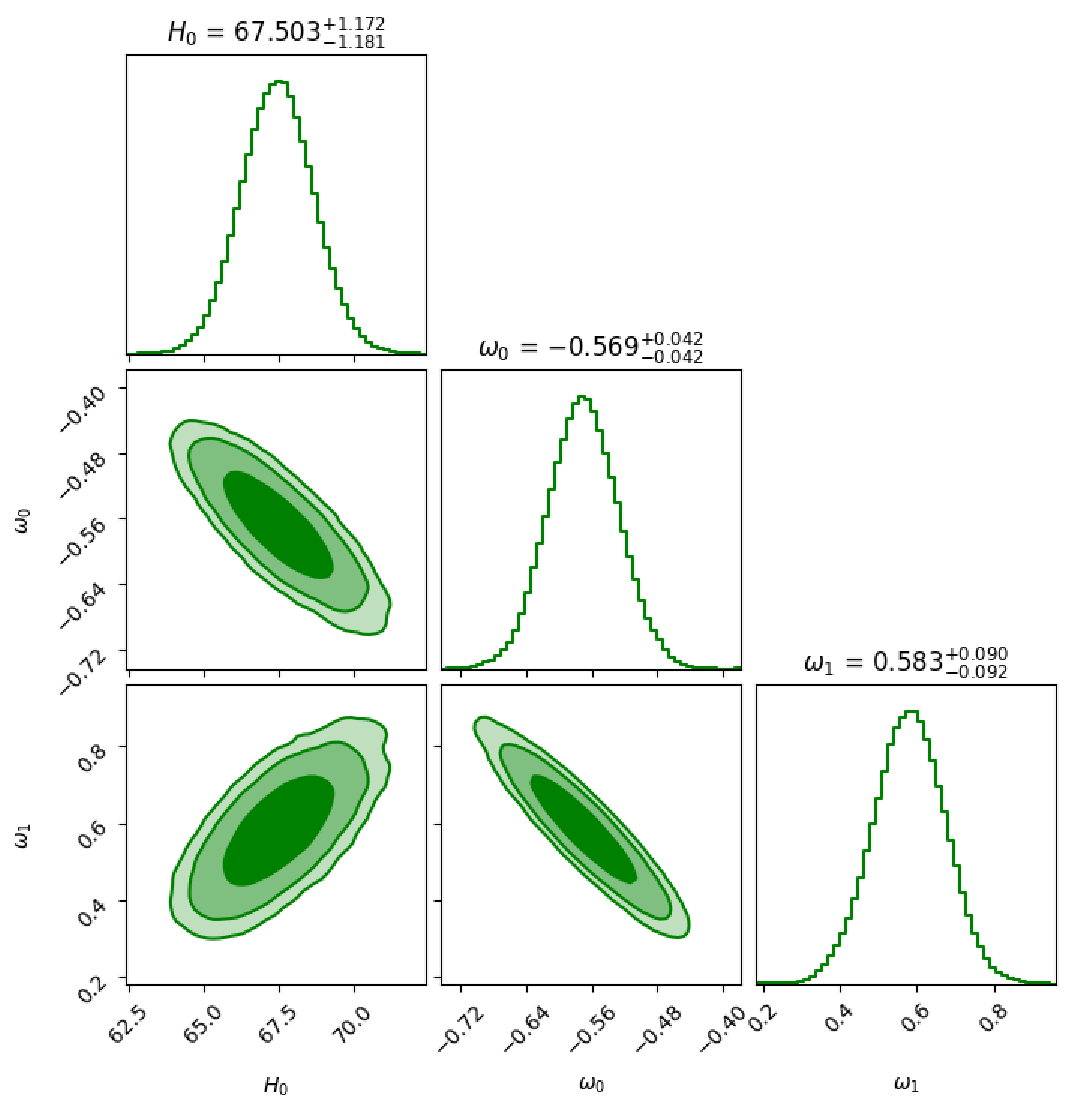} 
        \caption{}
        \label{fig:subfig13}
    \end{subfigure}

    \caption{ $1-\sigma$ and $2-\sigma$ likelihood contours from the analysis of (a) Hubble, (b) Hubble+BAO and (c) Hubble+BAO+Pantheon datasets.}
    \label{fig:f5}
\end{figure}

\begin{figure}[hbt!]
    \centering
    \begin{subfigure}[b]{0.45\textwidth}
        \includegraphics[scale=0.45]{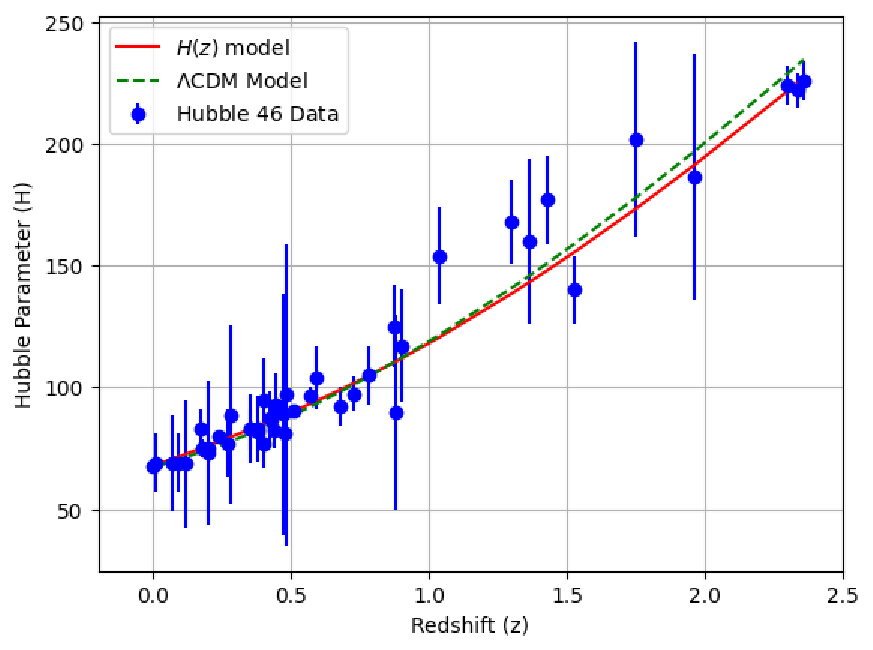}
        \caption{}
        \label{fig:subfig14}
    \end{subfigure}
    \hfill
    \begin{subfigure}[b]{0.45\textwidth}
        \includegraphics[scale=0.45]{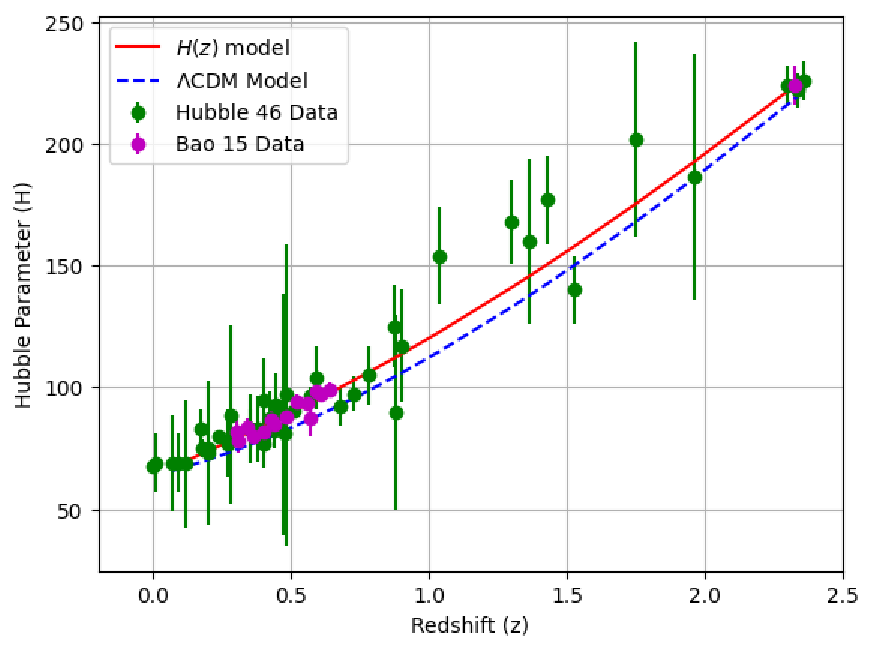}
        \caption{}
        \label{fig:subfig15}
    \end{subfigure}

    \caption{The error bar plot for our $H(z)$ model by using (a) Hubble and (b) Hubble+BAO datasets.}
    \label{fig:f6}
\end{figure}

The resulting Hubble parameter form (\ref{37}) is inserted in the equations (\ref{22}), (\ref{26}) and (\ref{31}) to produce contour plots that illustrate the levels of confidence for parameters $H_{0}$, $\omega_{0}$ and $\omega_{1}$ shown in Figure \ref{fig:f5}. With three dataset combinations as the basis for these visualizations, it is possible to compare how each combination limits parameter values and improves our comprehension of the model. Furthermore, the model's coherence with observational data throughout redshift is visually evaluated using error bar plots in Figure \ref{fig:f6}, which display the predicted variation in $H(z)$ values generated from best-fit parameters from each dataset.
\begin{table}[h!]
\centering
\caption{Observationally derived the estimated values of model parameters $H_{0}$, $\omega_{0}$ and $\omega_{1}$ }
\begin{tabular}{||p{4.5cm}|p{2.0cm}|p{2.0cm}|p{2.0cm}||}
\hline\hline
 Parameters & \hspace{0.6cm}$H_{0}$ & \hspace{0.9cm}$\omega_{0}$ & \hspace{0.9cm}$\omega_{1}$ \\
\hline\hline
$\hspace{0.3cm}Hubble$ & $67.525^{+1.201}_{-1.192}$ & $-0.571^{+0.043}_{-0.042}$ & $0.585^{+0.093}_{-0.094}$\\[1pt]
\hline
$Hubble+BAO$ & $67.525^{+1.221}_{-1.207}$ & $-0.570^{+0.044}_{-0.043}$ & $0.585^{+0.093}_{-0.095}$\\ [1pt]
\hline
$Hubble+BAO+Pantheon$ & $67.503^{+1.172}_{-1.181}$ & $-0.569^{+0.042}_{-0.042}$ & $0.583^{+0.090}_{-0.092}$\\ [1pt]
\hline\hline
\end{tabular}
\label{Tab:T6}
\end{table}
\\

From Table \ref{Tab:T6}, the values obtained for $H_{0}$ are consistent across the three datasets—Hubble, Hubble+BAO and Hubble+BAO+Pantheon—showing a slight variation with values centered around $67.5$ $km/s/Mpc$. The calculated values converge within a limited range, closely mirroring the CMB measurement from Planck (approximately $67.4$ $km/s/Mpc$), highlighting a remarkable consistency with well-established cosmological findings.

The values of $\omega_{0}$ across all datasets are around $-0.570$ slightly above the standard cosmological constant value $(\omega=-1)$. The slight offset from the expected value points to a quintessence-type dark energy, with $\omega$ exceeding $-1$,  signifying that the dark energy density diminishes more slowly than anticipated by the $\Lambda$CDM model, as the Universe evolves and expands. The strong agreement among $\omega_{0}$ values across various datasets, within error margins, demonstrates the parameter's robustness and stability. This consistency solidifies the model's capacity to capture dark energy behavior, without deviating significantly from observational constraints, thus reinforcing the model's validity and reliability. The value of $\omega_{1}$, which governs the redshift dependence of the dark energy equation of state, clusters around $0.585$, with minimal variations between the datasets. A positive value of $\omega_{1}$ indicates that the dark energy equation of state becomes progressively less negative at higher redshifts, enabling dark energy to evolve dynamically over cosmic time. This behavior aligns with theoretical frameworks where dark energy's influence on cosmic expansion varies across different epochs, particularly in scenarios where its impact diminishes at earlier times, mirroring the dominance of matter. The uniformity of $\omega_{1}$ across multiple datasets indicates that this dynamical model is statistically resilient and provides a reliable characterization of dark energy's evolution, thereby reinforcing the model's credibility and stability in describing the time-varying nature of dark energy.
\section{Cosmological implications of the model}\label{sec6}
\hspace{0.6cm} Here, we analyze the cosmological dynamics of Models \ref{sec5.1}, \ref{sec5.2} and \ref{sec5.3}, employing the observationally informed parameter values to investigate the models' performance in reproducing the observed Universe evolution and to gain insights into their underlying cosmological mechanisms.
\subsection{Deceleration parameter}\label{sec6.1}
\hspace{0.6cm} The deceleration parameter $(q)$ is a cosmic diagnostic tool that reveals the expansion history of the Universe, quantifying the rate of change of the Hubble parameter $(H)$ and offering insights into the dynamics of cosmic growth. $q=-1-\frac{\dot{H}}{H^{2}}$ is its mathematical expression, where $\dot{H}$ denotes its time derivative. In terms of the scale factor $a(t)$, the deceleration parameter $(q)$ takes the form: $q=-\frac{a\ddot{a}}{\dot{a}^{2}}$, where $\dot{a}$ and $\ddot{a}$ are the first and second derivatives of $a(t)$, respectively, providing a direct link between the expansion dynamics and the scale factor's behavior.\\
$1.$ \textsc{Deceleration} $(q>0)$: The Universe's expansion is slowing down, a phase dominated by matter or radiation.\\
$2.$ \textsc{Acceleration} $(q<0)$: The expansion is speeding up, indicating dark energy's influence or a similar accelerating force.\\
$3.$ \textsc{Constant Expansion} $(q=0)$: The Universe expands at a steady rate, with neither acceleration nor deceleration.\\
$4.$ \textsc{Super-Exponential Expansion} $(q<-1)$: An extremely rapid expansion, possibly due to exotic forms of dark energy, which could lead to the ``Big Rip" scenario.\\
$5.$ \textsc{de Sitter Expansion} $(q=-1)$: A steady exponential expansion associated with a cosmological constant, leading to a stable accelerating phase.\\

In line with the current cosmological understanding, the Universe's accelerated expansion corresponds to a negative deceleration parameter, $q$. For our $H(z)$ models, we derive the deceleration parameter expressions by utilising the equations (\ref{33}), (\ref{35}) and (\ref{37}) as follows:\\
$\bullet$ Model \ref{sec5.1}:
\begin{equation}\label{38}
q(z)=-1+\frac{3(1+\omega_{0}-\omega_{1})}{2}+\frac{3\omega_{1}}{2}(1+z), 
\end{equation}
$\bullet$ Model \ref{sec5.2}: 
\begin{equation}\label{39}
q(z)=-1+\frac{3(1+\omega_{0})}{2}+\frac{3\omega_{1}z(1+z)}{2(1+z^{2})},
\end{equation}
$\bullet$ Model \ref{sec5.3}:
\begin{equation}\label{40}
q(z)=-1+\frac{3(2+2\omega_{0}+\omega_{1})}{4}+\frac{3\omega_{1}z(1+z)}{4(1+z^{2})}-\frac{3\omega_{1}}{4(1+z^{2})}.
\end{equation}

\begin{figure}[htbp]
    \centering
    \begin{subfigure}{0.45\textwidth}
        \includegraphics[width=\linewidth]{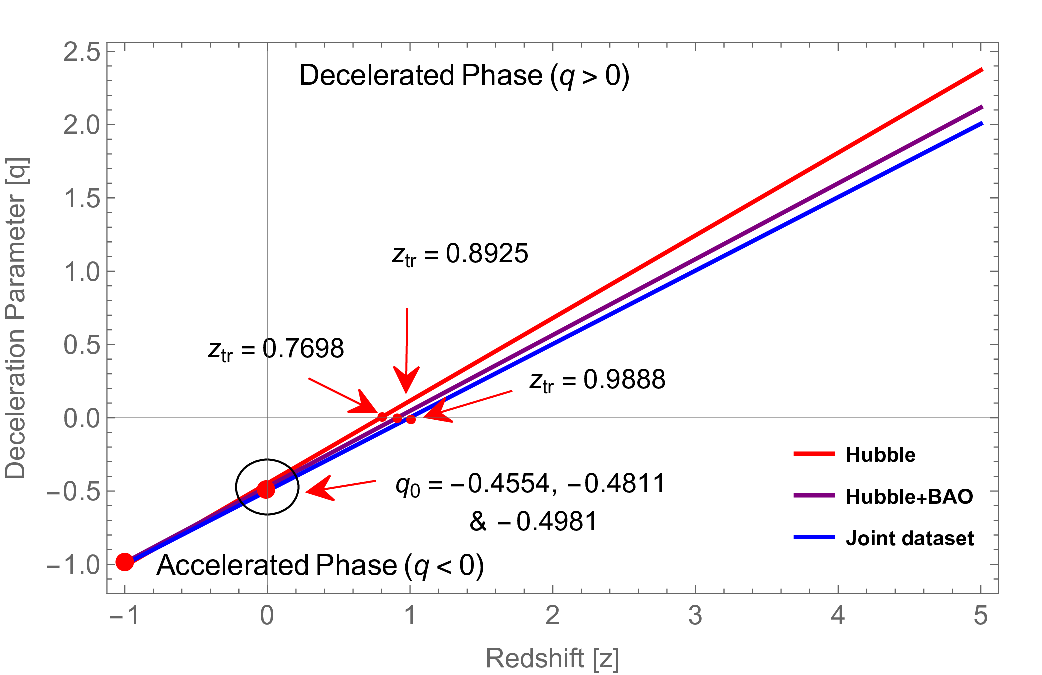} 
        \caption{}
        \label{fig:subfig16}
    \end{subfigure}
    \hfill
    \begin{subfigure}{0.45\textwidth}
        \includegraphics[width=\linewidth]{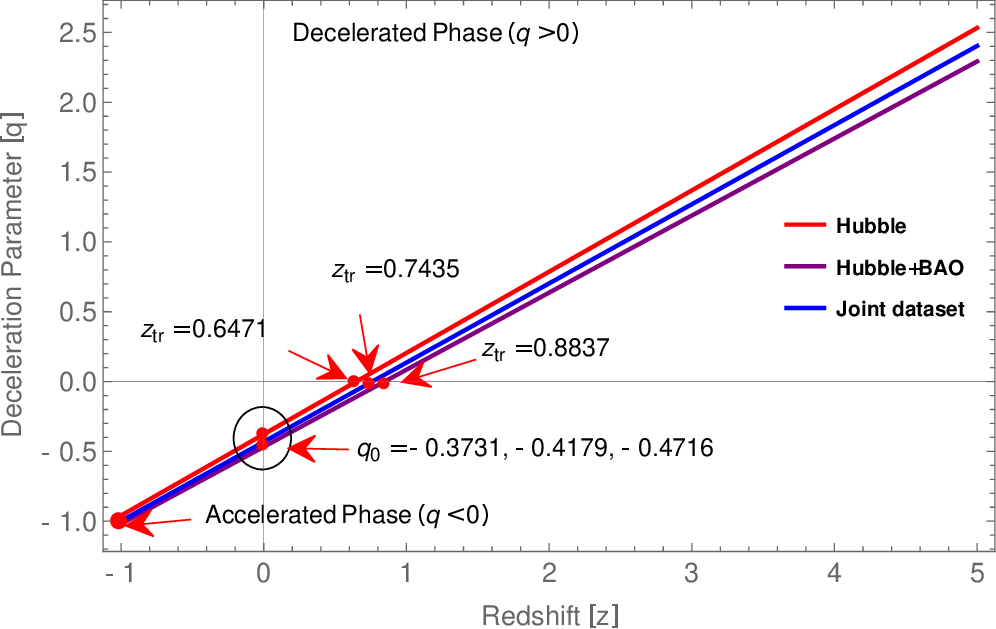} 
        \caption{}
        \label{fig:subfig17}
    \end{subfigure}

    \begin{subfigure}{0.45\textwidth}
        \includegraphics[width=\linewidth]{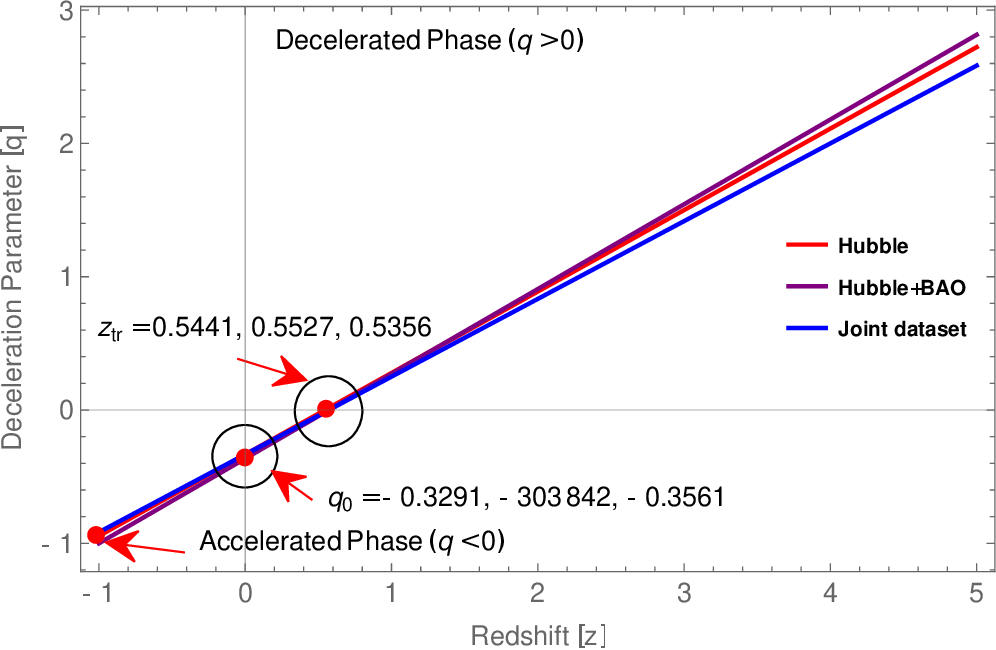} 
        \caption{}
        \label{fig:subfig18}
    \end{subfigure}

    \caption{The plot of $q$ vs. $z$ for the constrained parameter values of (a) Model \ref{sec5.1}, (b) Model \ref{sec5.2} and (c) Model \ref{sec5.3}.}
    \label{fig:f7}
\end{figure}
The plot \ref{fig:f7} of $q(z)$ vs. $z$ for the three models showcases the behavior of the deceleration parameter across cosmic time, illustrating the transition from a decelerating to an accelerating Universe. Each model starts with a positive $q$ value at higher redshifts, indicating deceleration, which corresponds to an era dominated by matter. As $z$ decreases, $q$ gradually transitions to negative values, marking the onset of cosmic acceleration driven by dark energy or a similar mechanism.

For Model \ref{sec5.1}, the transition points at redshifts $z_{tr}=0.7698,0.8925$ and $0.9888$ for the Hubble, Hubble+BAO and Hubble+BAO+Pantheon datasets respectively suggest a later transition from deceleration to acceleration. The present-day deceleration parameter values for these datasets, $q_{0}=-0.4554,-0.4811$ and $-0.4981$ indicate a strong accelerating phase, close to a cosmological consensus that suggests $q_{0}\approx0.5$, consistent with dark energy’s influence in the current epoch.

For Model \ref{sec5.2}, transition redshifts occur at $z_{tr}=0.6471,0.7435$ and $0.8837$, with present-day $q_{0}$ values of $-0.3737,-0.4179$ and $-0.4716$, respectively. This model suggests a slightly earlier transition to acceleration compared to Model \ref{sec5.1}, aligning with observations that imply gradual cosmic acceleration driven by dark energy. The lower $q_{0}$ values for the datasets are consistent with a stronger fit for acceleration. 

For Model \ref{sec5.3}, the transition points at $z_{tr}=0.5441,0.5527$ and $0.5356$ suggest the earliest shift to acceleration among the three models, with corresponding present-day $q_{0}$ values of $-0.3291,-0.3842$ and $-0.3561$. These values indicate a moderately accelerating Universe, though less rapid than Model \ref{sec5.1}, making Model \ref{sec5.3} potentially less compatible with certain observations suggesting stronger late-time acceleration.

In summary, all three models capture the expected trend of an early decelerating phase followed by late-time acceleration, aligning with current cosmological consensus on the influence of dark energy. Models \ref{sec5.1} and \ref{sec5.2} show particularly strong alignment with the observed range of $q_{0}$ values, whereas Model \ref{sec5.3} suggests a milder acceleration. The consistency of these transitions and present-day $q_{0}$ values with cosmological observations validates the robustness of each model, especially when considering the combined dataset (Hubble+BAO+Pantheon) which provides tighter constraints and yields values closer to established cosmological findings.
\subsection{Energy density and pressure}\label{sec6.2}
\hspace{0.6cm} In cosmology, energy density $(\rho)$ and pressure $(p)$ serve as key descriptors of the Universe's matter-energy content and its impact on expansion dynamics. The energy density captures the total energy per unit volume, covering contributions from matter, radiation and dark energy. Pressure represents the force exerted by these components within a given volume, impacting the rate of expansion or contraction. Linked through the Friedmann equations, these quantities govern the curvature and evolution of space-time. As dark energy becomes dominant, the pressure turns negative, driving an accelerated expansion phase, while energy density generally remains positive. This negative pressure, unlike that of matter or radiation, induces the Universe's late-time acceleration, marking a pivotal shift from earlier decelerated expansion.

Through equations (\ref{19}), (\ref{20}), (\ref{33}), (\ref{35}) and (\ref{37}), we determine the energy density and pressure associated with our models, leading to\\
$\bullet$ Model \ref{sec5.1}:
\begin{equation}\label{41}
\rho=3\alpha H_{0}^{2}(1+z)^{3(1+\omega_{0}-\omega_{1})}exp[3\omega_{1}z],
\end{equation}
\begin{eqnarray}\label{42}
p&=&-3\alpha H_{0}^{2}(1+z)^{3(1+\omega_{0}-\omega_{1})}exp[3\omega_{1}z]+2\alpha H_{0}^{2}exp[3\omega_{1}z]\bigg[\frac{3(1+\omega_{0}-\omega_{1})}{2}(1+z)^{3(1+\omega_{0}-\omega_{1})}\\\nonumber
&&+\frac{3\omega_{1}}{2}(1+z)^{4+3\omega_{0}-3\omega_{1}}\bigg].
\end{eqnarray}
$\bullet$ Model \ref{sec5.2}:
\begin{equation}\label{43}
\rho=3\alpha H_{0}^{2}(1+z)^{3(1+\omega_{0})}(1+z^{2})^{\frac{3\omega_{1}}{2}},
\end{equation}
\begin{eqnarray}\label{44}
p&=&-3\alpha H_{0}^{2}(1+z)^{3(1+\omega_{0})}(1+z^{2})^{\frac{3\omega_{1}}{2}}+2\alpha H_{0}^{2}\bigg[\frac{3(1+\omega_{0})}{2}(1+z)^{3(1+\omega_{0})}(1+z^{2})^{\frac{3\omega_{1}}{2}}\\\nonumber
&&+\frac{3\omega_{1}}{2}(1+z)^{4+3\omega_{0}-3\omega_{1}}\bigg].
\end{eqnarray}
$\bullet$ Model \ref{sec5.3}:
\begin{equation}\label{45}
\rho=3\alpha H_{0}^{2}(1+z)^{\frac{3(2+2\omega_{0}+\omega_{1})}{2}}(1+z^{2})^{\frac{3\omega_{1}}{4}}exp\bigg[-\frac{3\omega_{1}}{2}tan^{-1}z\bigg],
\end{equation}
\begin{eqnarray}\label{46}
p&=&-3\alpha H_{0}^{2}(1+z)^{\frac{3(2+2\omega_{0}+\omega_{1})}{2}}(1+z^{2})^{\frac{3\omega_{1}}{4}}exp\bigg[-\frac{3\omega_{1}}{2}tan^{-1}z\bigg]+2\alpha H_{0}^{2}(1+z)^{4+3\omega_{0}+\frac{\omega_{1}}{2}}\\\nonumber 
&&(1+z^{2})^{\frac{3\omega_{1}}{4}}exp\bigg[-\frac{3\omega_{1}}{2}tan^{-1}z\bigg]\bigg(\frac{3(2+2\omega_{0}+\omega_{1})}{4(1+z)}+\frac{3\omega_{1}z}{4(1+z^{2})}-\frac{3\omega_{1}}{4(1+z^{2})}\bigg).
\end{eqnarray}
\begin{figure}[htbp]
    \centering
    \begin{subfigure}{0.45\textwidth}
        \includegraphics[width=\linewidth]{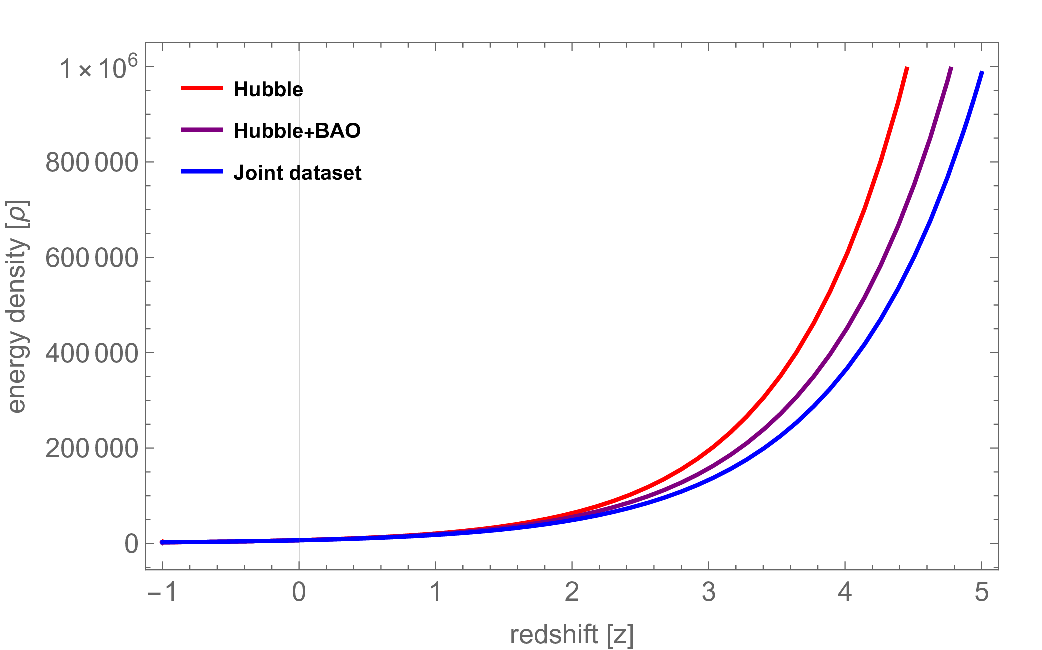} 
        \caption{}
        \label{fig:subfig19}
    \end{subfigure}
    \hfill
    \begin{subfigure}{0.45\textwidth}
        \includegraphics[width=\linewidth]{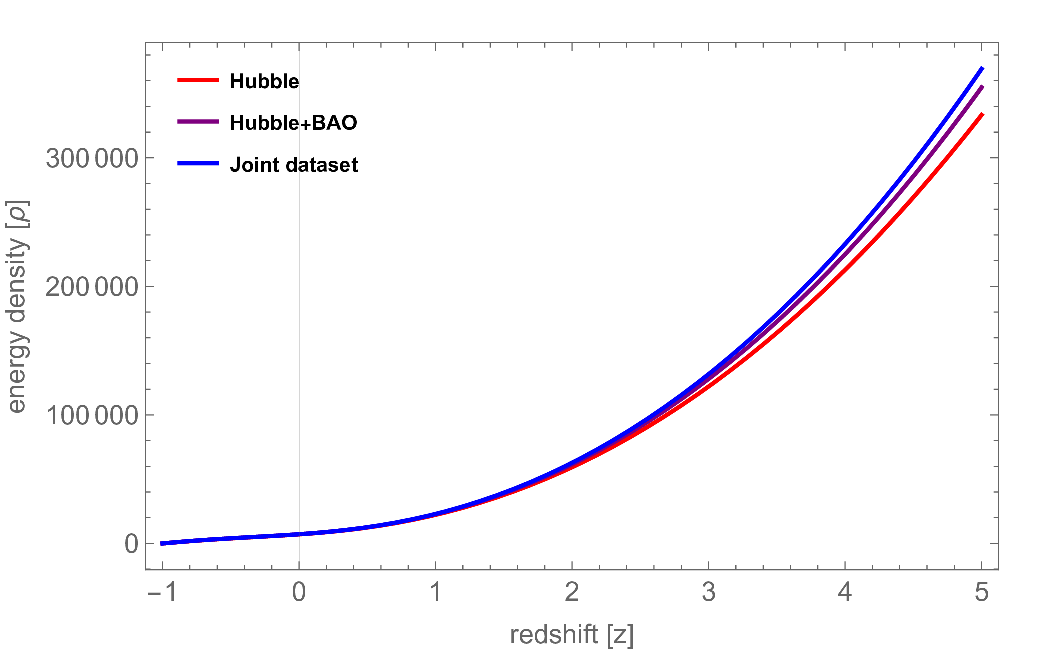} 
        \caption{}
        \label{fig:subfig20}
    \end{subfigure}

    \begin{subfigure}{0.45\textwidth}
        \includegraphics[width=\linewidth]{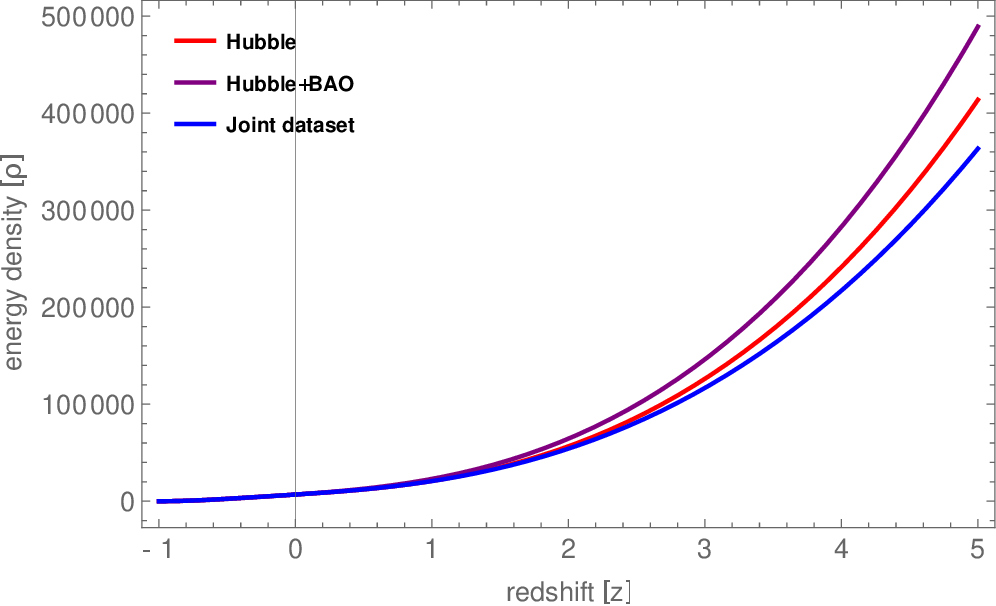} 
        \caption{}
        \label{fig:subfig21}
    \end{subfigure}

    \caption{The plot of $\rho$ vs. $z$ for the constrained parameter values of (a) Model \ref{sec5.1}, (b) Model \ref{sec5.2} and (c) Model \ref{sec5.3} with $\alpha=0.5$.}
    \label{fig:f8}
\end{figure}
\begin{figure}[htbp]
    \centering
    \begin{subfigure}{0.45\textwidth}
        \includegraphics[width=\linewidth]{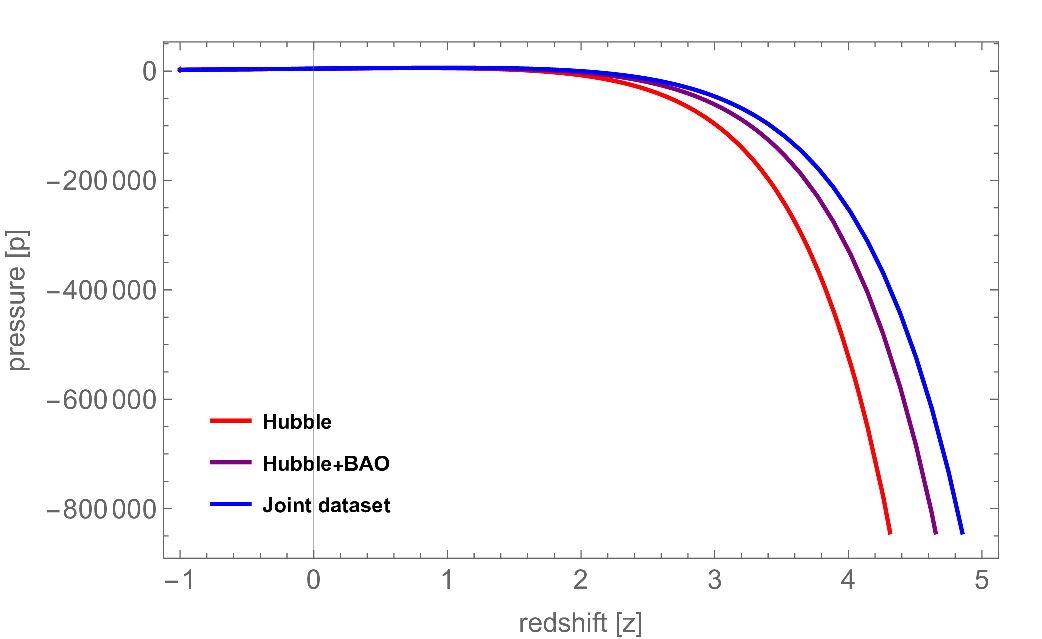} 
        \caption{}
        \label{fig:subfig22}
    \end{subfigure}
    \hfill
    \begin{subfigure}{0.45\textwidth}
        \includegraphics[width=\linewidth]{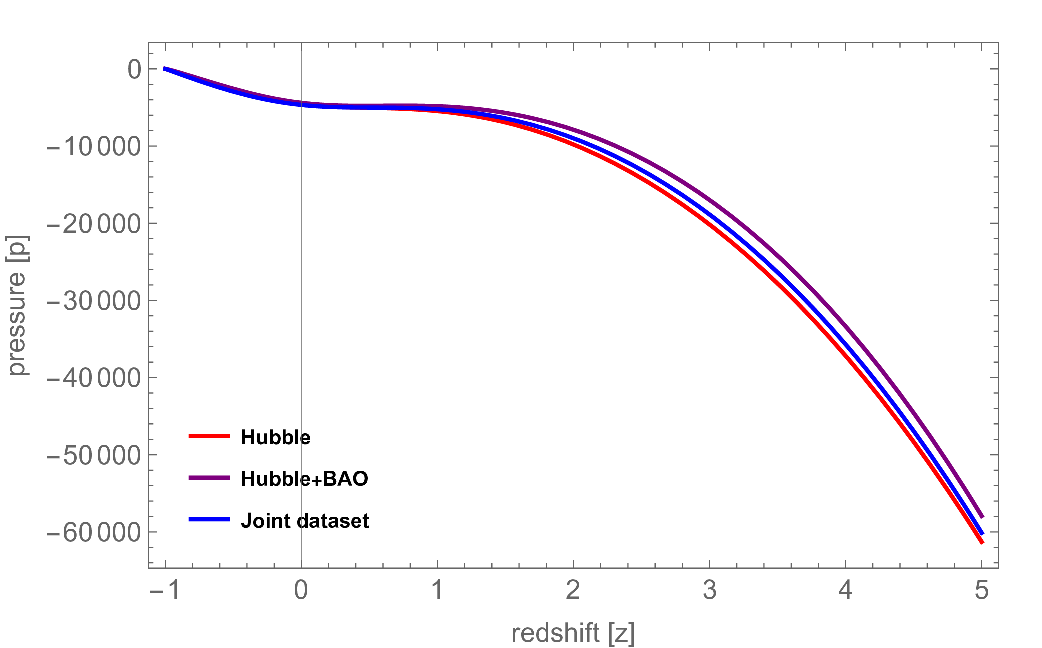} 
        \caption{}
        \label{fig:subfig23}
    \end{subfigure}

    \begin{subfigure}{0.45\textwidth}
        \includegraphics[width=\linewidth]{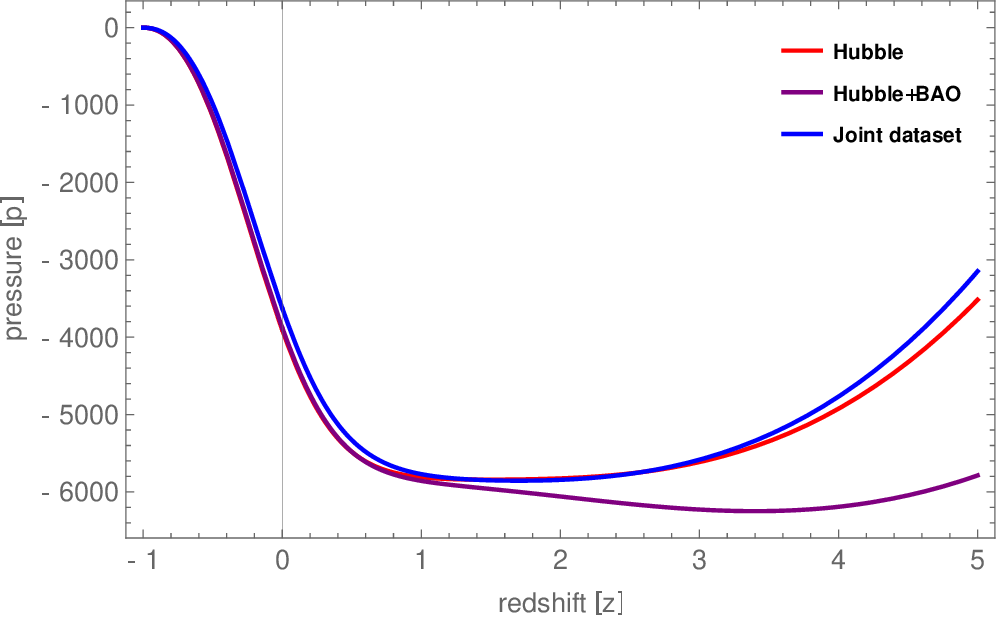} 
        \caption{}
        \label{fig:subfig24}
    \end{subfigure}

    \caption{The plot of $p$ vs. $z$ for the constrained parameter values of (a) Model \ref{sec5.1}, (b) Model \ref{sec5.2} and (c) Model \ref{sec5.3} with $\alpha=0.5$.}
    \label{fig:f9}
\end{figure}

In Figures \ref{fig:subfig19}, \ref{fig:subfig20} and \ref{fig:subfig21}, the plot of $\rho(z)$ against redshift $z$ illustrates that the energy density starts at high positive values at early times (high $z$) and gradually decreases as the Universe expands. As $z$ approaches $-1$ (representing the distant future), the energy density trends closer to zero. This pattern aligns with the expected shift from an early Universe with a high energy density to one in which energy density decreases over time, reflecting the dynamics of cosmic evolution.

In Figures \ref{fig:subfig22}, \ref{fig:subfig23} and \ref{fig:subfig24}, the plot of pressure $p(z)$ shows that it begins at large negative values for high redshift, signifying an early decelerating phase. As redshift decreases, the pressure gradually increases toward zero, which aligns with the accelerated expansion observed in the current Universe. This trend of negative pressure is essential in modeling the acceleration phase, driven by dark energy in the present cosmological model. With the parameter $\alpha = 0.5$, chosen to fit observational data, we see how this parameter influences both the energy density and pressure dynamics, reflecting a Universe with positive energy density and negative pressure—key indicators of dark energy’s role in accelerating expansion.
\subsection{Equation of state parameter}\label{sec6.3}
\hspace{0.6cm} The EoS parameter provides a vital link between pressure $p$ and energy density $\rho$ in cosmological models. It characterizes various phases of the Universe's evolution: for instance, during the dust-dominated era, the EoS parameter is $\omega=0$, indicating pressureless matter. In contrast, the radiation-dominated phase has an EoS parameter of $\omega=\frac{1}{3}$. The behavior of $\Lambda$CDM model is represented by $\omega=-1$, reflecting a constant energy density as the Universe expands. The EoS parameter also captures the accelerating phase of the Universe, characterized by $\omega<-\frac{1}{3}$. This includes the phantom regime, where $\omega<-1$,  suggesting an instability leading to potential future singularities, and the quintessence regime, where $-1<\omega<-\frac{1}{3}$, indicating a dynamic form of dark energy that evolves over time. This classification emphasizes the diverse roles that different forms of energy play in shaping the Universe's expansion history.
\begin{figure}[htbp]
    \centering
    \begin{subfigure}{0.45\textwidth}
        \includegraphics[width=\linewidth]{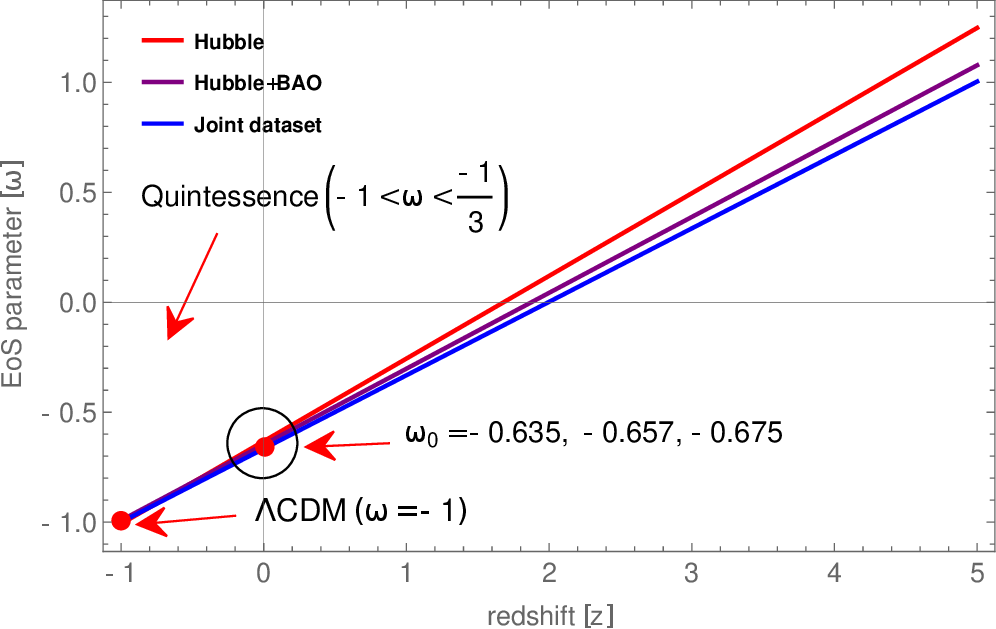} 
        \caption{}
        \label{fig:subfig25}
    \end{subfigure}
    \hfill
    \begin{subfigure}{0.45\textwidth}
        \includegraphics[width=\linewidth]{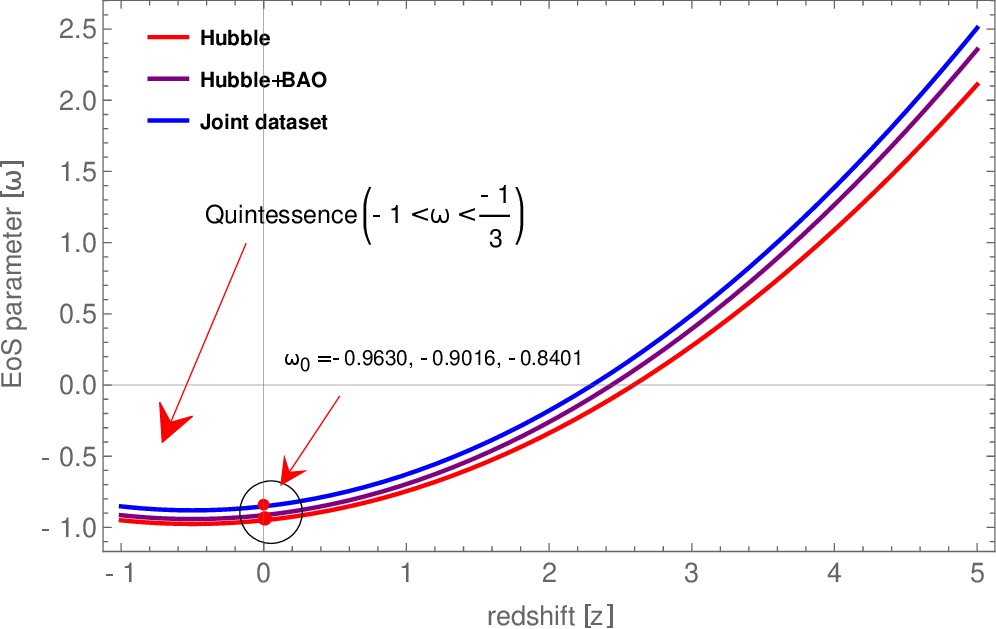} 
        \caption{}
        \label{fig:subfig26}
    \end{subfigure}

    \begin{subfigure}{0.45\textwidth}
        \includegraphics[width=\linewidth]{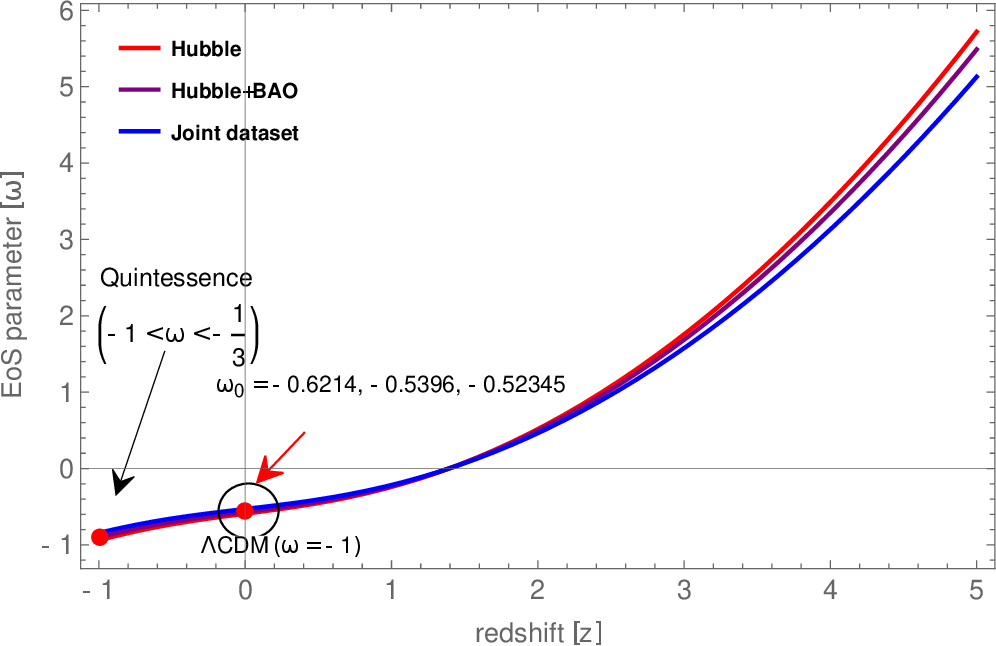} 
        \caption{}
        \label{fig:subfig27}
    \end{subfigure}

    \caption{The plot of $\omega$ vs. $z$ for the constrained parameter values of (a) Model \ref{sec5.1}, (b) Model \ref{sec5.2} and (c) Model \ref{sec5.3}.}
    \label{fig:f10}
\end{figure}

As depicted in Figures \ref{fig:subfig25}, \ref{fig:subfig26} and \ref{fig:subfig27}, the evolution of $\omega$ provides significant insights into the dynamic behavior of the Universe over time. Initially, at high redshift (early times), $\omega$ starts with a positive value indicating a matter-dominated phase. As redshift decreases, $\omega$ transitions to negative values, reflecting the increasing influence of dark energy and a shift towards accelerated expansion. In the far future, as $z$ approaches $-1$, $\omega$ asymptotically approaches $-1$, indicating that the Universe will evolve into a dark energy-dominated phase, reminiscent of the $\Lambda$CDM model. 

The present values of the EoS parameter $\omega_{0}$ derived from our models for the Hubble, Hubble+BAO and Hubble+BAO+Pantheon datasets are: For Model \ref{sec5.1}: $\omega_{0}=-0.635,-0.657$ and $-0.675$, For Model \ref{sec5.2}: $\omega_{0}=-0.9630,-0.9016$ and $-0.8401$ and For Model \ref{sec5.3}: $\omega_{0}=-0.6214,-0.5396$ and $-0.52345$. These values reflect a transition into dark energy dominance, with Model \ref{sec5.1} and \ref{sec5.3} values being relatively close to $-0.7$, which is commonly associated with quintessence models. The values from Model \ref{sec5.2} show a strong negative trend, suggesting a more pronounced dark energy component that could hint at phantom behavior in certain conditions \cite{Singh23}.
\subsection{Energy conditions}\label{sec6.4}
\hspace{0.6cm} Energy conditions in General Relativity (GR) and analogous spacetime theories establish essential guidelines for the stress-energy tensor of matter, enforcing non-negativity of energy density values. These conditions are probed using a three-fold approach: geometric, physical and effective, to uncover their essential characteristics. From a geometric perspective, these conditions are described using tensors like the Ricci and Weyl tensors, whereas in physical terms, they exert direct influence on the stress-energy tensor \cite{JK23}. Research conducted by \cite{Curel17,Kon20} explores a range of energy conditions, such as WEC, NEC, SEC and DEC, shedding light on their individual and collective roles in understanding the behavior of matter and energy. These energy conditions are formulated as mathematical inequalities, specifying constraints on energy density and pressure: WEC: $\rho\geq0$, $\rho+p\geq0$, NEC: $\rho+p\geq0$, SEC: $\rho+3p\geq 0$ and DEC: $\rho-p\geq 0$.
\begin{figure}[htbp]
    \centering
    \begin{subfigure}{0.4\textwidth}
        \includegraphics[width=\linewidth]{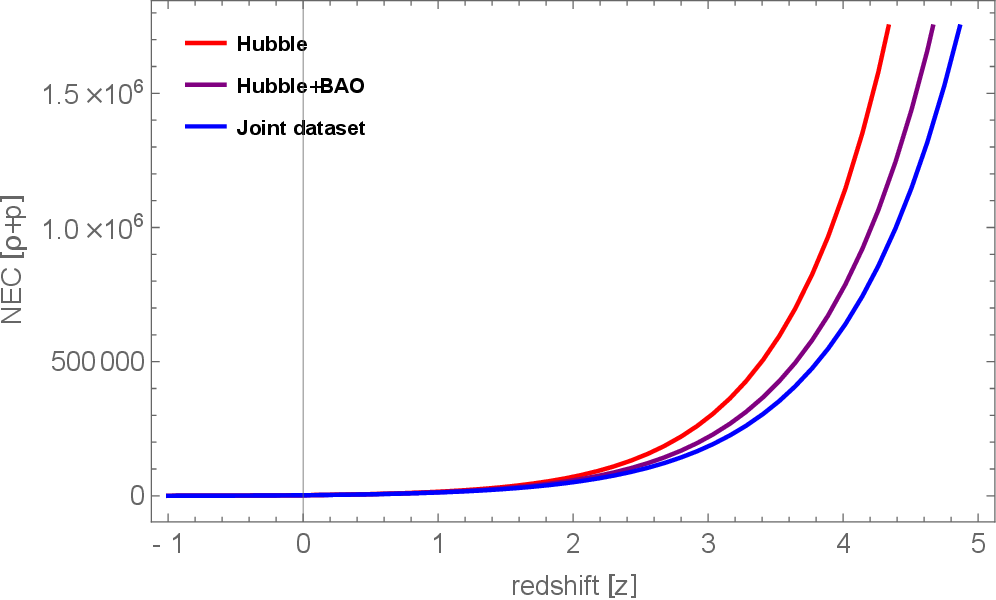} 
        \caption{}
        \label{fig:subfig28}
    \end{subfigure}
    \hfill
    \begin{subfigure}{0.4\textwidth}
        \includegraphics[width=\linewidth]{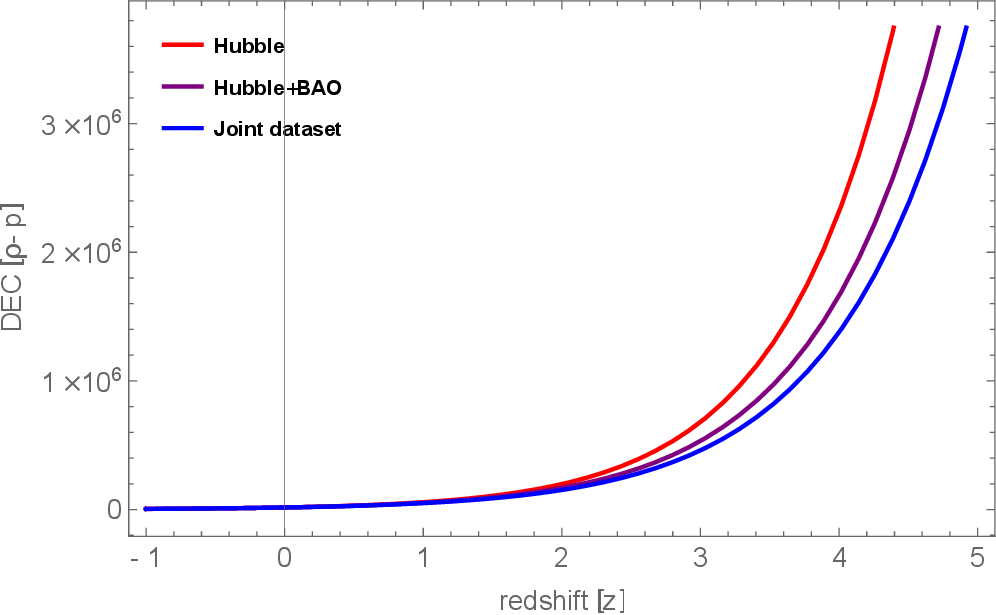} 
        \caption{}
        \label{fig:subfig29}
    \end{subfigure}

    \begin{subfigure}{0.4\textwidth}
        \includegraphics[width=\linewidth]{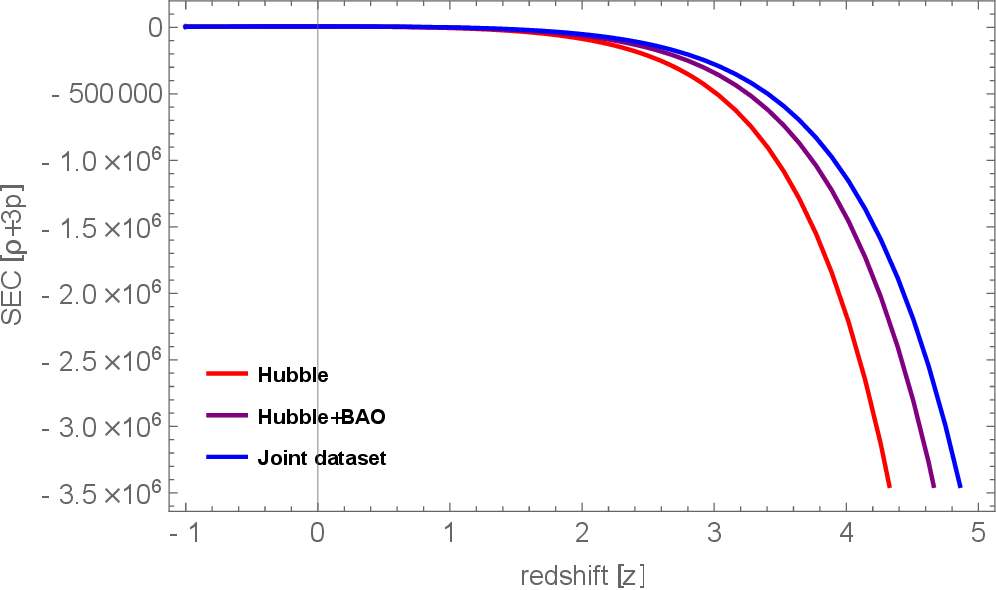} 
        \caption{}
        \label{fig:subfig30}
    \end{subfigure}

    \caption{Energy conditions vs. $z$ for the constrained parameter values of Model \ref{sec5.1} with $\alpha=0.5$.}
    \label{fig:f11}
\end{figure}
\begin{figure}[htbp]
    \centering
    \begin{subfigure}{0.4\textwidth}
        \includegraphics[width=\linewidth]{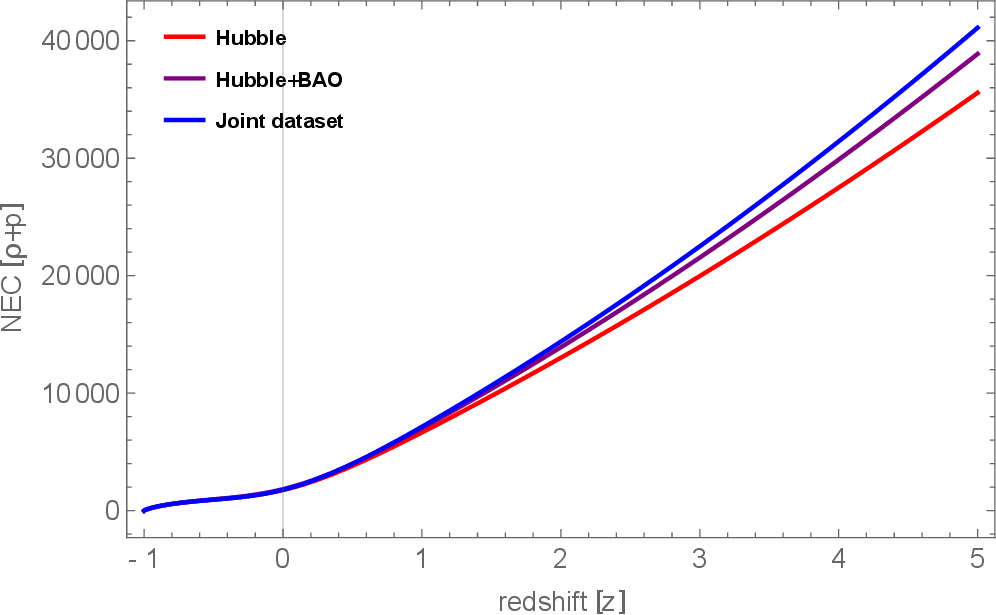} 
        \caption{}
        \label{fig:subfig31}
    \end{subfigure}
    \hfill
    \begin{subfigure}{0.4\textwidth}
        \includegraphics[width=\linewidth]{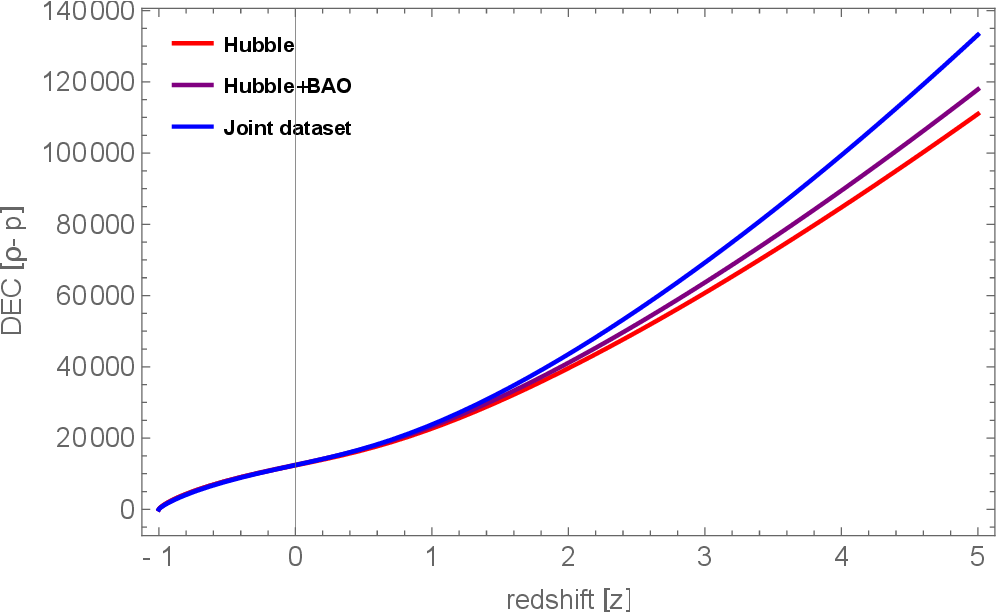} 
        \caption{}
        \label{fig:subfig32}
    \end{subfigure}

    \begin{subfigure}{0.4\textwidth}
        \includegraphics[width=\linewidth]{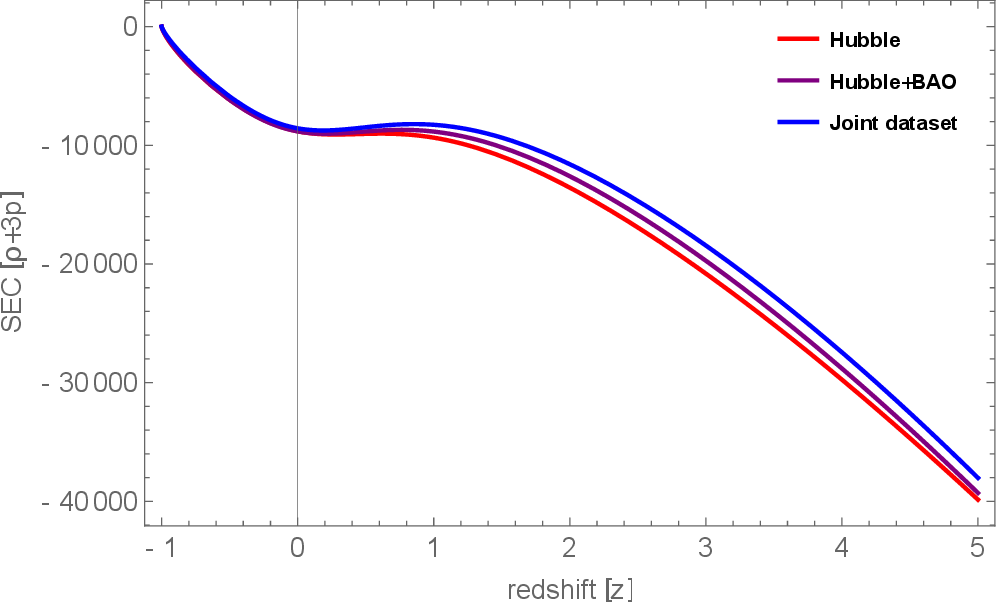} 
        \caption{}
        \label{fig:subfig33}
    \end{subfigure}

    \caption{Energy conditions vs. $z$ for the constrained parameter values of Model \ref{sec5.2} with $\alpha=0.5$.}
    \label{fig:f12}
\end{figure}
\begin{figure}[htbp]
    \centering
    \begin{subfigure}{0.4\textwidth}
        \includegraphics[width=\linewidth]{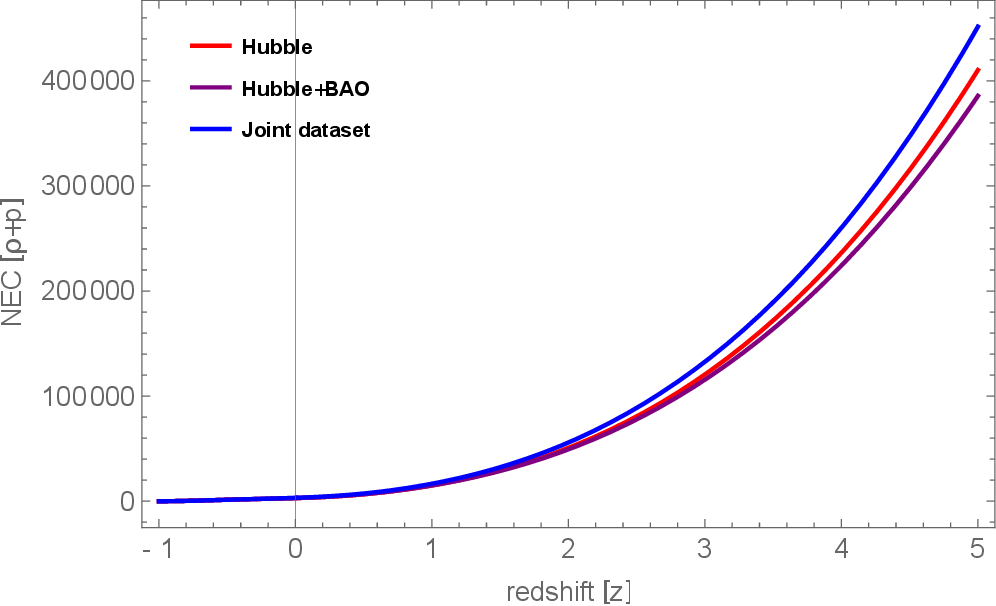} 
        \caption{}
        \label{fig:subfig34}
    \end{subfigure}
    \hfill
    \begin{subfigure}{0.4\textwidth}
        \includegraphics[width=\linewidth]{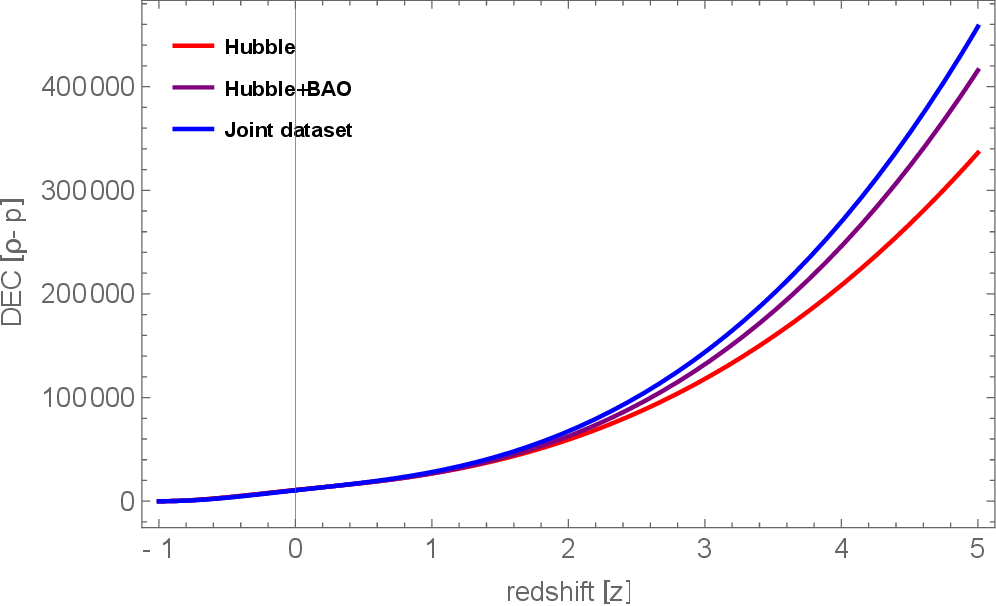} 
        \caption{}
        \label{fig:subfig35}
    \end{subfigure}

    \begin{subfigure}{0.4\textwidth}
        \includegraphics[width=\linewidth]{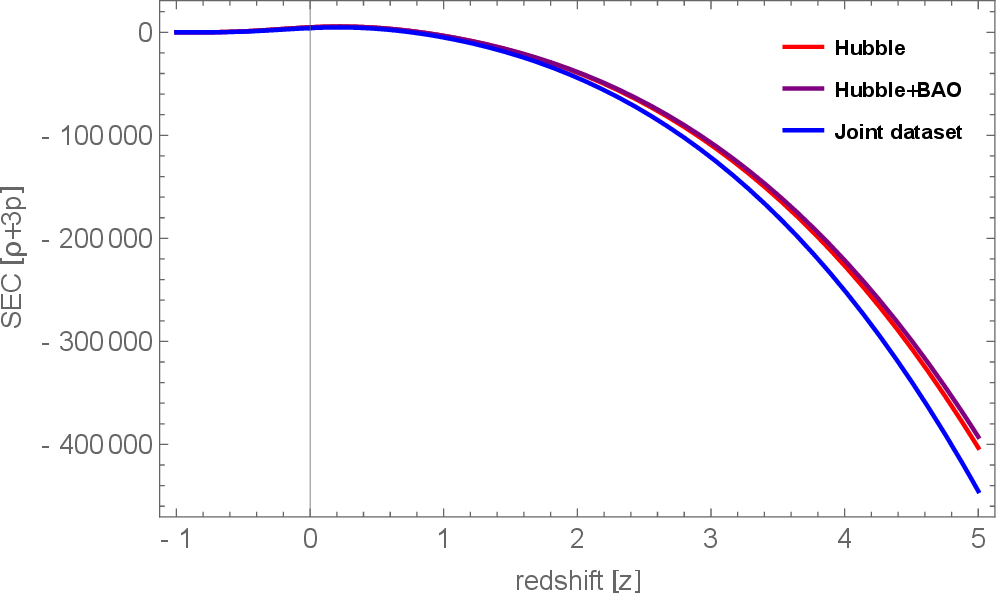} 
        \caption{}
        \label{fig:subfig36}
    \end{subfigure}

    \caption{Energy conditions vs. $z$ for the constrained parameter values of Model \ref{sec5.3} with $\alpha=0.5$.}
    \label{fig:f13}
\end{figure}

The expressions of the energy conditions for our models are derived by using the equations (\ref{41})-(\ref{46}) as follows:\\
$\bullet$ Model \ref{sec5.1}:
\begin{equation}\label{47}
\rho+p=2\alpha H_{0}^{2}exp[3\omega_{1}z]\bigg[\frac{3(1+\omega_{0}-\omega_{1})}{2}(1+z)^{3(1+\omega_{0}-\omega_{1})}+\frac{3\omega_{1}}{2}(1+z)^{4+3\omega_{0}-3\omega_{1}}\bigg]\geq0,
\end{equation}
\begin{eqnarray}\label{48}
\rho-p&=&6\alpha H_{0}^{2}(1+z)^{3(1+\omega_{0}-\omega_{1})}exp[3\omega_{1}z]-2\alpha H_{0}^{2}exp[3\omega_{1}z]\bigg[\frac{3(1+\omega_{0}-\omega_{1})}{2}(1+z)^{3(1+\omega_{0}-\omega_{1})}\\\nonumber
&&+\frac{3\omega_{1}}{2}(1+z)^{4+3\omega_{0}-3\omega_{1}}\bigg]\geq0,
\end{eqnarray}
\begin{eqnarray}\label{49}
\rho+3p&=&-6\alpha H_{0}^{2}(1+z)^{3(1+\omega_{0}-\omega_{1})}exp[3\omega_{1}z]+6\alpha H_{0}^{2}exp[3\omega_{1}z]\bigg[\frac{3(1+\omega_{0}-\omega_{1})}{2}(1+z)^{3(1+\omega_{0}-\omega_{1})}\\\nonumber
&&+\frac{3\omega_{1}}{2}(1+z)^{4+3\omega_{0}-3\omega_{1}}\bigg]\geq0,
\end{eqnarray}
$\bullet$ Model \ref{sec5.2}:
\begin{equation}\label{50}
\rho+p=2\alpha H_{0}^{2}\bigg[\frac{3(1+\omega_{0})}{2}(1+z)^{3(1+\omega_{0})}(1+z^{2})^{\frac{3\omega_{1}}{2}}+\frac{3\omega_{1}}{2}(1+z)^{4+3\omega_{0}-3\omega_{1}}\bigg]\geq0,
\end{equation}
\begin{eqnarray}\label{51}
\rho-p&=&6\alpha H_{0}^{2}(1+z)^{3(1+\omega_{0})}(1+z^{2})^{\frac{3\omega_{1}}{2}}-2\alpha H_{0}^{2}\bigg[\frac{3(1+\omega_{0})}{2}(1+z)^{3(1+\omega_{0})}(1+z^{2})^{\frac{3\omega_{1}}{2}}\\\nonumber
&&+\frac{3\omega_{1}}{2}(1+z)^{4+3\omega_{0}-3\omega_{1}}\bigg]\geq0,
\end{eqnarray}
\begin{eqnarray}\label{52}
\rho+3p&=&-6\alpha H_{0}^{2}(1+z)^{3(1+\omega_{0})}(1+z^{2})^{\frac{3\omega_{1}}{2}}+6\alpha H_{0}^{2}\bigg[\frac{3(1+\omega_{0})}{2}(1+z)^{3(1+\omega_{0})}(1+z^{2})^{\frac{3\omega_{1}}{2}}\\\nonumber
&&+\frac{3\omega_{1}}{2}(1+z)^{4+3\omega_{0}-3\omega_{1}}\bigg]\geq0,
\end{eqnarray}
$\bullet$ Model \ref{sec5.3}:
\begin{eqnarray}\label{53}
\rho+p&=&+2\alpha H_{0}^{2}(1+z)^{4+3\omega_{0}+\frac{\omega_{1}}{2}}\\\nonumber 
&&(1+z^{2})^{\frac{3\omega_{1}}{4}}exp\bigg[-\frac{3\omega_{1}}{2}tan^{-1}z\bigg]\bigg(\frac{3(2+2\omega_{0}+\omega_{1})}{4(1+z)}+\frac{3\omega_{1}z}{4(1+z^{2})}-\frac{3\omega_{1}}{4(1+z^{2})}\bigg)\geq0,
\end{eqnarray}
\begin{eqnarray}\label{54}
\rho-p&=&6\alpha H_{0}^{2}(1+z)^{\frac{3(2+2\omega_{0}+\omega_{1})}{2}}(1+z^{2})^{\frac{3\omega_{1}}{4}}exp\bigg[-\frac{3\omega_{1}}{2}tan^{-1}z\bigg]-2\alpha H_{0}^{2}(1+z)^{4+3\omega_{0}+\frac{\omega_{1}}{2}}\\\nonumber 
&&(1+z^{2})^{\frac{3\omega_{1}}{4}}exp\bigg[-\frac{3\omega_{1}}{2}tan^{-1}z\bigg]\bigg(\frac{3(2+2\omega_{0}+\omega_{1})}{4(1+z)}+\frac{3\omega_{1}z}{4(1+z^{2})}-\frac{3\omega_{1}}{4(1+z^{2})}\bigg)\geq0,
\end{eqnarray}
\begin{eqnarray}\label{55}
\rho+3p&=&-6\alpha H_{0}^{2}(1+z)^{\frac{3(2+2\omega_{0}+\omega_{1})}{2}}(1+z^{2})^{\frac{3\omega_{1}}{4}}exp\bigg[-\frac{3\omega_{1}}{2}tan^{-1}z\bigg]+6\alpha H_{0}^{2}(1+z)^{4+3\omega_{0}+\frac{\omega_{1}}{2}}\\\nonumber 
&&(1+z^{2})^{\frac{3\omega_{1}}{4}}exp\bigg[-\frac{3\omega_{1}}{2}tan^{-1}z\bigg]\bigg(\frac{3(2+2\omega_{0}+\omega_{1})}{4(1+z)}+\frac{3\omega_{1}z}{4(1+z^{2})}-\frac{3\omega_{1}}{4(1+z^{2})}\bigg)\geq0.
\end{eqnarray}

Cosmologists have found that violating the SEC is closely linked to the Universe's accelerating expansion, as this violation marks a departure from the sole dominance of attractive gravity and is frequently linked to the enigmatic presence of dark energy. The observational evidence suggests that while the NEC, WEC and DEC are consistently maintained, the SEC is frequently breached during periods of accelerated expansion. This pattern is consistent with the role of dark energy, which imposes negative pressure and drives the acceleration of the Universe. The Figures \ref{fig:f11}, \ref{fig:f12} and \ref{fig:f13} demonstrate that the SEC is negative at all redshifts $z$, in contrast to the NEC, WEC and DEC, which exhibit positive values, supporting a stable matter-energy configuration. This finding emphasizes that SEC violations are a marker of accelerated expansion and that NEC, WEC and DEC maintain the physical integrity and consistency of matter-energy distributions in the expanding Universe.
\subsection{Statefinder parameters}\label{sec6.5}
\hspace{0.6cm} The examination of geometric parameters represents a quest for understanding the fundamental nature of reality, inviting us to transcend the boundaries of the standard $\Lambda$CDM framework and contemplate the deeper truths of the Universe. Beyond the Hubble parameter $(H)$ and deceleration parameter $(q)$, other parameters play a crucial role in discriminating between competing cosmological models, offering a more nuanced understanding of the Universe's evolution. Higher-order derivatives of the scale factor $a(t)$ serve as a powerful tool for probing the Universe's evolution, yielding fresh insights into the cosmic expansion. The statefinder diagnostics, $\{r,s\}$, introduced by \cite{Sahni03,Sai03}, are geometric parameters that have proven to be effective probes for analyzing the behavior of different dark energy models, providing valuable insights into their underlying dynamics. This approach provides a means to trace the evolution of dark energy, offering a more refined understanding of its role in the Universe's expansion. The statefinder pair $\{r,s\}$ is defined as follows:
\begin{equation}\label{56}
  r=\frac{\dddot a}{aH^{3}}=2q^{2}+q-\frac{\dot{q}}{H},
\end{equation}
\begin{equation}\label{57}
  s=\frac{(r-1)}{3(q-\frac{1}{2})}. \bigg(q\neq\frac{1}{2}\bigg)
\end{equation}
By mapping dark energy models onto the $\{r,s\}$ phase space, their statefinder values serve as a fingerprint, distinguishing one model from another and in the case of the Chaplygin gas model, its parameters are confined to a specific region defined by $r>1$ and $s<0$. The $\Lambda$CDM model serves as a reference point, fixed at $r=1$, $s=0$. Quintessence models generally fall into the region with $r<1$ and $s>0$, reflecting their specific evolutionary traits. Holographic dark energy is represented by a fixed point at $r=1$ and $s=\frac{2}{3}$, while the standard cold dark matter (CDM) model aligns with $r=1$ and $s=1$. This phase space classification allows for clear distinctions between these models and aids in understanding their individual cosmic expansion behaviors.
\begin{figure}[hbt!]
    \centering
    \begin{subfigure}[b]{0.3\textwidth}
        \includegraphics[scale=0.4]{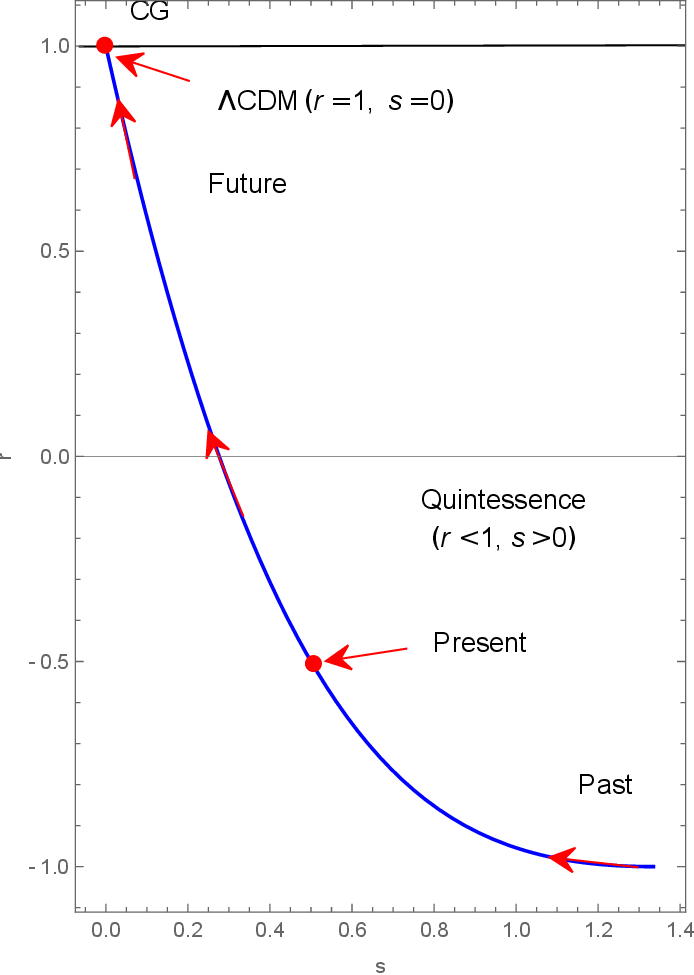}
        \caption{}
        \label{fig:subfig37}
    \end{subfigure}
    \hfill
    \begin{subfigure}[b]{0.28\textwidth}
        \includegraphics[scale=0.4]{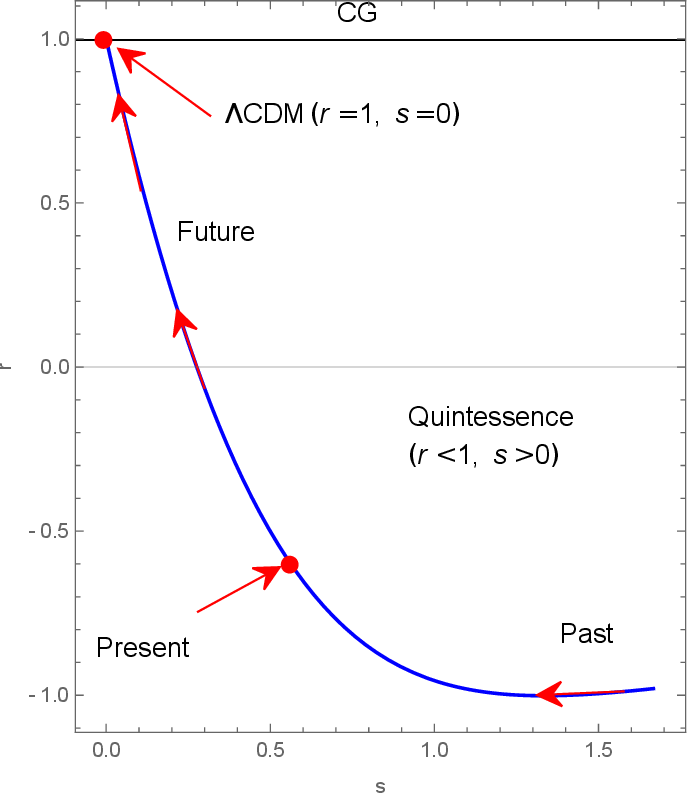}
        \caption{}
        \label{fig:subfig38}
    \end{subfigure}
    \hfill
    \begin{subfigure}[b]{0.26\textwidth}
        \includegraphics[scale=0.4]{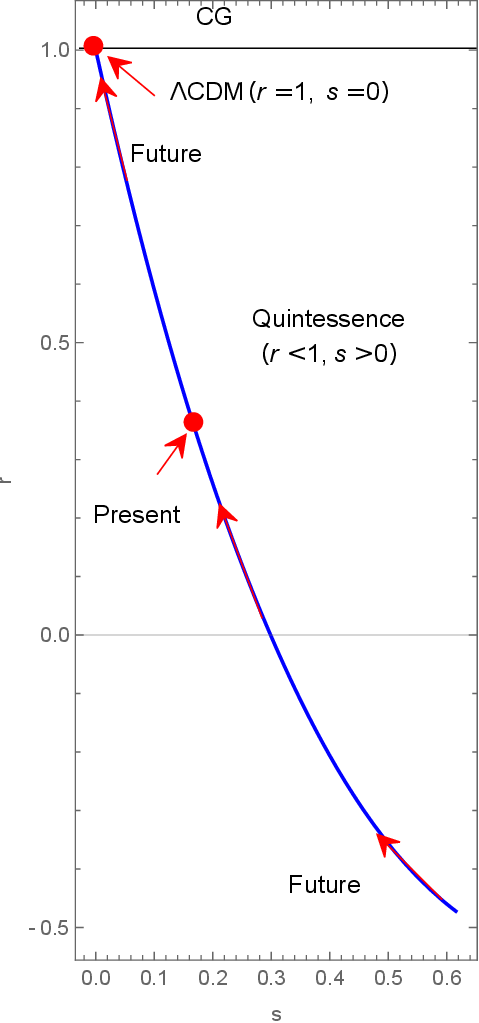}
        \caption{}
        \label{fig:subfig39}
    \end{subfigure}

    \caption{r-s plane for (a) Model \ref{sec5.1}, (b) Model \ref{sec5.2} and (c) Model \ref{sec5.3} by using the constrained parameter values from Hubble+BAO+Pantheon datasets.}
    \label{fig:f14}
\end{figure}
\begin{figure}[hbt!]
    \centering
    \begin{subfigure}[b]{0.3\textwidth}
        \includegraphics[scale=0.4]{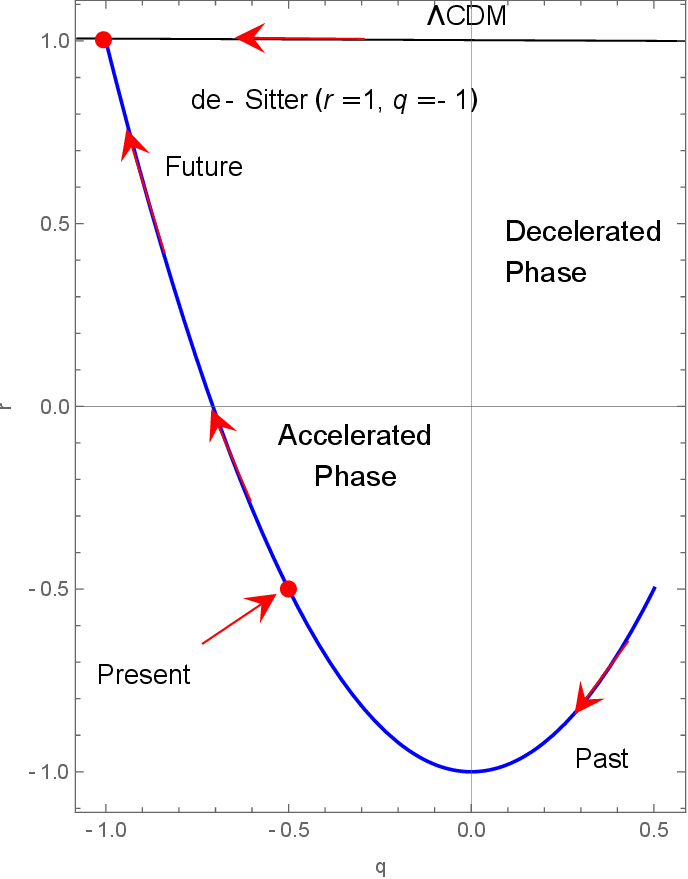}
        \caption{}
        \label{fig:subfig40}
    \end{subfigure}
    \hfill
    \begin{subfigure}[b]{0.28\textwidth}
        \includegraphics[scale=0.4]{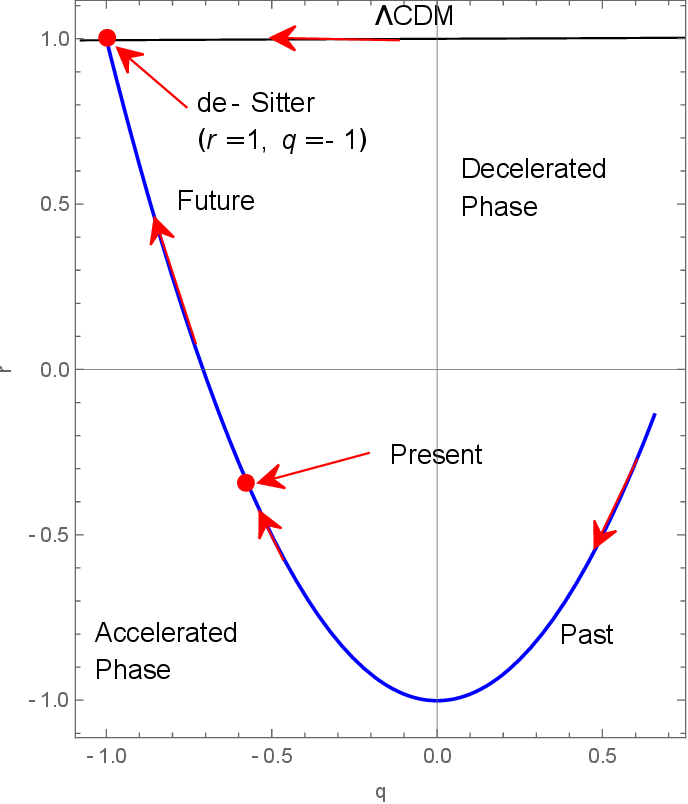}
        \caption{}
        \label{fig:subfig41}
    \end{subfigure}
    \hfill
    \begin{subfigure}[b]{0.26\textwidth}
        \includegraphics[scale=0.4]{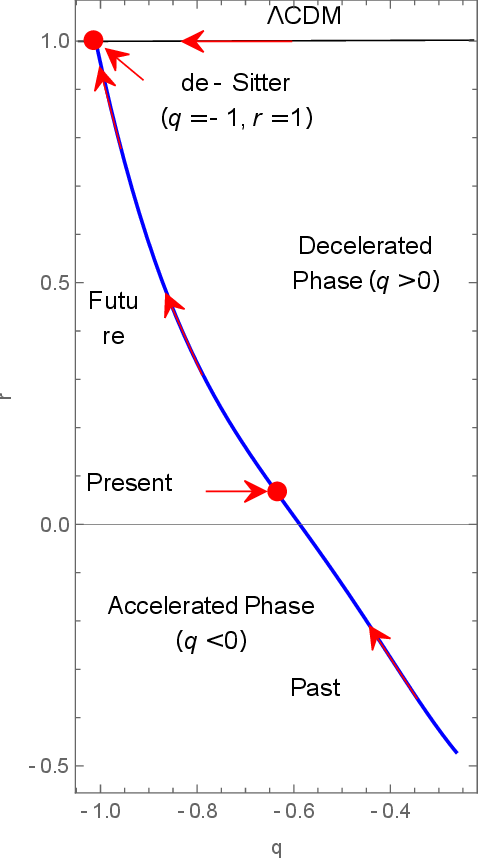}
        \caption{}
        \label{fig:subfig42}
    \end{subfigure}

    \caption{r-q plane for (a) Model \ref{sec5.1}, (b) Model \ref{sec5.2} and (c) Model \ref{sec5.3} by using the constrained parameter values from Hubble+BAO+Pantheon datasets.}
    \label{fig:f15}
\end{figure}

The statefinder parameters $\{r,s\}$ in Figure \ref{fig:f14} trace out a trajectory that decodes the evolution of our $f(Q,C)$ model across multiple cosmic epochs, offering a glimpse into its dynamical behavior. At the earliest phase, the model resides in a Quintessence-like regime, defined by $r<1$ and $s>0$, implying that the Universe's expansion is driven by a dynamic and evolving dark energy entity. The model's evolution leads it toward the $\Lambda$CDM fixed point at $r=1$ and $s=0$, representing a cosmological constant-dominated regime, analogous to a de Sitter Universe, where the expansion becomes exponential and dark energy drives the acceleration. This transition indicates the model’s ability to replicate current observations. The statefinder parameters $\{r_{0},s_{0}\}$ have been computed for three model variants, yielding distinct present values: for Model \ref{sec5.1}, with $\{r_{0},s_{0}\}=(-0.502,0.501)$, the parameters fall within the Quintessence-like regime suggesting a time-varying dark energy component, for Model \ref{sec5.2}, with $\{r_{0},s_{0}\}=(-0.600092,0.56262)$ also place it firmly in the Quintessence region. The slightly lower $r$ and higher $s$ values compared to Model \ref{sec5.1} imply an even stronger dynamic dark energy component. This matches some recent observational analyses that suggest dark energy may not be perfectly constant but could have mild dynamics, particularly at recent redshifts. For Model \ref{sec5.3}, with $\{r_{0},s_{0}\}=(0.0671,0.2709)$, this model’s values are closer to the $\Lambda$CDM point but still indicate slight deviations. This suggests a model that is nearly constant but allows for some dynamic behavior in dark energy. These calculated values of $\{r_{0},s_{0}\}$ fall within ranges that are consistent with current observational constraints, as they capture a spectrum from Quintessence-like dynamics (for Models \ref{sec5.1} and \ref{sec5.2}) to a $\Lambda$CDM-like scenario for Model \ref{sec5.3}. 

To interpret our results on the $r-q$ plane, we analyze the trajectories of Figure \ref{fig:f15} for Models \ref{sec5.1}, \ref{sec5.2} and \ref{sec5.3} and highlight how they align with or differ from other typical dark energy behaviors. The curves for models \ref{sec5.1} and \ref{sec5.2} start in the region where $q>0$ and $r<0$. This starting position contrasts with the standard cold dark matter (SCDM) phase. However, as these models evolve, their trajectories move downward through the $r<1$ and $q<0$ region, which aligns with the Quintessence-like phase, marking a transition from decelerated to accelerated expansion. This trajectory highlights the dynamic dark energy component in these models, supporting an evolving equation of state rather than a constant $\Lambda$-like behavior. At the end, the trajectories approach $r=1$ and $q=-1$ aligns our models with a stable, accelerating expansion phase, similar to a de Sitter Universe, though with variations depending on each model's trajectory. But for Model \ref{sec5.3}, the curve begins in the region where $q<0$ and $r<0$, suggesting an initial state dominated by dark energy-induced acceleration, differing from both SCDM and early Quintessence phases. The initial position of Model \ref{sec5.3} implies that it may describe a Universe with early dark energy effects that immediately accelerate expansion. As it evolves, Model \ref{sec5.3}’s trajectory approaches the de Sitter-like point $(q=-1, r=1)$, indicating an asymptotic approach to a steady-state expansion phase.
\subsection{Stability analysis}\label{sec6.6}
\hspace{0.6cm} To assess the stability of a dark energy model, we examine the behavior of the sound speed squared, $c_{s}^{2}$. The model's stability is guaranteed when $c_{s}^{2}>0$, as this ensures that small perturbations will fade away rather than escalate, leading to a stable evolutionary trajectory. Conversely, if $c_{s}^{2}<0$, the model is considered unstable, as perturbations are likely to amplify, leading to possible instabilities in the dark energy component. This approach enables researchers to gauge the stability of DE models by testing their response to minor fluctuations, offering a straightforward yet insightful way to determine whether they can maintain consistent behavior without evolving into unstable or divergent states. The formula for finding $c_{s}^{2}$ is: $\frac{dp}{d\rho}$. 
\begin{figure}[htbp]
    \centering
    \begin{subfigure}{0.4\textwidth}
        \includegraphics[width=\linewidth]{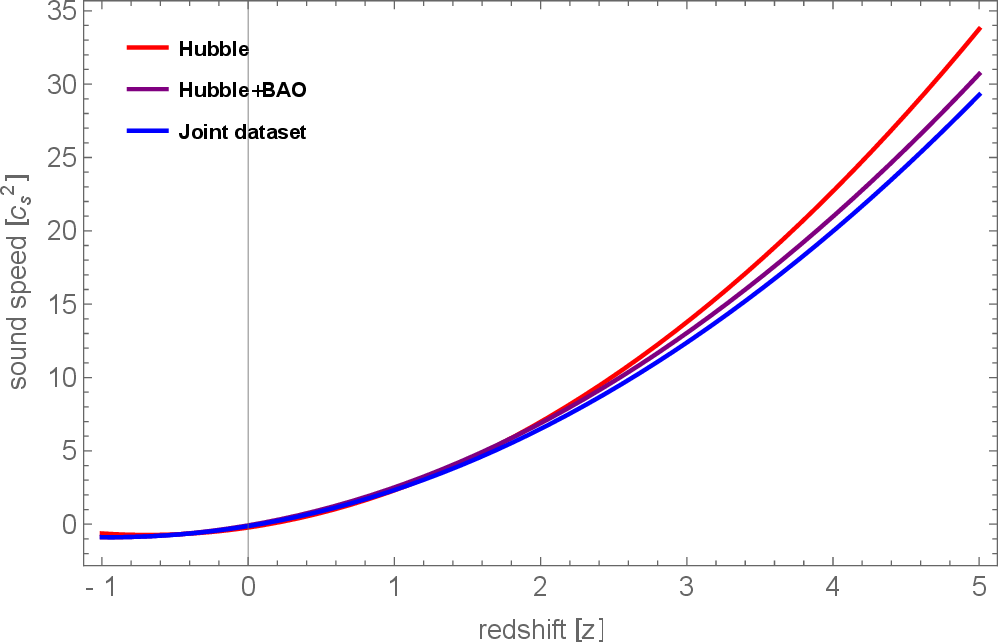} 
        \caption{}
        \label{fig:subfig43}
    \end{subfigure}
    \hfill
    \begin{subfigure}{0.4\textwidth}
        \includegraphics[width=\linewidth]{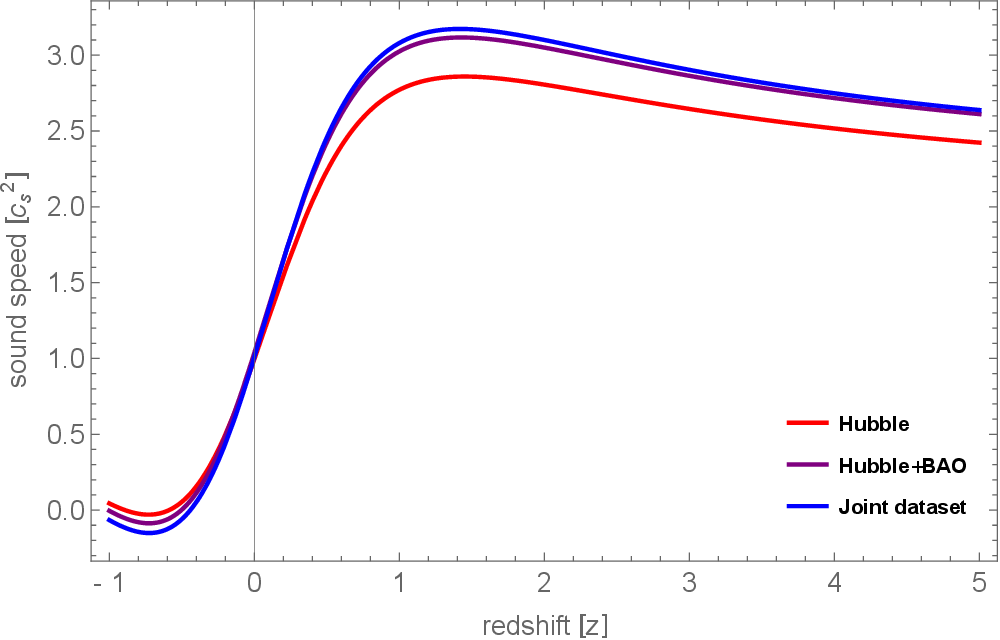} 
        \caption{}
        \label{fig:subfig44}
    \end{subfigure}

    \begin{subfigure}{0.4\textwidth}
        \includegraphics[width=\linewidth]{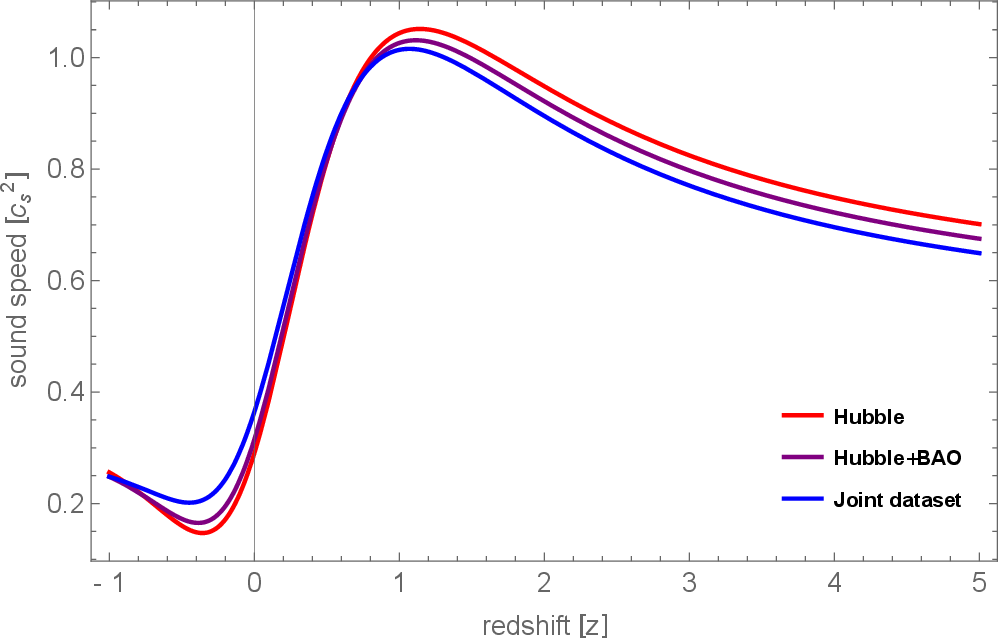} 
        \caption{}
        \label{fig:subfig45}
    \end{subfigure}

    \caption{$c_{s}^{2}$ vs. $z$ for the constrained parameter values of (a) Model \ref{sec5.1}, (b) Model \ref{sec5.2} and (c) Model \ref{sec5.3} with $\alpha=0.5$.}
    \label{fig:f16}
\end{figure}

For our models, we obtain the expressions of $c_{s}^{2}$ by using the equations (\ref{41})-(\ref{46}) in the following manner:\\
$\bullet$ Model \ref{sec5.1}:
\begin{equation}\label{58}
c_{s}^{2}=-1+\frac{2\omega_{0}(1+z)\bigg[\frac{3(1+\omega_{0}-\omega_{1})}{2}+\frac{3\omega_{1}(1+z)}{2}\bigg]+\frac{2}{3}\bigg[\frac{9(1+\omega_{0}-\omega_{1})^{2}}{2}+\frac{3\omega_{1}(4+3\omega_{0}-3\omega_{1})}{2}
(1+z)\bigg]}{3(1+\omega_{0}-\omega_{1})+3\omega_{1}(1+z)},
\end{equation}
$\bullet$ Model \ref{sec5.2}:
\begin{equation}\label{59}
c_{s}^{2}=-1+\frac{\frac{2}{3}\bigg[\frac{9(1+\omega_{0})^{2}}{2}+\frac{9(1+\omega_{0})\omega_{1}(1+z)}{4(1+z^{2})}+\frac{3(4+3\omega_{0})\omega_{1}z(1+z)}{2(1+z^{2})}+\frac{3\omega_{1}(1+z)^{2}}{2(1+z^{2})}
+\frac{3\omega_{1}(1+z)^{2}}{2(1+z^{2})}\bigg(\frac{3\omega_{1}-2}{2}\bigg)\bigg]}{3(1+\omega_{0})+\frac{3\omega_{1}(1+z)}{2(1+z^{2})}},
\end{equation}
$\bullet$ Model \ref{sec5.3}:
\begin{align}\label{60}
c_{s}^{2} &= -1 + \bigg[\frac{\frac{2}{3} \bigg[ \frac{3(2 + 2\omega_{0} + \omega_{1})}{4(1 + z)} \frac{3\omega_{1} z}{4(1 + z^{2})} - \frac{3\omega_{1}}{4(1 + z^{2})} \bigg]}{(3 + 3\omega_{0} + \frac{3\omega_{1}}{2}) + \frac{9\omega_{1}(1 + z)}{4(1 + z^{2})}} \nonumber \\
&\quad \times \bigg[(4 + 3\omega_{0} + \frac{3\omega_{1}}{2})(1 + z) + \frac{2(1 + z)^{2}}{3(1 + z^{2})} - \frac{\omega_{1}(1 + z)^{2}}{(1 + z^{2})^{2}} \bigg] \nonumber \\
&\quad + \frac{4(1 + z)^{2}}{9} \bigg[-\frac{3(2 + 2\omega_{0} + \omega_{1})}{4(1 + z)^{2}} + \frac{3\omega_{1}(1 - z^{2})}{4(1 + z^{2})^{2}} + \frac{3\omega_{1} z}{2(1 + z^{2})^{2}} \bigg]\bigg] \Bigg/ \bigg[(3 + 3\omega_{0} + \frac{3\omega_{1}}{2}) + \frac{9\omega_{1}(1 + z)}{4(1 + z^{2})}\bigg]
\end{align}

In our analysis of the sound speed $c_{s}^{2}$ for the different dark energy models, we observe distinct behaviors as the redshift $z$ changes. For both Models \ref{sec5.1} and \ref{sec5.2}, the sound speed starts at a higher value and gradually decreases as $z$ decreases, approaching zero. The high initial sound speed suggests that these models exhibit good stability in the early Universe when the energy density is higher. As the Universe expands and redshift increases, the decrease in $c_{s}^{2}$ suggests that the effective pressure of the dark energy diminishes, leading to a slower propagation of perturbations. In contrast, Model \ref{sec5.3} starts with $c_{s}^{2}<1$  and exhibits an initial increase as $z$ decreases, reaching a peak at $z\approx1.04$ before decreasing again towards zero. This behavior implies that Model \ref{sec5.3} experiences a dynamic shift in stability during the cosmic expansion. All models satisfy the range $0<c_{s}^{2}<1$ confirming that they maintain causal propagation of perturbations and adhere to physical constraints. Current analyses indicate that models exhibiting a stable sound speed are favored in explaining the accelerated expansion of the Universe, especially those approaching the cosmological constant regime in a late-time context.
\section{Conclusion}\label{sec7}
\hspace{0.6cm} In this work, we investigated a modified gravity model $f(Q,C)$ with specific EoS parameterizations, aiming to better understand the nature of dark energy and the accelerated expansion of the Universe. Our approach involved analyzing different observational datasets and physical parameters associated with cosmological evolution, yielding the following key insights:

We explored three different parameterizations for the EoS: $(i)$ $\omega=\omega_{0}+\omega_{1}z$, $(ii)$ $\omega=\omega_{0}+\frac{\omega_{1}z(1+z)}{1+z^{2}}$ and $(iii)$ $\omega=\omega_{0}+\frac{\omega_{1}z^{2}}{1+z^{2}}$. By deriving the Hubble parameter $H(z)$ from the field equations for each model, we constrained the parameters using the Hubble, Hubble+BAO and Hubble+BAO+Pantheon datasets. These constraints enabled us to fine-tune our models for compatibility with observed cosmic expansion data, providing a foundational understanding of each model's behavior over time. The constrained values of the parameters are given in Tables \ref{Tab:T4}, \ref{Tab:T5} and \ref{Tab:T6}.

The results show a notable transition in the Universe's expansion, as the deceleration parameter $q(z)$ indicates a shift from deceleration to acceleration, a finding depicted in Figure \ref{fig:f7}. The present values of $q_{0}$ for each model confirmed this shift: Model \ref{sec5.1}: $q_{0}=-0.4554,-0.4811$ and $-0.4981$, Model \ref{sec5.2}: $q_{0}=-0.3737,-0.4179$ and $-0.4716$, Model \ref{sec5.3}: $q_{0}=-0.3291,-0.3842$ and $-0.3561$. These negative values at $z=0$ indicate an accelerating Universe across all models, consistent with current observations. This transition underpins the model's viability in explaining the observed late-time acceleration.

Our analysis of the energy density $\rho$, pressure $p$ and EoS parameter $\omega$ reveals a dual role for dark energy: a positive density that fuels the cosmic expansion, and a negative pressure that accelerates this expansion, especially in recent epochs, highlighting the intricate dynamics of dark energy's influence on the Universe's growth. The results for $\omega_{0}$ across different datasets and models reveal a spectrum of dark energy dynamics, with Models \ref{sec5.1} and \ref{sec5.2} exhibiting more pronounced variations in dark energy behavior, whereas Model \ref{sec5.3} aligns more closely with a cosmological constant scenario, indicating a more static dark energy nature.

Energy condition analysis reveals a stable matter-energy configuration in each model, with SEC violation supporting accelerated expansion and NEC, WEC and DEC satisfaction ensuring stability, aligning with theoretical requirements.

The statefinder analysis reveals that Model \ref{sec5.1} and \ref{sec5.2} exhibit Quintessence-like dark energy behavior with time-varying dynamics where $\{r_{0},s_{0}\}=(-0.502,0.501)$ and $\{r_{0},s_{0}\}=(-0.600092,0.56262)$, while $\{r_{0},s_{0}\}=(0.0671,0.2709)$ for Model \ref{sec5.3} shows relatively steady dark energy behavior with slight deviations towards $\Lambda$CDM, consistent with recent studies suggesting slight variations in dark energy over time, confirming each model's compatibility with observational data.

Sound speed analysis reveals stable dark energy behavior in all models within the range $0<c_{s}^{2}<1$, with Models \ref{sec5.1} and \ref{sec5.2} showing decreasing sound speed and Model \ref{sec5.3} exhibiting initial low sound speed, followed by stability, consistent with $\Lambda$CDM-like behavior.

This study showcases the adaptability of $f(Q,C)$ gravity in accommodating various dark energy scenarios and transitions in cosmic acceleration. Models \ref{sec5.1} and \ref{sec5.2}, exhibiting Quintessence-like characteristics, suggest a dynamic dark energy component, whereas Model \ref{sec5.3}'s proximity to $\Lambda$CDM supports a more static dark energy interpretation. Together, these models provide a robust framework for understanding the evolving nature of dark energy and the Universe's accelerating expansion, opening up new avenues for exploring the intricacies of cosmic acceleration.

\end{document}